\documentclass[11pt]{article}
\usepackage{jheppubmod}
\pdfoutput=1
\usepackage{mathtools}
\mathtoolsset{showonlyrefs}
\usepackage{psfrag}
\usepackage{array}
\usepackage{amssymb}
\usepackage{amsmath}
\usepackage{amsthm}
\usepackage{graphicx}
\usepackage{caption}
\usepackage[labelsep=quad]{subcaption}
\usepackage{epstopdf}
	
\usepackage{epsfig}
\usepackage[punctsep]{collref}

\newcommand{\psine}{\Psi^{\text{ne}}} 

\def\psinesub[#1]{\psine_{#1}}

\newcommand{\cop}[1]{#1}
\newcommand{\al}{\cop{A}} 
\newcommand{\alset}{{\cal A}} 
\newcommand{\hpsi}{{\cal H}_{\Psi}}

\newcommand{\alsetcaus}{{\cal A}^{\cal C}}
\newcommand{\albarsetcaus}{{\bar{\cal A}}^{\cal C}}
\newcommand{\op}{{\cal O}} 
\newcommand{\hcft}{\cop{H}} 

\def\ukrus{U_K}
\def\vkrus{V_K}

\def\tD{\widetilde{D}}
\def\Oright[#1]{\op_{#1}} 
\def\Oleft[#1]{\op_{L#1}} 
\def\anc{\cop{a}}

\def\an[#1]{\anc_{#1}} 
\def\aleft[#1]{\cop{a}_{L#1}} 
\def\arelr[#1]{\cop{a}^{\text{rel}}_{R#1}} 
\def\tphi{\widetilde{\cop{\phi}}}
\def\tzeta{\widetilde{\zeta}}

\def\ta{\widetilde{\cop{a}}}

\def\alen[#1]{\al_{\text{L}, #1}} 

\def\enrange[#1,#2]{{\cal R}_{#1}}
\def\dimrange[#1,#2]{{\cal D}_{#1}}
\def\hilb[#1]{{\cal H}_{#1}}

\def\projrange[#1, #2]{\cop{P}_{#1}}
\def\projh[#1]{\cop{P}_{\hilb[#1]}}

\def\proj[#1]{\cop{P}_{#1}}

\def\ang{\ell}

\def\rhor{r_0}
\def\tcaus{t_{\cal C}}
\def\tout{\tau}
\def\cutoffT{\vartheta}
\def\ircutoff{|\cutoffT|}
\def\pcaus{{\cal P}_{\cal C}}
\def\bcaus{{\cal B}_{\cal C}}
\def\wcaus{\wedge_{\cal C}}
\def\vcaus{\vee_{\cal C}}

\def\phicaus{\phi_{\cal C}}
\def\phicausprime{\phi_{\cal C'}}

\def\phihat{\widehat{\phi}}

\def\alcaus{\al^{\cal C}}
\def\alcausd{\al^{\cal C^{\dagger}}}
\def\ucaus{U^{\cal C}}
\def\ucauscomp{\widehat{V}^{\cal C}}
\def\ucauscompd{\widehat{V}^{\cal C^{\dagger}}}
\def\ump{U_{\text{MP}}}
\def\ucausd{U^{\cal C^{\dagger}}}
\def\albarcaus{\bar{\al}^{\cal C}}
\def\albarcausd{\bar{\al}^{\cal C^{\dagger}}}

\def\dmax{{\cal D}_{m}}
\def\dzero{D^0(\omega, \ang)}
\def\dminzero{D^0(-\omega, \ang)}

\def\tarelr[#1]{\widetilde{a}^{\text{rel}}_{R#1}}
\def\coeff[#1]{\alpha_{#1}}

\def\pb[#1,#2]{\{#1, #2\}}
\def\deb[#1,#2]{[#1,#2]_{\text{D.B.}}}
\def\tO{\widetilde{\op}}

\def\tr{{\rm Tr}}

\def\Or[#1]{{\text{O}}\left({#1}\right)}
\def\dotl[#1,#2]{\left\langle #1,\, #2 \right\rangle}
\def\dotlb[#1,#2]{\left\langle #1,\, #2 \right\rangle}
\def\dotlm[#1,#2]{\left[ #1,\, #2 \right]}
\def\dotp[#1,#2]{(\vect{#1} \cdot\vect{#2})}
\def\aff[#1,#2]{\hat{#1}(#2)}

\def\n4sym{{\cal N}=4 SYM}
\def\>{\rangle}
\def\<{\langle}
\def\weight[#1,#2,#3]{\{(#1),#2,#3\}}
\def\ads[#1]{$\text{AdS}_{#1}$}

\def\rtor{{r_*}}
\def\rtorp{r'_*}
\def\rtorsin{{\rtor}_s}
\def\rtors[#1]{r_{*#1}}

\collectsep[]{;}
\def\vbd{V_{\text{bd}}}
\def\vhor{V_{\text{hor}}}
\def\ghor{G_{\text{hor}}}
\newcommand{\be}{\begin{equation}}
\newcommand{\ee}{\end{equation}}
\newcommand{\ba}{\begin{align}}
\newcommand{\ea}{\end{align}}
\newcommand{\bs}{\begin{split}}

\def\sess\end{split}
\newcommand{\vect}[1]{{\vec{#1}}}

\title{Smooth Causal Patches for AdS Black Holes}
\author{Suvrat Raju}
\affiliation{International Centre for Theoretical Sciences, Tata Institute of Fundamental Research, Shivakote, Bengaluru 560089, India.}
\emailAdd{suvrat@icts.res.in}
\date{}
\abstract{
We review the paradox of low energy excitations about an AdS black hole. An appropriately chosen unitary operator in the boundary theory can create a locally strong excitation near the black hole horizon, whose global energy is small as a result of the gravitational redshift. The paradox is that 
this seems to violate a general rule of statistical mechanics, which states that an operator with energy parametrically smaller than $k T$ cannot create a significant excitation in a thermal system. When we carefully examine the position dependence of the boundary unitary operator that produces the excitation and the bulk observable necessary to detect the anomalously large effect, we find that they do not both fit in a single causal patch. This follows from a remarkable property of position space AdS correlators that we establish explicitly, and resolves the paradox in a generic state of the system, since no combination of observers can both create the excitation and observe its effect. As a special case of our analysis, we show how this resolves the ``Born rule'' paradox of \href{http://www.arxiv.org/abs/1506.01337}{\texttt{arXiv:1506.01337}} and we verify our solution using an independent calculation. We then consider boundary states that are finely tuned to display a spontaneous excitation outside the causal patch of the infalling observer,  and we propose a version of causal patch complementarity in AdS/CFT that resolves the paradox for such states as well.}
\keywords{AdS/CFT, Black hole interiors, Causal patches}
\begin{document}
\maketitle
\section{Introduction and summary of results}
Does a typical pure state with high enough energy in a holographic CFT correspond to a classical black hole geometry with a smooth interior? This question can be made precise as follows. Consider a CFT with a holographic dual  \cite{Maldacena:1997re,Gubser:1998bc,Witten:1998qj}, and in such a CFT pick a typical state $|\Psi \rangle$ from the set of states with energy much larger than the central charge of the CFT. A typical state is a linear combination of energy eigenstates with coefficients that have a random magnitude and phase. In such a state, can we map the local field degrees of freedom at a bulk point $x_i$ to some CFT operator $\phi(x_i)$? Moreover, if we compute multi-local correlators in the state $|\Psi \rangle$,
\be
\label{bulkcorrelator}
\langle \Psi | \phi(x_1) \ldots \phi(x_n) | \Psi \rangle,
\ee
do these correlators match those that one would obtain by quantizing a bulk field about the classical black hole background?

If the points $x_i$ are all outside the black hole horizon, then this question was answered constructively in \cite{Hamilton:2005ju,Hamilton:2006az,Hamilton:2007wj,Hamilton:2006fh,Papadodimas:2012aq}, by finding an explicit set of operators in the CFT, $\phi(x_i)$, whose correlators in a typical state match bulk correlators about a black hole background.  In \cite{ Papadodimas:2013wnh, Papadodimas:2013jku} we also proposed a representation of the black hole interior for typical states in the CFT. These constructions, taken together, provide strong evidence that the answer to the question above is affirmative --- typical high energy states of the CFT do correspond to a smooth black hole solution.

However, a novel feature of the construction of  \cite{ Papadodimas:2013wnh, Papadodimas:2013jku} was that operators inside the black hole were {\em state-dependent}. This feature was  necessary because the authors of  \cite{Almheiri:2012rt, Almheiri:2013hfa,Marolf:2013dba} pointed out that when some of the points $x_i$ were behind the horizon, it was impossible to find state-independent operators, $\phi(x_i)$, whose correlators matched with the correlators of a bulk black hole geometry with a smooth interior.  The authors of \cite{Almheiri:2012rt, Almheiri:2013hfa,Marolf:2013dba} suggested that typical pure states in the CFT have an exterior geometry that looks exactly like a classical black hole, but have no interior; state-dependence provides an alternative that preserves a smooth interior.

\paragraph{\bf The paradox of low energy excitations \\}
However, state-dependence does not automatically resolve all paradoxes associated with the black hole, and in \cite{Marolf:2015dia}, Marolf and Polchinski, building on previous work by Harlow \cite{Harlow:2014yoa}, presented one such paradox. They constructed a unitary operator, $\ump$, in the boundary theory that rotates the phases of the modes outside the black hole at a very small cost in energy. But since this phase rotation breaks the entanglement between the interior and the exterior, one might naively believe that when some points $x_i$ are behind the horizon, and others are outside, the correlator \eqref{bulkcorrelator} is modified sharply. 

This is unusual because in a typical state of a general statistical system, the change in the value of an observable, $\al_{\alpha}$, under the action of a  unitary excitation $U$, is bounded by
\be
\label{bornruleintro}
\delta \langle \al_{\alpha} \rangle \equiv |\langle \Psi | U^{\dagger} \al_{\alpha} U |\Psi \rangle - \langle \Psi | \al_{\alpha} |\Psi \rangle|  \leq 2 \sqrt{\beta \delta E} \sigma_{\alpha},
\ee
where $\delta E$ is the change in the expectation value of the Hamiltonian induced by the unitary, and the inequality holds in the limit where $\beta \delta E \ll 1$. Here $\sigma_{\alpha}$ is a measure of the fluctuations of the operator $\al_{\alpha}$. We will review and prove this inequality in section \ref{secbornrule}.

If we take $\al_{\alpha}$ to be the correlator in \eqref{bulkcorrelator} with some points behind the horizon and the unitary to be $\ump$ then it appears that this inequality is violated. This is a sign of what is a more general paradox in AdS/CFT. Consider the bulk dual to the state $U |\Psi \rangle$, where $U$ is some unitary operator, and use the bulk equations of motion to analyze the geometry in the past and the future. Then, as a result of the gravitational redshift it may happen that even though $U$ only injects  a small amount of global energy into the CFT state, the geometry obtained by the procedure above has a locally large excitation near the horizon. This is similar to the setup of \cite{Shenker:2013pqa,Shenker:2013yza} although one  difference is that we are interested in excitations that inject an energy $\beta \delta E \ll 1$ whereas \cite{Shenker:2013pqa,Shenker:2013yza} considered excitations where $\beta \delta E \sim 1$. Then by considering observables along the worldline of an infalling observer it might appear that we could violate \eqref{bornruleintro}. 

This paradox necessarily involves an infalling observer.  Intuitively, this is because an observer who stays outside the horizon, and tries to detect the strong excitation, has to go close to the horizon and then return and so cannot differentiate the excitation from the Unruh effect that he experiences \cite{Susskind:1993mu}. More formally, the bulk-boundary map involves only state-independent operators outside the horizon so all correlators evaluated purely outside the horizon, including the boundary correlators considered in \cite{Shenker:2013pqa,Shenker:2013yza},  automatically obey the constraints of \eqref{bornruleintro}. 

\paragraph{\bf Plan of the resolution \\}
In this paper, we address the paradox for the infalling observer as follows. First, we consider bulk observables that are confined to a single causal patch. We did not impose this restriction in our previous work  \cite{Papadodimas:2013wnh,Papadodimas:2013jku, Papadodimas:2015jra, Papadodimas:2015xma} but it is a natural restriction because while a theory might predict some values for observables that do not fit in any causal patch, these observables cannot actually be measured by any combination of observers.  

In {\em addition}, we divide the allowed excitations into two classes that we deal with separately.
\begin{enumerate}
\item
First in section \ref{secfullads}, we focus on the class of excitations that can be produced by  deforming the CFT Hamiltonian at time $t$  through 
\be
\label{simpledeform}
\hcft \rightarrow \hcft + J(t) \al_{\gamma}^J(t),
\ee
where $\al_{\gamma}^J(t)$ is some simple operator on the boundary that is localized at the same time $t$.\footnote{Note that we only turn on sources that are dual to boundary operators. This does not involve any loss of generality since all gauge invariant operators in the bulk theory that can be added to the Hamiltonian are contained in the boundary. Indeed, as we discuss below, a bulk source would violate local conservation of energy. Energy can only be thrown in from the boundary and then, in the bulk, it can only be redistributed subject to the Wheeler-de Witt constraints.} This  class of excitations is more restrictive than the excitations previously considered in \cite{Papadodimas:2013wnh,Papadodimas:2013jku,Papadodimas:2015jra, Papadodimas:2015xma}. In those papers, while we insisted that the bulk observer could only perform experiments involving simple operators, we did {\em not} insist that the deformation of the Hamiltonian be local. For example, we did not previously exclude excitations of the form $H \rightarrow H + \delta(t) \int d x\al^J_{\gamma}(x) \cos (\omega x)$. Such an excitation would involve acting with the {\em Fourier mode} of $\al^J_{\gamma}$. However, physically an observer at time $t$ does not have access to boundary operators at a later (or earlier) time. In fact, it is not difficult to show that if an observer could act with excitations from a later time, she could engineer violations of the second law
of thermodynamics. So we exclude excitations that are nonlocal in time in section \ref{secfullads}.  With this restriction, we will show that {\em bulk correlators in a causal patch obey \eqref{bornruleintro} for arbitrary excitations produced by sources of the form \eqref{simpledeform}. }
\item
In section \ref{seccomplement} we allow more general excitations where the Hamiltonian at time $t$ is deformed by operators that may be localized in the future of $t$. This returns us to the class of excitations considered in \cite{Papadodimas:2013wnh,Papadodimas:2013jku,Papadodimas:2015jra,Papadodimas:2015xma}. This analysis is closely related to the situation where the excitation is produced {\em spontaneously} in the theory and the observer simply happens to find herself in a state that is about to undergo an excitation after some time. 
However, as explained above, this larger set of excitations and states is of limited physical interest: black holes with spontaneous macroscopic excitations occupy an {\em exponentially small} volume in the Hilbert space, which contains states that violate the second law of thermodynamics. 
Nevertheless, in section \ref{seccomplement} we consider this possibility and advance a tentative proposal of ``causal patch complementarity'' to ensure that \eqref{bornruleintro} is preserved even in this situation.
\end{enumerate}

\paragraph{\bf Local sources and unitaries  on the boundary of a causal patch \\}
As explained above, the central thrust of the resolution advanced in this paper is that when the excitation is produced by a boundary source dual to local CFT operators, and when the observable in question is confined to a single causal patch,  we do not observe any violation of \eqref{bornruleintro}. However, as we now describe, it is more convenient to frame this result in terms of unitary excitations on the boundary of the causal patch that contains the observable.

The correct Heisenberg operators at time t, which we should use upon adding  a term $J(t) \al_{\gamma}^J(t)$ to the Hamiltonian, are given by
\be
\label{schwingerops}
\phi^{J}(t, \rtor, \Omega) = \overline{{\cal T}}\{e^{i \int_{\cutoffT}^t J(x) \al_{\gamma}(x) d x}\}  \phi(t, \rtor, \Omega) {\cal T}\{e^{-i \int_{\cutoffT}^t J(x) \al_{\gamma}(x) d x}\}.
\ee
where we have separately displayed the time, radial and spherical coordinates in the local field, and ${\cal T}$ indicates time-ordering whereas $\overline{\cal T}$ indicates anti-time-ordering.  We have assumed that the source has no support before a past cutoff $\cutoffT$. This formula follows from Schwinger's in-in formalism \cite{Weinberg:2005vy}. 

The operators \eqref{schwingerops} are somewhat inconvenient to work with directly. So, rather than working with the deformed Hamiltonian and the operators \eqref{schwingerops}, we instead consider correlators of operators evolved using the {\em undeformed Hamiltonian} but sandwiched in a state excited by a unitary of the form 
\be
\label{auxiliaryU}
U = \exp\big[-i \int_{\cutoffT}^{\tcaus} \al_{\alpha}(t) d t \big].
\ee
This means that we consider correlators 
\[
\langle \Psi | U^{\dagger} \phi(x_1) \ldots \phi(x_n) U | \Psi \rangle,
\]
where we restrict the Kruskal ``$\ukrus$'' coordinate,  corresponding to the points $x_i$,  to $\ukrus < e^{-{2 \pi \tcaus \over \beta}}$. (We use conventions where $\ukrus < 0$ outside the horizon, and $\ukrus > 0$ inside the horizon.) An infalling observer who jumps in at the time $\tcaus$ cannot reach any value of $\ukrus$ larger than this limit, even if he travels at the speed of light. 

Even though the problem above looks like it is different from the situation where we have turned on a source in the boundary, these two situations are closely related when the bulk observables fit in a single causal patch. We explain the logic briefly here and  direct the reader to subsection \ref{secsourcesunitaries} for further details.

It is a simple consequence of bulk causality that when we turn on a source dual to some local operator on the boundary, only the part of the source that is in the causal past of a bulk point can affect  the local field operator there.  In particular, if we consider a correlator with insertions in a single causal patch, and the largest $\ukrus$-value of the insertions is $e^{-{2 \pi \tcaus \over \beta}}$, then we can just ignore the part of the source that acts after time $\tcaus$. 

This is why the class of observables obtained by considering correlators in a causal patch in the presence of an {\em arbitrary} simple local source --- even one that continues for a long time in the future --- is contained in the class of observables obtained by considering correlators sandwiched between a unitary of the form \eqref{auxiliaryU} and its adjoint that live on the boundary of the causal patch. The precise map between observables in the two possible descriptions is given in section \ref{secsourcesunitaries}.

In the discussion below, we will often refer to the case where the ``unitary and the observable both fit in the same causal patch.'' The reader should remember this contains the case where we turn on arbitrary local sources and examine a bulk observable that fits in some single causal patch. 

\paragraph{\bf Some technical details of the resolution \\}
The central technical result that we prove in this paper is that given a unitary and a product of field operators, all of which fit in a single causal patch, we have
\be
\label{centralresult}
\langle \Psi | U^{\dagger} \phi(x_1) \ldots \phi(x_n) U | \Psi \rangle - \langle \Psi | \phi(x_1) \ldots \phi(x_n)  | \Psi \rangle \leq 2 \sqrt{\beta \delta E} \sigma,
\ee
where $\delta E$ is the energy injected by the unitary $U$ and $\sigma$ is precisely the measure of fluctuations in this correlator that appears in \eqref{bornruleintro}. This is a very non-trivial and surprising property of AdS correlation functions, and section \eqref{secfullads} is devoted to establishing this result. We reiterate that by the argument above, \eqref{centralresult} implies that correlators of the operators \eqref{schwingerops} restricted to a single causal patch also obey the constraints of \eqref{bornruleintro}. 

The conclusion, therefore, is as follows: no combination of observers with finite powers can start with an equilibrium state, {\em create} an excitation $U$ and detect a correlator that violates \eqref{bornruleintro}. Since typical states of the form $|\Psi \rangle$ take up all but an exponentially small volume of the Hilbert space, and it is very reasonable to restrict to experiments  where we turn on a local source and observe its effect, this already resolves the paradox in almost all cases of interest.
  
In the original formulation of the paradox in \cite{Marolf:2015dia}, this resolution is not evident because both the excitation and the observable to be measured are framed in frequency-space. The fact that the excitation and the observable do not fit in a causal patch becomes apparent only when we examine these operators in position space.

\paragraph{\bf A specific example \\}
In section \ref{secparadox} we explicitly examine the specific unitary $\ump$ and a set of near-horizon correlators through an independent calculation that does not rely on the results of section \ref{secfullads}. We consider a position space version of $\ump$,
\be
\ump = \exp\Big[i \int_{\cutoffT}^{\tcaus} d t_1 d t_2 \int d \Omega_1 d \Omega_2  \op(t_1, \Omega_1) \, \op(t_2, \Omega_2) G(t_1, \Omega_1) G^*(t_2, \Omega_2) \Big],
\ee
where $G(t, \Omega) \propto e^{i \omega_0 t} Y_{\ang}(\Omega)$ and $\op(t, \Omega)$ is a single trace primary operator on the boundary. Nevertheless,  we show that if we restrict correlators in the state $\ump |\Psi \rangle$ to the causal patch that contains $\ump$, which, in particular, implies that we consider fields localized at points so that $\ukrus <  e^{-{2 \pi \tcaus \over \beta}}$ then such correlators do not differ significantly from correlators in the state  $|\Psi \rangle$.   

\paragraph{\bf Some speculations on causal patch complementarity \\}
In section \ref{seccomplement} we turn to a general class of excitations where the state is excited by an operator from the future before the bulk observer jumps in. Alternately, we can think of the situation where  an observer jumps into a CFT state that is about to undergo a spontaneous excitation, but hits the singularity before the light ray from the excitation on the boundary can reach him. 

Then it seems that, in principle, a superobserver with the ability  to prepare arbitrary states in the CFT could compare the experience of the infaller in such states to the experience of the infaller in states where the excitation is absent altogether and observe a violation of \eqref{bornruleintro}.

However, the result of section \ref{secfullads} lead us to an interesting idea:  it is possible to modify the standard map between bulk and boundary operators in such a way as to preserve \eqref{bornruleintro} even in such experiments. This is the hypothesis of ``causal patch complementarity.''

Before we explain the mathematical details, a rough analogy may help to explain this idea. In quantum field theory, a correlation function that is classically well defined requires the additional specification of a cutoff.  The idea of causal patch complementarity is that bulk correlators in spontaneously excited states require additional data --- about the perspective of the causal patch from which they are viewed --- before they are fixed completely. 

The precise proposal is as follows. For a large class of unitaries $U$ that we describe more precisely in section \ref{subsecstability}, we suggest that the field operators, $\phicaus$, that are appropriate for an infalling observer in such a state and in a particular causal patch satisfy
\be
\label{patchproposal}
\langle \Psi | U^{\dagger} \phicaus(x_1) \ldots \phicaus(x_n) U | \Psi \rangle = \langle \Psi | \ucausd \phi(x_1) \ldots \phi(x_n) \ucaus |\Psi \rangle,
\ee
where on the right hand side we have the usual field operators constructed using the standard bulk-boundary transfer function. The unitary $\ucaus$ is an ``attenuated'' version of the unitary $U$. It has the property that it is localized on the intersection of the causal patch with the boundary (which we term $\bcaus$) and for any other simple operator $\alcaus_{\alpha}$ that lives on $\bcaus$ we have $\langle \Psi | \ucausd \alcaus_{\alpha} \ucaus | \Psi \rangle = \langle \Psi|  U^{\dagger} \alcaus_{\alpha} U |\Psi \rangle$; so the unitaries $U$ and $\ucaus$ are indistinguishable using operators only on $\bcaus$.  We show that the description using the fields $\phicaus$ is consistent within the causal patch ${\cal C}$ and also that it obeys \eqref{bornruleintro}.  But it is not the same as the description that would be used by another observer who stays on the boundary long enough to see the excitation $U$. We suggest that these two descriptions are complementary. Our exploration of this proposal is not yet complete but we describe it in this paper, since it seems to point in some very interesting directions.

\paragraph{\bf Plan of the Paper \\}
This paper is structured as follows. In section \ref{secbornrule}, we  derive the inequality \eqref{bornruleintro} that shows that in a general statistical  system, it is not possible to definitively excite the system with energy less than $k T$. Section \ref{secbornrule} follows the ideas of \cite{Marolf:2015dia}, but  is self-contained, and may be considered a derivation of a simple result in statistical mechanics. In section \ref{secsetup} we describe our setup in holography and define our notation and various concepts that we use later. In section \ref{secfullads}, we prove that causal-patch correlators in AdS obey the constraints of statistical mechanics \eqref{bornruleintro}. Section \ref{secfullads} is technical, but contains the central results of this paper. In section \ref{secparadox}, we turn to a specific example to elucidate the  calculations of section \ref{secfullads}. We examine the ``Born rule'' paradox of \cite{Marolf:2015dia} and simply do a direct calculation, without using the results of section \ref{secfullads}, to show that this paradox disappears if we restrict the excitation and correlators to a single causal patch.  Finally, in section \ref{seccomplement}, we consider correlators in a causal patch but in a state that is excited with a unitary outside that patch; we explore the use of field operators that obey  \eqref{patchproposal} to describe physics in a particular causal patch in this state, and we show that this ensures that these correlators are also stable,  in that they are close to correlators in a state where the excitation is absent.

We emphasize, once again, that our calculations in section \ref{secfullads} and section \ref{secparadox} are logically independent of the discussion in section \ref{seccomplement}. The proof that no violation of \eqref{bornruleintro} can be seen if we perform experiments where we turn on a local source in an equilibrium state and then perform bulk observations in a causal patch, which is the setting for any reasonable experiment,  is an independent result that does not rely on our hypothesis of causal patch complementarity. 

The recent literature on the reconstruction of the bulk from the boundary, and the associated paradoxes, starting with the work of Mathur \cite{Mathur:2009hf} and AMPS \cite{Almheiri:2012rt},  is rather extensive \cite{Bousso:2012as,  Susskind:2012uw,  Mathur:2012jk, Chowdhury:2012vd, Susskind:2012rm, Bena:2012zi, Giveon:2012kp, Banks:2012nn, Ori:2012jx, Hossenfelder:2012mr, Hwang:2012nn, Avery:2012tf, Larjo:2012jt, Rama:2012fm, Page:2012zc,  Saravani:2012is, Jacobson:2012gh, Susskind:2013tg,  Kim:2013fv, Park:2013rm, Hsu:2013cw, Giddings:2013kcj,  Lee:2013vga, Avery:2013exa,   Kang:2013wda, Chowdhury:2013tza,Page:2013mqa,  Axenides:2013iwa,  Gary:2013oja, Chowdhury:2013mka, delaFuente:2013nba,  Barbon:2013nta,  Lloyd:2013bza, Hsu:2013fra, Page:2013dx,Giddings:2013noa,Mathur:2013qda,  Mathur:2013gua, Balasubramanian:2014gla,Freivogel:2014dca,Akhoury:2013bia,Lashkari:2014pna,Barbon:2014rma, Kabat:2016rsx,Guica:2015zpf,Haehl:2015foa, Lin:2015lfa, Bao:2015nca, Emelyanov:2015wna,  Roy:2015pga, Engelhardt:2015fwa, Emelyanov:2015nca, Verlinde:2013vja,Verlinde:2012cy,Verlinde:2013uja, Verlinde:2013qya, Nomura:2012sw,Brustein:2012jn,Brustein:2013ena,Brustein:2013xga,Brustein:2013qma,Silverstein:2014yza, Nomura:2012cx,Nomura:2012ex,Nomura:2013gna, braunstein2009v1, Harlow:2013tf,Susskind:2013lpa,Sivaramakrishnan:2016qdv}. We reviewed some of these developments in detail in \cite{Papadodimas:2015jra}.  The application of causal patches to the information paradox has also been explored in  \cite{Freivogel:2014fqa,Ilgin:2013iba,Banks:2014xja} and causal patches are also relevant in the  cosmological context \cite{Bousso:2006ev,Bousso:2006ge}.

A summary of the results of section \ref{secfullads}, with minimal technical details,  is presented in \cite{gravresearch2017}, which the reader may find helpful.

\section{Low energy excitations in a thermal system\label{secbornrule}}
In this section, we will derive a bound on the effect of a low energy excitation in a thermal system. Our discussion is not specific to AdS/CFT, and this can be taken to be a simple and general result in statistical mechanics. This derivation is motivated by the analysis of \cite{Marolf:2015dia}. 

We consider a  statistical system with a large number of degrees of freedom. In such a system, we consider the set of energy eigenstates around some  energy, $E$
in the energy range $E \pm \Delta$, with $\Delta \ll E$, and we denote the Hilbert space spanned by these states by $\hilb[E]$. In a generic statistical system, we expect that the dimension of this space is given by $\text{dim}(\hilb[E]) = e^{S}$ with $S \propto E$. We now consider a state $|\Psi \rangle$ that we pick using the Haar measure on this Hilbert space. Then we remind the reader that by using the eigenstate thermalization hypothesis \cite{Deutsch,srednicki1994chaos,srednicki1999approach}, for any coarse-grained observable $\al_{\alpha}$, we have
\be
\label{typicalthermal}
\langle \Psi | \al_{\alpha} | \Psi \rangle = {1 \over Z(\beta)}\tr(e^{-\beta H} \al_{\alpha}) + \Or[{1 \over \sqrt{S}}],
\ee
where $\beta = {\partial S \over \partial E}$ is the inverse of the effective temperature of the state and $Z(\beta)$ is the partition function at that temperature. So, for any coarse grained probe, the pure state $|\Psi \rangle$ is effectively thermal.

We now consider a low energy excitation of such a system. By this, we mean a unitary operator $U$ that has the property that for a generic state $|\Psi \rangle$, we have
\be
\label{lowenergyexc}
\langle \Psi | U^{\dagger} H U |\Psi \rangle - \langle \Psi | H | \Psi \rangle =  \delta E.
\ee
Note that a unitary cannot, on average, decrease the energy of a typical state and so---leaving aside very special unitaries that commute with the Hamiltonian---we have  $\delta E > 0$. We are interested in the situation where $\delta E \ll \beta$, and also $\delta E \ll \Delta$; so ${\delta E \over E} = \Or[{1 \over S}]$.   

We can also write \eqref{lowenergyexc} in the form
\be 
U: \hilb[E] \rightarrow \hilb[E + \delta E],
\ee
which just shows that the unitary maps us from the microcanonical ensemble at energy $E$ into an ensemble at energy $E + \delta E$. The map $U$ is an injective map. On the other hand, it is not surjective because the larger ensemble has dimension 
\be
\text{dim}(\hilb[E + \delta E]) = e^{S + {\partial S \over \partial E} \delta E} = e^{S + \beta \delta E}.
\ee
The image of $\hilb[E]$ under $U$ forms a subspace of $\hilb[E+\delta E]$ and we denote the projector onto this subspace by $\proj[U_E]$.  Now, let $|\Psi' \rangle$ be a typical state picked using the Haar measure on $\hilb[E+\delta E]$. We note that
\be
\begin{split}
&\langle \Psi' | \proj[U_E] |\Psi' \rangle  = 1 - \beta \delta E  + \Or[(\beta \delta E)^2], \\ 
&\langle \Psi' |(1 - \proj[U_E]) |\Psi' \rangle = \beta \delta E + \Or[(\beta \delta E)^{2}].
\end{split}
\ee
For use below, we define 
\be
\begin{split}
&|\Psi_E \rangle = \langle \Psi' | \proj[U_E] |\Psi' \rangle^{-1 \over 2} \proj[U_E] |\Psi' \rangle, \\
&|\Psi_o\rangle = \langle \Psi' | 1 - \proj[U_E] |\Psi' \rangle^{-1 \over 2} (1 - \proj[U_E]) |\Psi' \rangle, 
\end{split}
\ee
so that we can decompose 
\be
\label{decomprime}
|\Psi' \rangle = (1 - {\beta \delta E \over 2}) |\Psi_E \rangle + (\beta \delta E)^{1 \over 2} |\Psi_o\rangle + \Or[(\beta \delta E)^{3 \over 2}].
\ee

We also note that statistically, $|\Psi' \rangle$  is very similar to a typical state picked from $\hilb[E]$. This is because the difference in the temperature of the two ensembles is given by 
\be
\label{changetemp}
\delta \beta = {\partial \beta \over \partial E} \delta E = -\beta^2 {\delta E \over C_V} = 0+\Or[{1 \over S}],
\ee
where $C_V$ is the specific heat at constant volume, which is also macroscopic and scales like $S$. Then, using \eqref{typicalthermal} and \eqref{changetemp}, we see that
\be
\langle \Psi' | \al_{\alpha} |\Psi' \rangle = \langle \Psi | \al_{\alpha} | \Psi \rangle + \Or[{1 \over \sqrt{S}}].
\ee
This means that the expectation value of a coarse grained observable in a typical state from the larger ensemble is the same as the expectation value in the smaller ensemble. 

On the other hand, we can also calculate this expectation value using the decomposition \eqref{decomprime}, which leads to
\be
\langle \Psi' | \al_{\alpha} |\Psi' \rangle - \langle \Psi_E | \al_{\alpha} |\Psi_E \rangle = \sqrt{\beta \delta E} \left(\langle \Psi_E | \al_{\alpha} |\Psi_o \rangle  + \langle \Psi_o | \al_{\alpha}  | \Psi_E \rangle \right) + \Or[\beta \delta E].
\ee 
Since $|\Psi_E \rangle$ is a state obtained by acting with the unitary on a typical state of $\hilb[E]$, the absolute value of the left hand side can also be replaced by the change in the expectation value of $\al_{\alpha}$ under the action of the unitary:
\be
\delta \langle \al_{\alpha} \rangle \equiv |\langle \Psi | U^{\dagger} \al_{\alpha} U |\Psi \rangle - \langle \Psi | \al_{\alpha} |\Psi \rangle|.
\ee
We can also bound the magnitude of the right hand side by noticing that
\be
\begin{split}
&|\langle \Psi_E | \al_{\alpha} | \Psi_o \rangle|^2 = \langle \Psi_E | \al_{\alpha} | \Psi_o \rangle \langle \Psi_o | \al_{\alpha}^{\dagger} | \Psi_E \rangle \\ &= \langle \Psi_E | \al_{\alpha} \left( |\Psi_o \rangle \langle \Psi_o| + | \Psi_E \rangle \langle \Psi_E |  \right) \al_{\alpha}^{\dagger} | \Psi_E \rangle - |\langle \Psi_E | \al_{\alpha} | \Psi_E \rangle|^2 \\ &\leq \langle \Psi_E | \al_{\alpha} \al_{\alpha}^{\dagger} | \Psi_E \rangle - |\langle \Psi_E | \al_{\alpha} | \Psi_E \rangle|^2 .
\end{split}
\ee
In the intermediate step, we used the fact that  $|\Psi_E \rangle \langle \Psi_E | + |\Psi_o \rangle \langle \Psi_o | $ is a projector when the two vectors are orthogonal and that for any projector, $P$, 
\be
\langle \Psi_E | \al_{\alpha} P \al_{\alpha}^{\dagger} |\Psi_E \rangle = \langle \Psi_E | \al_{\alpha} P P \al_{\alpha}^{\dagger} | \Psi_E \rangle = |P \al_{\alpha}^{\dagger} |\Psi_E \rangle|^2 \leq |\al^{\dagger}_{\alpha} | \Psi_E \rangle|^2.
\ee

We can, of course, derive a similar equality showing that
\be
|\langle \Psi_E | \al_{\alpha} | \Psi_o \rangle|^2 \leq  \langle \Psi_o | \al^{\dagger}_{\alpha} \al_{\alpha} | \Psi_o\rangle - |\langle \Psi_o | \al_{\alpha} | \Psi_o \rangle|^2 .
\ee
We expect correlators in $|\Psi_o \rangle$ to be thermal and the analysis above tells us that, to leading order in $\beta \delta E$,  correlators in $|\Psi_E \rangle$ are also thermal. 

Putting these results together, we find that
\be
\label{bornruleineq}
\delta \langle \al_{\alpha} \rangle \leq 2 \sqrt{\beta \delta E} \sigma_{\alpha},
\ee
for $\beta \delta E \ll 1$, 
where the ``deviation'' of $\al_{\alpha}$ is defined through
\be
\label{stddev}
\sigma_{\alpha}^2 \equiv {1 \over Z(\beta)} \text{min} \Bigg[\tr\left(e^{-\beta H} \al_{\alpha}^{\dagger} \al_{\alpha}  \right), \tr\left(e^{-\beta H} \al_{\alpha} \al_{\alpha}^{\dagger} \right) \Bigg]  -\left| {1 \over Z(\beta)} \tr\left(e^{-\beta H} \al_{\alpha} \right) \right|^2.
\ee

Therefore we see that if $\beta \delta E \ll 1$, the change in the expectation value of $\al_{\alpha}$ is very small. In other words, in terms of the expectation value of coarse-grained operators, the state $U |\Psi \rangle$ is very similar to the state $|\Psi \rangle$ for a typical state $|\Psi \rangle \in \hilb[E]$. This result can be stated in the form of a slogan:  it is impossible to definitively excite a thermal system with energy less than $k T$.

We emphasize that the result above applies to a typical state $|\Psi \rangle$. Given a specific state $|\Psi \rangle$, we can always tailor a unitary operator, $U_{\psi}$ to excite that particular state so that $U_{\psi} |\Psi \rangle$ has very different properties from the original state. But the same unitary $U_{\psi}$ will not work in other states that also belong to the space $\hilb[E]$. 

The result that we have proved here is entirely general and applies to any statistical system. We will see in the next few sections how at first sight, it appears that black holes could violate this bound \cite{Marolf:2015dia} but that, in fact, these putative violations are unobservable. In a gravitational context, we need to be careful about the position dependence of the unitary excitation $U$ and the observable $\al_{\alpha}$ that could lead to a violation of \eqref{bornruleineq}.
 When we take this into account, we find that these two never fit into a single causal patch, and so no observer or collection of observers can detect this potential violation.

\section{Holographic setup and causal patches\label{secsetup}}
We are interested in examining \eqref{bornruleineq} in the context of AdS black holes. In this section, we describe our setup. As far as possible, we have maintained the conventions used in our previous papers \cite{Papadodimas:2013wnh,Papadodimas:2013jku, Papadodimas:2015jra}. The notion of causal patches, however, has not appeared previously in our work, and so we describe them in our context in some detail in subsection \ref{subsecausal}. 
The proof that AdS correlators, after we account for the constraints of causality, obey  \eqref{bornruleineq} is provided in the next section.

\subsection{Review of conventions \label{subsecsetupreview}}
We will consider a CFT with a holographic dual, and in this dual geometry we are interested in a large AdS Schwarzschild black hole. The metric is given by 
\be
\label{bhmetric}
ds^2 = g_{\mu \nu} d x^{\mu} d x^{\nu} = -f(r) dt^2 + {dr^2 \over f(r)} + r^2 d \Omega^2,
\ee
with
\be
\label{frdef}
f(r) = 1 - {c_d M \over r^{d-2}} + r^2,
\ee
with $c_d = {8 (d-1)^{-1} \pi^{{2-d}\over 2} \Gamma(d/2)}$, and where $M$ is the mass of the black hole.
It is convenient to go to tortoise coordinates
\be
\label{tortoisedef}
dr_* = {d r \over f(r)}.
\ee
We set $\rtor \rightarrow 0$ at the boundary, and this leads to $\rtor \rightarrow -\infty$ at the horizon. Note that near the boundary, we have $\rtor = {1 \over r}$. Near the horizon, we have $f(r) \rightarrow {4 \pi \over \beta}(r - \rhor)$, and therefore in this region we have $r_* \sim {\beta \over 4 \pi} \ln(r - \rhor)$, where $\beta$ is the inverse temperature of the black hole. 

To cross the horizon, we move to Kruskal coordinates defined by
\be
\label{kruskaldef}
\ukrus = -e^{{2 \pi\over \beta} (\rtor - t) }; \quad \vkrus = e^{{2 \pi \over \beta}(\rtor + t) }, \quad \text{outside~the~horizon}.
\ee
Once we are behind the horizon, we can introduce a second Schwarzschild patch using the definitions
\be
\ukrus = e^{{2 \pi\over \beta} (\rtor - t) }; \quad \vkrus = e^{{2 \pi \over \beta}(\rtor + t) }, \quad \text{inside~the~black~hole}.
\ee

In this paper we are interested in single-sided black holes, and in observers who fall into the black hole given by the metric \eqref{bhmetric} starting around time $t=0$ on the boundary. Therefore, the region of the Penrose diagram deep in the interior of the black hole, corresponding to $\ukrus \gg 1$ is not of interest to us, since it is not accessible to any such observer.  We are also only concerned with the region where $\vkrus > 0$. 

The black hole metric \eqref{bhmetric} is dual to a state $|\Psi \rangle$ in the boundary CFT, with $\langle \Psi | H | \Psi  \rangle \propto N^2$, where $N^2$ is the central charge of the CFT and $H$ is the CFT Hamiltonian. This is a pure state in the CFT, which is not entangled with any other system. The expectation is that a typical  state picked with the Haar measure on the microcanonical ensemble centered around an energy much larger than the central charge will correspond to the AdS-Schwarzschild geometry above. 

\subsection{Causal patches \label{subsecausal}}
Causal patches will play an important role in this paper. We define them as follows. Consider a point, $\pcaus$, on the singularity. The causal patch is then defined as the set of all bulk points that are in the causal past of this point. To picturize this set, consider past directed null rays emanating from the singularity at all possible angles. This light cone intersects the boundary of AdS on a sphere, at a time that we denote by $\tcaus$.  We now extend this region into the past. In the past, the black hole may have formed through collapse or through another process and we are not interested in the details of the formation. So we cut off the causal patch at a value of $\vkrus = \exp[{2 \pi \cutoffT \over \beta }]$; this intersects the boundary at a time $\cutoffT$ in the past.  The intersection of the causal patch with the boundary is denoted by $\bcaus$. This extends for a time $|\bcaus| = \tcaus - \cutoffT$ along the boundary. It will be convenient for the reader to think of $\cutoffT$ as a large negative number and we discuss its magnitude in the next subsection.  We denote the entire causal patch by ${\cal C}$.  In an AdS Schwarzschild black hole, we have a one-parameter family of patches that can all be labelled by $\tcaus$.

The physical motivation for considering causal patches is as follows. Consider the worldlines of an army of observers who all jump in from the boundary and end their lives on the singularity at $\pcaus$. Then the causal patch, as we have defined it, is the set of all spacetime events that could influence these observers before their wordlines terminate on the singularity. 

Strictly speaking, as explained in \cite{Freivogel:2014fqa}, in $d > 2$, an observer who enters the horizon at some angular point $\Omega$ can receive signals from all points on the sphere only when he is exactly at the singularity. However, by considering the set of all wordlines that end at a point on the singularity rather than the causal past of a single observer, we obtain a spherically symmetric patch, which simplifies the algebra in the calculations that follow. In other contexts, it may be important to consider the causal patch corresponding to a single observer rather than an army of observers, but, in our context, this distinction does not seem to be important. 

\begin{figure}[!h]
\begin{center}
\begin{subfigure}[t]{0.4\textwidth}
\begin{center}
\includegraphics[height=0.3\textheight]{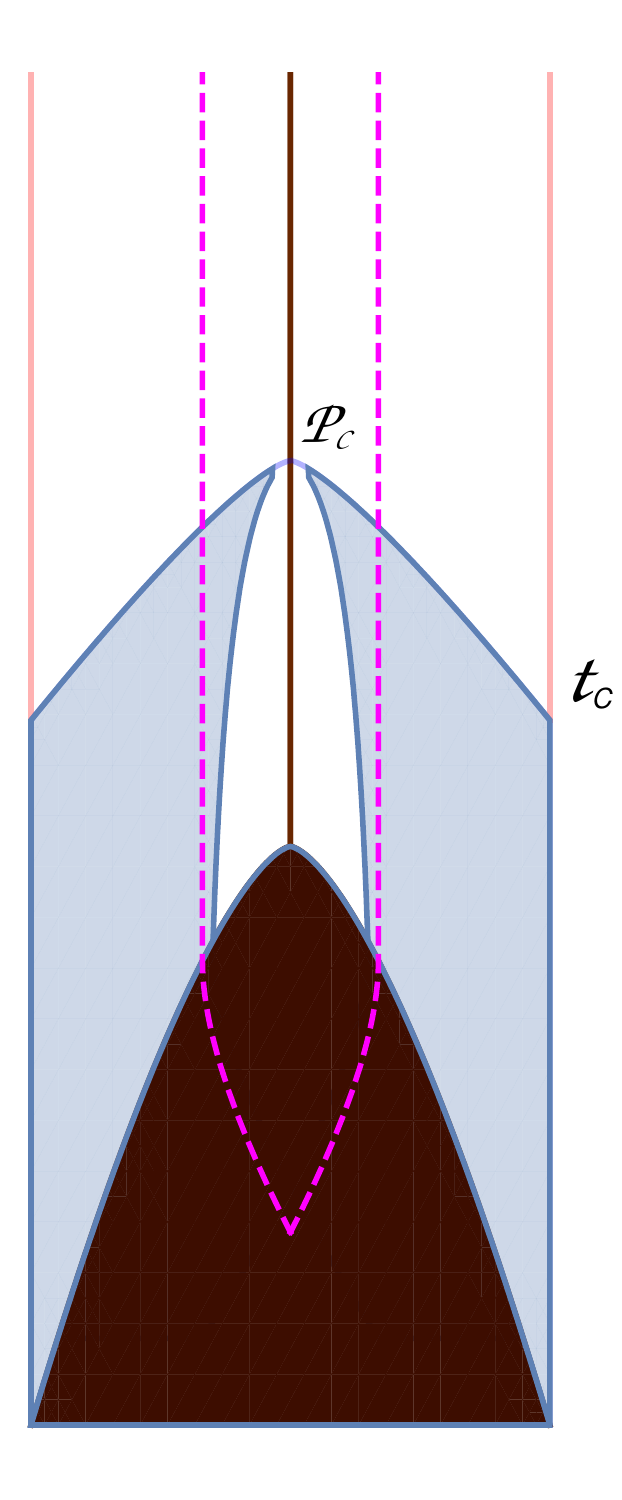}
\end{center}
\end{subfigure}
\qquad \qquad \qquad
\begin{subfigure}[t]{0.4\textwidth}
\begin{center}
\includegraphics[height=0.3\textheight]{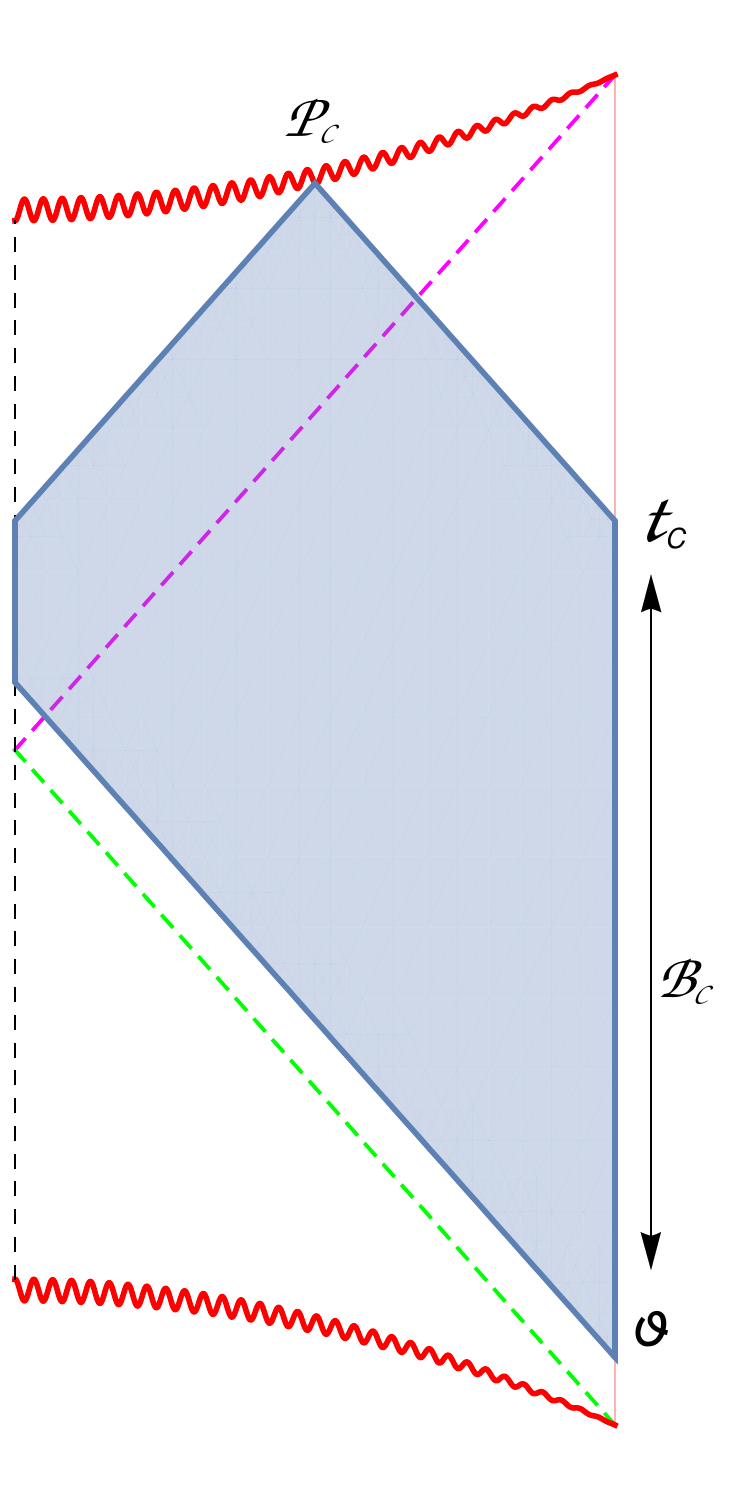}
\end{center}
\end{subfigure}
\caption{\em Two views of a causal patch. On the left, we show a causal patch in a Datt-Oppenheimer-Snyder collapse in AdS. The collapsing star is shaded in brown.  On the right, we show a causal patch in an eternal single-sided geometry and we also show the early time cutoff at $\cutoffT$. In both subfigures, the patch is demarcated in blue-gray and the future horizon is shown by a dashed magenta line. \label{figcausalpatch}}
\end{center}
\end{figure}

Figure \ref{figcausalpatch} illustrates our perspective on the causal patch. The subfigure on the left shows a causal patch in a black hole formed from a Datt-Oppenheimer-Snyder type collapse  \cite{datt1999class,datt1938klasse,oppenheimer1939continued} in AdS. Note that in the left subfigure, light rays are not 45 degree lines. So the inner boundary of the patch inside the horizon is a null geodesic but not a straight line.    We are interested in the geometry of the causal patches at late time, without being concerned about the details of the collapsing shell. So, it is more convenient to think of an eternal geometry. Such a geometry can be obtained even in the single sided case by considering an orbifold of the eternal black hole geometry. This is shown on the right, where we also show the early time cutoff $\cutoffT$.

\begin{figure}
\begin{center}
\includegraphics[height=0.3\textheight]{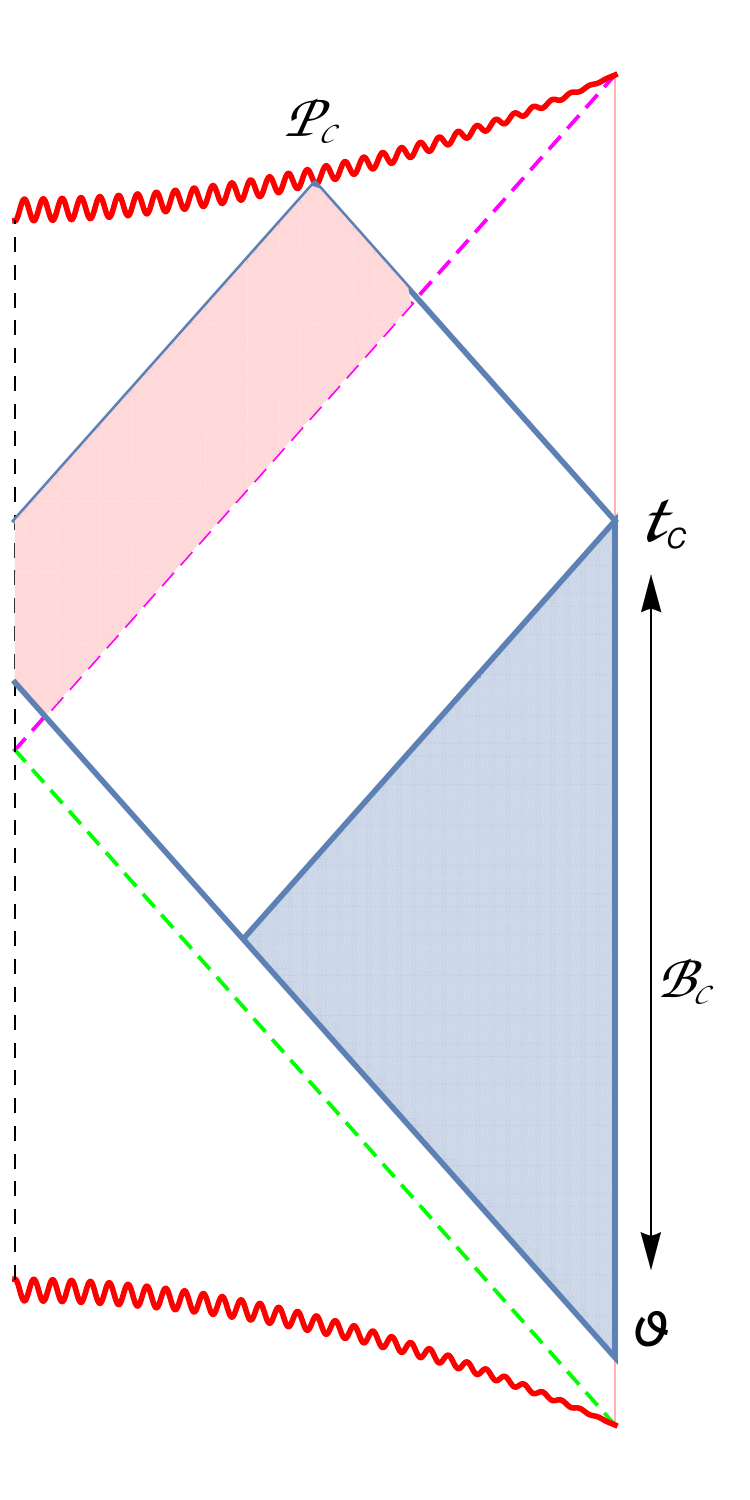}
\caption{\em The causal wedge, $\wcaus$ is shaded in blue outside the horizon; the intersection of the causal patch with the interior, denoted by $\vcaus$ is shaded pink inside the horizon. \label{figvwwedge}}
\end{center}
\end{figure}

For future use, it is also useful to define two additional geometrical regions. We denote the causal wedge of $\bcaus$ --- all points in the bulk with the property that  $\bcaus$ contains some part of their causal past and also some part of their causal future --- by $\wcaus$. The intersection of the causal patch with the interior of the black hole is denoted by  $\vcaus$. These regions are shown in figure \ref{figvwwedge}.

We pause to emphasize two points. First, it is possible to consider the causal patch corresponding to the observer who stays outside the black hole at all times. This causal patch, which we can call ${\cal C}_{\infty}$,  encompasses the entire region outside the black hole.  Second, note that these distinct causal patches only appear in a black hole background. In empty AdS, for example, there is a single causal patch that encompasses all observers. 

As we explained above, a causal patch is specified by a point on the singularity. Note that the singularity is located at $r = 0$, and using \eqref{tortoisedef}, it is located at a value of $\rtorsin$, given by 
\be
\rtorsin = \lim_{\delta \rightarrow 0^+} \Big[\int_{\rhor - \delta}^{0} {d r \over f(r)} - \int_{\rhor + \delta}^{\infty}{d r \over f(r)} \Big].
\ee
As explained in \cite{Fidkowski:2003nf}, for $d > 2$, we have ${\rtor}_s < 0$. For $d = 2$, we have ${\rtor}_s = 0$. 

In Kruskal coordinates, the singularity is located at $\ukrus \vkrus = e^{{4 \pi \over \beta} {\rtor}_s}$. A point on the singularity can be labelled by a value of time $t_{\text{sin}}$ on the singularity. Corresponding to this value we have a causal patch, whose boundaries are $\ukrus =e^{{2 \pi \over \beta} (\rtorsin - t_{\text{sin}})}$ and $\vkrus = {e^{{2 \pi \over \beta} (\rtorsin + t_{\text{sin}})}}$. This causal patch intersects the boundary at the time $\tcaus = \rtorsin + t_{\text{sin}}$.   Since $\rtorsin \leq 0$ as we pointed out above, we see that  $t_{\text{sin}}  \geq \tcaus$.

\subsection{CFT definitions and conventions}
A central role in our analysis is played by generalized free-fields, which are light operators in the theory whose correlators factorize. We denote such a field by $\op(t, \Omega)$. For example, in the N=4 SYM theory, both $\tr(F^2)$ and the stress tensor $T_{\mu \nu}$ are generalized free fields, although in this paper to lighten the notation we will drop the tensor indices on all such operators. Just as in \cite{Banerjee:2016mhh}, we consider angular momentum modes in such fields 
\begin{equation} 
\label{modegff}
\op_{\ell}(t) =  \int \, d \Omega \,\op(t,\Omega) Y_{\ell}^*(\Omega),
\end{equation}
where the integral is over the boundary ${S^{d-1}}$ and $Y_{\ell}(\Omega)$ are the spherical harmonics. We then consider the polynomial algebra in these modes
\be
\label{alsetdef}
\alset =  \text{span of}\{\op_{\ell_1}(t_1), \op_{\ell_2}(t_2)\op_{\ell_3}(t_3),  \ldots ,\op_{\ell_4}(t_4)\op_{\ell_5}(t_5) \cdots \op_{\ell_{\dmax}} (t_{\dmax}) \}.
\ee
Here we have introduced a cutoff $\dmax$ in the highest angular momentum mode that is allowed to appear, with $\dmax \ll N$. As we mentioned above, it is also important to place a long time cutoff on the boundary. This can be done by restricting the times  $t_i$ that appear above to lie in $t_i \in [-|\cutoffT|, |\cutoffT|]$. Here $\cutoffT$ can be taken to be a power of $N$ but not exponentially large $|\cutoffT| \ll e^{N^2}$. As discussed in \cite{Papadodimas:2015xma,Papadodimas:2015jra}, unless we place such a cutoff on the length of the longest possible time-interval in the CFT, there is no meaningful way to speak of a dual geometry. 

The reader will notice that our notation differs from the notation of \cite{Papadodimas:2013wnh,Papadodimas:2013jku} in that we have defined the algebra using operators that are local in time, rather than using operators of definite frequency. At an abstract level these two definitions are equivalent. But our notation reflects one of the messages of this paper: a  position space analysis often provides insights that may be missed by a purely frequency-space analysis. 

An element of the algebra is denoted by $\al_{\alpha} \in {\cal A}$. At times, we would like to consider an element that is localized at a given point in time, and we denote this by $\al_{\alpha}(t)$. Such an operator is a polynomial in local operators, all of which are at the time $t$. The modes defined in \eqref{modegff} are examples of such operators.

We will also consider non-equilibrium states that are related to the simple equilibrium state $|\Psi \rangle$. These excited states are of the form 
\be
\label{psinedef}
|\psine \rangle = U|\Psi \rangle,
\ee
where $U$  is a unitary formed by exponentiating  a Hermitian element of the algebra localized around some time $T$. Sometimes, rather than thinking of the non-equilibrium state as a state in the original theory,  it is convenient to think of it as being produced by actively deforming the Hamiltonian on the boundary.  This is an elementary point, but since it may be the source of some confusion, we now explain it in some detail.

\subsubsection{Source deformed correlators as correlators in autonomously excited states \label{secsourcesunitaries}}

To actively produce a non-equilibrium state, we proceed as follows.  We allow the CFT to thermalize so that it goes into an equilibrium state $|\Psi \rangle$. Then we deform the CFT Hamiltonian by turning on a source
\be
\label{hdeformed}
H^{J}(t) =  H + J(t) \al^J_{\alpha}(t),
\ee
where 
$\al_{\alpha}(t)$ is a some element of the algebra made out of local operators at time $t$. $H$ is the original CFT Hamiltonian. $\al^J_{\alpha}$ are the new Heisenberg operators and they are related to the original Heisenberg operators through
\be
\label{timeordal}
\al^{J}_{\gamma}(t) = \overline{{\cal T}}\{e^{i \int_{\cutoffT}^t J(x) \al_{\alpha}(x) d x}\} \al_{\gamma}(t) {\cal T}\{e^{-i \int_{\cutoffT}^t J(x) \al_{\alpha}(x) d x}\},
\ee
where ${\cal T}$  (${\overline{\cal T}}$) is the (anti-)time-ordering symbol, and the superscript distinguishes these operators from the original operators.  
If we consider the expectation value of a single source-deformed Heisenberg observable localized at a time $T$, this is given just by sandwiching the expression above in a state
\be
\langle \Psi |  \al^{J}_{\gamma}(T)  |\Psi \rangle = \langle \Psi | \overline{{\cal T}}\{e^{i \int_{\cutoffT}^T J(x) \al_{\alpha}(x) d x}\} \al_{\gamma}(T) {\cal T}\{e^{-i \int_{\cutoffT}^T J(x) \al_{\alpha}(x) d x}\} |\Psi \rangle.
\ee
Therefore, modifying the Hamiltonian through \eqref{hdeformed} functions {\em as if} we were evaluating correlators in the non-equilibrium state $|\psine \rangle = {\cal T}\{e^{-i \int_{\cutoffT}^T J(x) \al_{\alpha}(x) d x}\} |\psi \rangle$  after the time $T$. Before that time, the state is just the ordinary equilibrium state.

On the other hand, it is also possible to think of $|\psine \rangle$ as an autonomous state i.e. a state in the theory with the undeformed Hamiltonian that simply happens to have a spontaneous excitation around the time $T$.  Note that even in this picture, at times $t \ll T$, the state looks close to equilibrium.
For example, if we think of the boundary as a fluid, then this is a state where the fluid is initially in equilibrium, but then the microscopic elements of the fluid conspire to throw up an excitation around $T$ that again decays away. An observer on the boundary in such a state would observe a violation of the second law of thermodynamics, but such states are part of the Hilbert space of the CFT.\footnote{Note that in switching descriptions from the picture where we evaluate the expectation value of a source-deformed observable to the picture where we evaluate the expectation value of the observable in  a spontaneously excited state in the undeformed theory,  the parameter $T$ which was a property of the observable in the source-deformed description now enters into the description of the state. This should not cause any confusion. The source itself can continue to act after time $T$ but this part of the source does not enter into $|\psine \rangle$.}

This equivalence between the two descriptions generalizes to the case where the observable in question is a product of localized operators in the bulk.  As a special case of \eqref{timeordal}, the bulk fields are deformed to
\be
\phi^{J}(t, \rtor, \Omega) = \overline{{\cal T}}\{e^{i \int_{\cutoffT}^t J(x) \al_{\alpha}(x) d x}\}  \phi(t, \rtor, \Omega) {\cal T}\{e^{-i \int_{\cutoffT}^t J(x) \al_{\alpha}(x) d x}\}.
\ee
But now, note that in the time-ordered unitary that appears on the right, the later part of the integral that runs from $[t + \rtor, t]$ involves operators that commute with the bulk field. Since $\al_{\alpha}(t)$ is localized on the boundary, by bulk locality, we have $[\phi(t, \rtor, \Omega), \al_{\alpha}(x)] = 0$ if $t+\rtor < x < t - \rtor$. We need only the first part of this inequality since in the expression above,  $x \leq t$.  The adjoint of the part of the time-ordered unitary that runs from $[t + \rtor, t]$ appears on the extreme right of the anti-time-ordered unitary. Therefore,
\be
\phi^{J}(t, \rtor, \Omega) = W^{\dagger}(t+ \rtor) \phi(t, \rtor, \Omega) W(t + \rtor),
\ee
where
\be
W(t + \rtor) \equiv {\cal T}\{e^{-i \int_{\cutoffT}^{t+\rtor} J(x) \al_{\alpha}(x) d x}\}.
\ee
But this means that a correlator of the deformed field can be written as
\be
\begin{split}
&\langle \Psi | \phi^{J}(t_1, \rtor_1, \Omega_1) \ldots \phi^{J}(t_n, \rtor_n, \Omega_n) | \Psi \rangle \\ &= 
 \langle \Psi | W(t_1 + \rtor_1)^{\dagger} \phi(t_1, \rtor_1, \Omega_1)  W(t_1 + \rtor_1) \ldots W(t_n + \rtor_n)^{\dagger} \phi(t_n, \rtor_n, \Omega_n)  W(t_n + \rtor_n) | \Psi \rangle.
\end{split}
\ee

Furthermore, we can massage this correlator so that it looks precisely of the form \eqref{bornruleineq} as follows. Define
\be
U = \exp\big[-i \int_{\cutoffT}^{\tcaus} \al_{\alpha}(t) d t \big],
\ee
as in \eqref{auxiliaryU}. Then we find that
\be
\langle \Psi | \phi^{J}(t_1, \rtor_1, \Omega_1) \ldots \phi^{J}(t_n, \rtor_n, \Omega_n) | \Psi \rangle = \langle \Psi | U^{\dagger} \al_{\gamma} U | \Psi \rangle
\ee
where 
\be
\label{algammadef}
\begin{split}
\al_{\gamma} &=  \exp \big[-i \int_{\cutoffT}^{\tcaus} \al_{\alpha}(t) d t \big] W(t_1 + \rtor_1)^{\dagger} \phi(t_1, \rtor_1, \Omega_1)  W(t_1 + \rtor_1) \ldots \\
&\times  W(t_n + \rtor_n)^{\dagger} \phi(t_n, \rtor_n, \Omega_n)  W(t_n + \rtor_n) \exp \big[i \int_{\cutoffT}^{\tcaus} \al_{\alpha}(t) d t \big].
\end{split}
\ee
Now, we note that  $(t, \rtor, \Omega) \in {\cal C} \Rightarrow (t + \rtor, 0, \Omega') \in {\cal C}, \forall \Omega'$. Therefore, the operator $\al_{\gamma}$ is entirely localized in the causal patch and $U$ is localized on the boundary of the patch. 

The conclusion is that  {\em a correlator with source-deformed fields can be reduced to a correlator of ordinary fields with insertions only in ${\cal C}$ and an excitation on the boundary of ${\cal C}$.}

In operational terms, this yields the following protocol to check \eqref{bornruleineq} in the situation where we create an excitation {\em actively} by deforming the Hamiltonian through a local source. The infalling observer can measure $\langle \Psi| \phi^J(t_1, \rtor_1, \Omega_1) \ldots \phi^{J}(t_n, \rtor_n, \Omega_n) | \Psi \rangle$ and compare this excited expectation value with $\langle \Psi| \al_{\gamma} | \Psi \rangle$ in the unexcited state, where $\al_{\gamma}$ is specified in \eqref{algammadef}.

On physical grounds, it is more natural to think of of $|\psine \rangle$ as being produced by turning on a local source on the boundary. But, from a technical perspective, it is more convenient to think of the state as an autonomous state since then we do not have to worry about the time-ordering in \eqref{timeordal}. 

In section \ref{secfullads}  we will be able to prove that AdS correlators in a causal patch are stable, in that they obey the constraints \eqref{bornruleineq} of statistical mechanics, in all possible autonomous states of the form $|\psine \rangle$ where the unitary and the observables fit in the same causal patch. As we have explained above,  this also implies that these correlators are stable if an observer or an army of observers on the boundary turn on a source and subsequently jump into the bulk and cross the horizon. We reiterate that this applies to arbitrary sources dual to local boundary operators --- even those that continue to act beyond the causal patch. 

\paragraph{No bulk sources \\}
While we have considered the effect of sources dual to local operators on the boundary above, we emphasize that we  do not consider situations where sources are turned on in the bulk. A bulk source would violate local energy conservation. So, in a theory of quantum gravity, it does not make sense to modify the bulk Hamiltonian and only boundary sources are allowed.

This is a subtle point.  If $\phi(f)$ is a smeared Hermitian bulk operator then correlators where local insertions are sandwiched between bulk unitaries, $e^{i \phi(f)}$ and $e^{-i \phi(f)}$,  can be expanded in a series of Wightman functions. But the bulk observer can only measure time ordered correlators, and not arbitrary Wightman functions. So the allowed bulk observables are
\be
\langle \psine | {\cal T}\Big[\phi(t_1, \rtor_1, \Omega_1) \ldots \phi(t_n, \rtor_n, \Omega_n) \Big] | \psine \rangle =
\langle \Psi | U^{\dagger} {\cal T} \Big[ \phi(t_1, \rtor_1, \Omega_1) \ldots \phi(t_n, \rtor_n, \Omega_n)  \Big] U | \Psi \rangle.
\ee
Notice that the unitaries on the left and the right are outside the time-ordering, since they may be produced by spontaneous excitations or boundary sources. 

We will not need this time-ordering in most of this paper since the precise results that we describe below --- such as the proof of \eqref{centralresult} --- hold for all Wightman functions, and not just for time-ordered bulk correlators. Nevertheless, the physical restriction to time-ordered bulk observables is important to keep in mind to avoid the appearance of {\em additional paradoxes} corresponding to bulk sources and bulk unitaries.

\subsection{Transfer function \label{subsectranscaus}}
An important role in the program of reconstructing the bulk from the boundary
is played by transfer functions that map bulk operators to boundary operators. This is reviewed in  detail in \cite{Papadodimas:2012aq}. Here, we review the essential formulas and direct the reader to \cite{Papadodimas:2012aq} and references there for further details. 

Outside the horizon (corresponding to $U < 0, e^{{2 \pi \cutoffT \over \beta}} < V < e^{{2 \pi \over \beta} \tcaus}$), the bulk field can be mapped to boundary operators via the mapping
\be
\label{transout}
\phi(t, \rtor, \Omega) = \sum_{\ang, \omega}{\dzero \op_{\omega, \ang}} \zeta_{\omega, \ang}(\rtor) e^{-i \omega t} Y_{\ang}(\Omega) + \text{h.c},
\ee
whereas for the part of the causal patch inside the horizon (corresponding to $0 < U < e^{{2 \pi \over \beta}(2\rtorsin-\tcaus)}, e^{2 \pi \cutoffT \over \beta} < V < e^{{2 \pi \over \beta} \tcaus}$, the mapping between bulk and boundary operators is given by
\be
\label{transin}
\phi(t, \rtor, \Omega) = \sum_{\ang, \omega}   { Y_\ang(\Omega) \dzero \tzeta^{-}_{\omega, \ang} (\rtor)} \left( \op_{\omega, \ang} e^{2 i \delta_{\omega,\ang}} e^{-i \omega t}  +  \tO_{\omega, \ang} e^{i \omega t} \right) + \text{h.c}.
\ee

We have introduced several pieces of notation above, and we explain them now.  The radial functions $\zeta_{\omega, \ang}(\rtor)$ and $\tzeta^{-}_{\omega, \ang}(\rtor)$ ensure that the full mode satisfies the wave equation both inside and outside the horizon. These functions are chosen to satisfy specific boundary conditions. So, if the bulk field has mass $m$, then
\be
\begin{split}
& (\Box - m^2) \zeta_{\omega, \ang}(\rtor) e^{-i \omega t} Y_{\ang}(\Omega) = 0, \\
&\zeta_{\omega, \ang}(\rtor) \underset{\rtor \rightarrow -\infty}{\longrightarrow} e^{i \omega \rtor} + e^{2 i \delta_{\omega, \ang}} e^{-i \omega \rtor},
\end{split}
\ee
whereas the mode behind the horizon satisfies
\be
\tzeta^{-}_{\omega, \ang}(\rtor) \underset{\rtor \rightarrow -\infty}{\longrightarrow}  e^{-i \omega \rtor}, 
\ee
and this time the limit $\rtor \rightarrow -\infty$ is taken from behind the horizon.

The phase $\delta_{\omega, \ang}$ and the function $D^0(\omega, \ang)$ are fixed by ensuring that $\zeta$ is normalizable at the boundary and the normalizable part has coefficient,
\be
\zeta_{\omega, \ang}(\rtor) \underset{\rtor \rightarrow 0}{\longrightarrow} {1 \over \dzero} (\rtor)^{\Delta},
\ee
where $\Delta$ is the dimension of the operator $\op$. These radial equations  are discussed in some detail in section \ref{secfullads}.

Finally $\op_{\omega, \ang}$ are modes of the boundary operators and $\tO_{\omega, \ang}$ are the ``mirror'' operators, which we will review below. 
As we pointed out in \cite{Papadodimas:2013jku}, the modes of the boundary operators cannot be defined as infinite time Fourier transforms; instead it is important to define these modes as Fourier series coefficients corresponding to a compact region on the boundary. 
\be
\label{opmodedef}
\op_{\omega, \ang} =  \int d \Omega \int_{-\ircutoff}^{\ircutoff} dt \, \op(t, \Omega) Y_{\ang}^*(\Omega) e^{i \omega t},
\ee
where $\ircutoff$ is the long time cutoff introduced above. Corresponding to this we see that $\omega$ must be quantized in units of ${\pi n \over \ircutoff}$. We wish to avoid having this cutoff appear in all our formulas. To do this we will use the following notational tricks. First, it is understood that whenever we write a sum over $\omega$, we really mean
\be
\label{sumtrick}
\sum_{\omega} f(\omega) \equiv {1 \over 2 \ircutoff} \sum_{n>0} f(n),
\ee
where $\omega = {\pi n \over \ircutoff}$ as above and additionally the sum over frequencies runs only over positive frequencies.
Second, when we write a delta function we mean
\be
\label{deltatrick}
\delta_{\omega, \omega'} \equiv 2 \ircutoff \delta_{n n'}.
\ee
The reader will note that we have already tacitly used this notation in the formulas above.

We now turn to the definition of the operators $\tO$. These operators were  described in \cite{Papadodimas:2013wnh,Papadodimas:2013jku,Papadodimas:2015jra}, and we refer the reader to those papers for details. Here, we simply recall that these operators are defined through a set of linear equations. Given an equilibrium state, we define the mirror operators to satisfy
\be
\label{tocausdef}
\tO_{\omega, \ang} \al_{\alpha} | \Psi \rangle = e^{-{\beta \omega \over 2}} \al_{\alpha} \op^{\dagger}_{\omega, \ang}  |\Psi \rangle.
\ee
This completely specifies the action of the operators on the subspace $\hpsi = \alset |\Psi \rangle$; this is the subspace formed by low energy excitations of the equilibrium state and is the only part of the Hilbert space that is relevant for effective field theory experiments done about the state $|\Psi \rangle$.

\section{Proof of the stability of AdS correlators in a causal patch   \label{secfullads}}

In this section, we will now prove that AdS correlators obey the inequality \eqref{bornruleineq} whenever the excitation $U$ and the correlation function to be measured fit in a single causal patch. This automatically implies that in the physically important setting where we start with an equilibrium state, and turn on a local source, we do not observe any violation of \eqref{bornruleineq}. This is because as we see from \eqref{schwingerops} an operator at any point in the bulk cannot be affected by any part of the source that is not in its causal past; therefore it cannot be affected by any part of the source that is outside its causal patch.\footnote{Note that as mentioned above, if we think in terms of sources, our proof of the stability of AdS correlators is valid for {\em arbitrary sources}, without any loss of generality. The restriction that the boundary excitation are limited to the same causal patch as the bulk points is implemented automatically through bulk causality when the excitation is created through a deformation of the boundary Hamiltonian.}

Since our proof is somewhat technical, we summarize the main ideas before starting the proof.  It is 
useful to understand why the inequality is in danger of being violated in the first place. Bulk operators defined through \eqref{transout} are simply rewritings of ordinary operators on the boundary. So they must automatically obey the inequality \eqref{bornruleineq} which applies to any ordinary operator. On the other hand, the bulk operators defined in \eqref{transin} depend on the operators $\tO_{\omega, \ang}$ that are defined through the 
linear equations \eqref{tocausdef}. We remind the reader that these operators are {\em state-dependent}. This means that it is impossible to find a globally defined operator $\tO_{\omega, \ang}$ that obeys the equations \eqref{tocausdef} about all equilibrium states $|\Psi \rangle$. The danger is that this state-dependence may cause these operators to behave very differently in an excited state $U |\Psi \rangle$ than in a typical equilibrium state $|\Psi \rangle$, even if $U$ is a low energy excitation.

But the modes $\tO_{\omega, \ang}$ are an auxiliary device for constructing the local field operators $\phi(t, \rtor, \Omega)$. What we will show in this section is that when these operators are localized in the same causal patch as the excitation $U$, they obey the inequality \eqref{bornruleineq} because the anomalous transformations of the individual modes cancel among each other.
 
In more detail, we proceed as follows.  Even though the AdS potential causes back-scattering, it is possible to define a set of right-moving and left-moving modes in AdS. The right movers are the modes that behave like  $\phi_r \sim e^{i \omega(\rtor - t)}$ near the horizon, and the left movers are those that behave like $\phi_l \sim e^{-i \omega (\rtor + t)} e^{2 i \delta_{\omega, \ang}}$ near the horizon. 
 As we cross the horizon
the left movers just continue smoothly from the region outside the horizon to the region inside the horizon. But the right movers behind the horizon, $\tphi_r$, now contain the mirror modes rather than the ordinary modes. 

\begin{figure}[!h]
\begin{center}
\includegraphics[height=0.3\textheight]{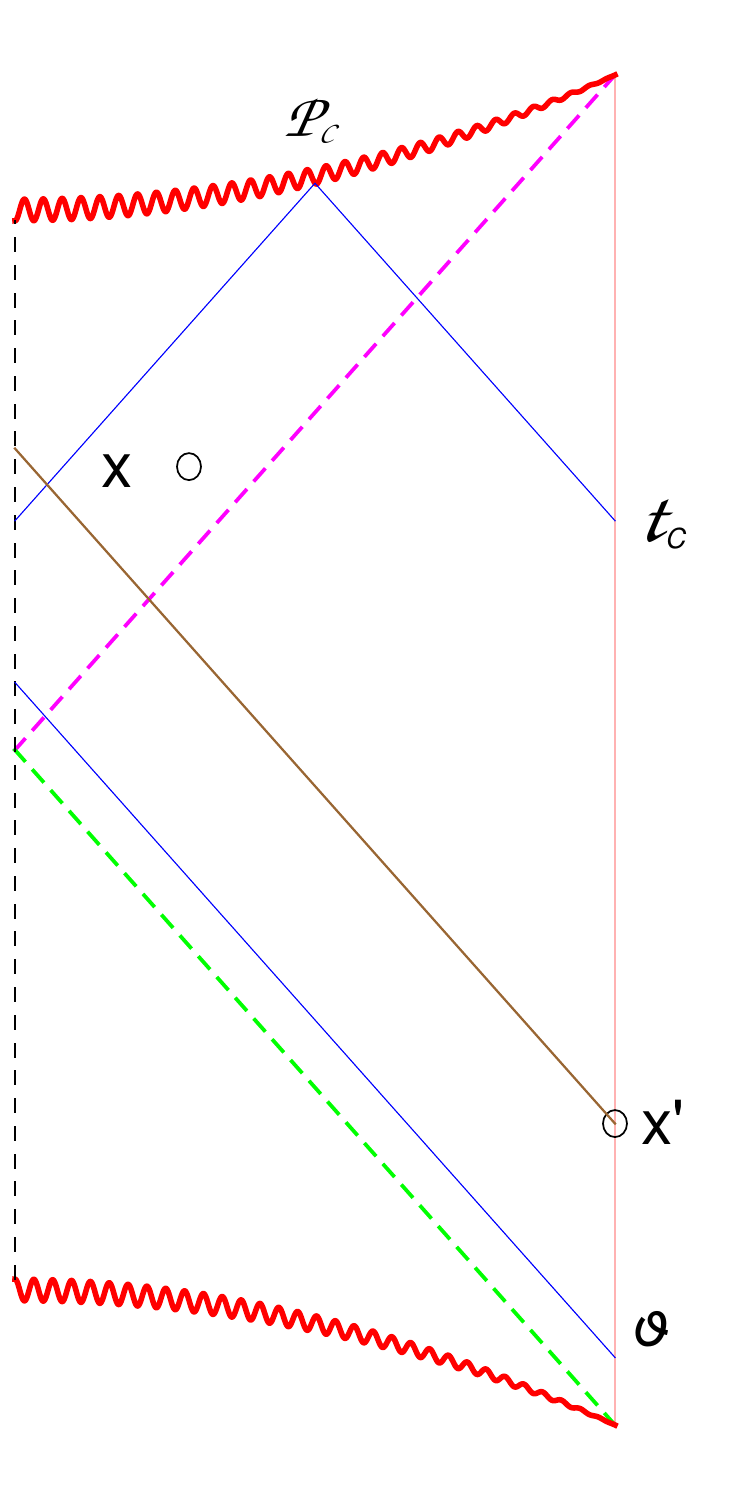}
\caption{\em A boundary excitation at $x'$ commutes with the right moving hatted field at $x$ even though $x$ is in the causal future of $x'$. The boundaries of the causal patch are in blue. The light cone from $x'$ is marked off in brown. \label{figwedgehorizon}}  
\end{center}
\end{figure}

The operators $\tphi_r$ are defined so that in an equilibrium state, they can be replaced,  up to a conjugation by $e^{\beta H \over 2}$, with an ordinary  operator $\phihat$.
The calculation that we perform is to show that the commutator $[\phihat(x), \phi(x')]$ vanishes when $x$ is placed in $\vcaus$ --- which is the part of the causal patch inside the horizon  --- and $x'$ is placed near the boundary on $\bcaus$.  (See Figure \ref{figwedgehorizon}.) 

Now, consider two states --- an equilibrium state $|\Psi \rangle$ and a near-equilibrium state $|\psine \rangle  = \ucaus |\Psi \rangle$, where $\ucaus$ is localized on $\bcaus$.  As a consequence of this vanishing commutator, a correlator involving $\tphi_r(x)$ can be replaced--- up to the conjugation by $e^{\beta H \over 2}$ that leads to an error of only $\Or[\beta \delta E]$--- by a correlator of ordinary operators involving just $\phihat(x)$ {\em both} in $|\Psi \rangle$ and $|\psine \rangle$.  If the commutator had not vanished, this would not have been possible since $\tphi_r(x)$ commutes with $\ucaus$ and so it can only be substituted by another operator that also commutes with $\ucaus$.  Now, since this ordinary operator obeys the inequality \eqref{bornruleineq}, we see that the correlator involving $\tphi_r(x)$ also obeys \eqref{bornruleineq} as long as $x$ is within the causal patch. 

To compute the commutator,  first, we analyze the analytic properties of the normalizable mode outside the horizon. This is most conveniently done by building the mode starting at the boundary and then solving inwards in subsection \ref{subsecbdry}. Next, we define and analyze the analytic properties of the right-moving component of this mode; this is most conveniently done by
specifying the mode at the horizon and solving outwards towards the boundary in subsection \ref{subsechor}. Finally we consider the analytic properties of components of the field behind the horizon in subsection \ref{subsecbehindhor}, which are the same as those of the right-moving components outside the horizon. In subsection \ref{subseccommut}, we determine the commutator $[\phihat(x), \phi(x')]$. A knowledge of the analytic properties of the mode functions is enough to show the vanishing of the commutator in the region of interest. In the last subsection, we put together all this information to show that AdS correlators are stable under low energy excitations.
This means that they obey the inequality \eqref{bornruleineq} and a low energy excitation cannot change the correlator significantly.

\subsection{Analysis from the boundary side \label{subsecbdry}}
To analyze the AdS wave-functions, we will follow the techniques of \cite{Jensen:1985in}. In the context of AdS/CFT these wave-functions have been discussed in the context of Schwarzschild-AdS quasinormal modes \cite{Horowitz:1999jd,Berti:2009kk,Kokkotas:1999bd}. We review and rephrase this discussion in the context that we need. 

The Klein-Gordon wave equation is given by 
\be
{1 \over \sqrt{-g}} \partial_{\mu} g^{\mu \nu} \sqrt{-g} \partial_{\nu} \phi -  m^2 \phi = 0,
\ee
where  the AdS$_{d+1}$ black hole metric is given by \eqref{bhmetric}.
As above it  is convenient to go to the tortoise coordinates defined in \eqref{tortoisedef}.
We remind the reader that $\rtor \rightarrow 0$ at the boundary and $\rtor \rightarrow -\infty$ at the horizon. 

In these coordinates, we have $\sqrt{-g} = f(r) r^{d -1}$. We also have $g^{\rtor \rtor} = -g^{t t} = {1 \over f(r)}$. Then the wave equation becomes 
\be
\label{tortwaveeqn}
{1 \over f(r) r^{d-1}} \partial_{\rtor} r^{d - 1} \partial_{\rtor} \phi  - {1 \over f(r)} \partial_t^2 \phi + {1 \over r^2} \Box_{\Omega} \phi - m^2 \phi  = 0,
\ee
where $r$ is now understood to be a function of $r_*$ and $\Box_{\Omega}$ is the Laplacian on the sphere.  

We can separate variables and write an ansatz for the solution of \eqref{tortwaveeqn} as
\be
\phi_{\omega, \ang}(t, \rtor, \Omega) = r^{1 - d \over 2} \chi_{\omega, \ang} (\rtor) e^{-i \omega t} Y_\ang(\Omega).
\ee
Note that 
\be
\resizebox{\hsize}{!}{$
\partial_{\rtor} r^{d-1} \partial_{\rtor} r^{1 - d \over 2} \chi_{\omega, \ang}(\rtor) = r^{d-1 \over 2} \partial_{\rtor}^2  \chi_{\omega, \ang}(\rtor)  + {1 \over 4} \left[{(d-3) } f(r) +  2 r  f'(r)  \right](1 - d) f(r) r^{d-5 \over 2}  \chi_{\omega, \ang}(\rtor),
$}
\ee 
and the differential equation for $\chi_{\omega, \ang}$ now becomes
\be
{\partial^2 \chi_{\omega, \ang}(\rtor) \over \partial \rtor^2} = V(\rtor) \chi_{\omega, \ang}(\rtor),
\ee
where
\be
V(\rtor) = -\left[\omega^2 + {(d-3)(1-d) \over 4} {f(r)^2 \over r^2} - m^2 f(r) - l(l+d-2){f(r) \over r^2} + {1 - d \over 2} {f'(r) f(r) \over r} \right].
\ee
Near the boundary, it is also convenient to write this equation as
\be
\label{bdryeqn}
{\partial^2 \chi_{\omega, \ang}(\rtor) \over \partial \rtor^2} + \gamma^2  \chi_{\omega, \ang}  -  {(m^2 + {d^2-1 \over 4}) \over \rtor^2} \chi_{\omega, \ang} = \vbd(\rtor),
\ee
where
\be
\gamma^2 \equiv \omega^2 - l(l+d - 2)) - {m^2  + (d-2)(d-1)\over 3},
\ee
and
\be
\begin{split}
\vbd(\rtor) = &l(l+d-2) \left({f(r) \over r^2} - 1 \right) + {(d-3)(d-1) \over 4}  \left({f^2(r) \over r^2} - {1 \over \rtor^2} - {4 \over 3} \right) \\ &+  m^2 \left(f(r) - {1 \over \rtor^2} - {1 \over 3} \right)  + {d - 1 \over 2} \left({f'(r) f(r) \over r}  - {2 \over \rtor^2}  - {2 \over 3} \right) .
\end{split}
\ee
We see that $\vbd(\rtor)$ is now finite at all values of $\rtor \in (-\infty, 0]$, which is the range that we are interested in, and that near the boundary $\vbd(\rtor) = \Or[\rtor^2]$. 

We can solve this equation using Green's functions both near the boundary and near the horizon. The starting point for this is to  write a ``free'' equation near the boundary (neglecting $\vbd$):
\be
\label{freebdryeqn}
{\partial^2 \chi^0_{\omega, \ang}(\rtor) \over \partial \rtor^2}  + \gamma^2 \chi^0_{\omega, \ang}(\rtor)  -  {(m^2 + {d^2-1 \over 4}) \over \rtor^2} \chi^0_{\omega, \ang}(\rtor) = 0.
\ee
Note that at the boundary, since the potential $\vbd \rightarrow 0$, we have $\chi^{0}_{\omega, \ang}(\rtor) \underset{\rtor \rightarrow 0}{\longrightarrow} \chi_{\omega, \ang}(\rtor)$. Therefore, imposing normalizable boundary conditions at the boundary, 
the only allowed solution is
\be
\chi^0_{\omega, \ang}(\rtor)  = {1 \over \gamma^{\nu}} \sqrt{\rtor} J_{\nu}(\gamma \rtor),
\ee
with $\nu = \sqrt{m^2 + {d^2 \over 4}}$. We have normalized the solution so that near the boundary, where $\rtor \rightarrow {1 \over r} \rightarrow 0$, the behaviour of the original field is
\be
\phi_{\omega, \ang}(t, \rtor, \Omega) =  e^{-i \omega t} Y_\ang(\Omega) r^{1 - d \over 2} \chi_{\omega, \ang}(\rtor)  \underset{\rtor \rightarrow 0}{\longrightarrow}  e^{-i \omega t} Y_\ang(\Omega) \rtor^{{d \over 2} + \nu}.
\ee

We are ultimately interested in the  analytic properties of the mode as we vary $\omega$. From the series expansion of the Bessel function given above, we see that 
\be
\label{besselexpansion}
{1 \over \gamma^{\nu}} J_{\nu}(\gamma \rtor) = \sum_{s = 0}^{\infty} {(-1)^s \over \Gamma(s+1) \Gamma(\nu+s+1)} \gamma^{2 s} ({\rtor \over 2})^{\nu + 2 s}.
\ee
In particular we see that in the expression for $\chi_{\omega, \ang}^{0}(\rtor)$,  $\gamma$ appears only raised to positive even integer powers. Therefore we see that $\chi^{(0)}_{\omega, \ang}(\rtor)$ has no poles or branch cuts in $\omega$ at any finite $\rtor$ at any finite value of $\omega$. 

We now show that the full solution with the potential $\vbd$ included also has the same analytic structure in $\omega$.  To construct the solution away from the boundary, we need the Green's function for the equation \eqref{bdryeqn}. This can be constructed as follows. Consider the second solution to the free equation  \eqref{freebdryeqn}. This can be written as $ \gamma^{\nu} \sqrt{\rtor} Y_{\nu}(\gamma \rtor)$. We can then write the Green's function as 
\be
\label{greenbessel}
G(\rtorp, \rtor) = {\pi \over 2} \sqrt{\rtor \rtorp} \left(J_{\nu}(\gamma \rtor) Y_{\nu}(\gamma \rtorp)  - Y_{\nu} (\gamma \rtor) J_{\nu} (\gamma \rtorp) \right) \theta(\rtorp - \rtor).
\ee

To examine the analytic properties of $G(\rtorp, \rtor)$ in $\omega$ we note that 
\be
Y_{\nu}(x)  = {\cos(\nu \pi) J_{\nu} (x) - J_{-\nu} (x) \over \sin \nu \pi}.
\ee
This form is valid for all non-integer $\nu$ and for integer $\nu$ it should be understood appropriately as a limit taken from non-integer $\nu$. We see then that a term proportional to $J_{\nu}(\gamma \rtor) J_{\nu} (\gamma \rtorp)$ cancels between the two terms of \eqref{greenbessel}, leading to a series expansion
\be
\label{greenseries}
\begin{split}
G(\rtorp, \rtor) = {\pi  \sqrt{\rtor \rtorp} \over 2 \sin(\nu \pi)}  \sum_{s,t=0}^{\infty} {(-1)^{s+t}  \gamma^{2 (s + t)} ({\rtorp \over 2})^{2 s + \nu} ({\rtor \over 2})^{2 t - \nu} \over \Gamma(s+1) \Gamma(\nu+s+1) \Gamma(t + 1) \Gamma( t -\nu + 1)} - (\rtorp \leftrightarrow \rtor),
\end{split}
\ee
for $\rtorp > \rtor$. 
The  leading non-analytic term $\gamma^{\nu}$ cancels in the product of $J_{\nu}(\gamma \rtor)$ and $J_{-\nu}(\gamma \rtorp)$ and thereafter only positive integer powers of  $\gamma^2$ occur. Therefore we conclude that \eqref{greenbessel} is also an analytic function of $\omega$ with no poles or branch-cuts at finite $\omega$. 

Using this Green's function and the solution $\chi_{\omega, \ang}^0$ we can proceed to construct the full solution. We write
\be
\label{serieschin}
\chi_{\omega, \ang}^{(n)}(\rtor) = \int_0^{\rtor} \chi_{\omega, \ang}^{(n-1)}(\rtorp) G(\rtorp, \rtor) \vbd(\rtorp) d \rtorp,
\ee
where the range of integration can be restricted to $\rtor$ because the Green's function vanishes beyond that. We then write the full solution as
\be
\label{chisoln}
\chi_{\omega, \ang}(\rtor) = \sum_{n=0}^{\infty} \chi_{\omega, \ang}^{(n)} (\rtor).
\ee

Now, we note that in \eqref{serieschin} the integrand is finite at all values of $\rtorp$ and each $\chi_{\omega, \ang}^{(n)}(\rtor)$ dies off at least as fast as $(\rtor)^{\nu + {1 \over 2} + 4 n}$ near the boundary. We can prove these facts together using induction.  These properties are clearly true for the iteration from $\chi_{\omega, \ang}^{(0)}$ to  $\chi_{\omega, \ang}^{(1)}(\rtor)$. Now assume that we have performed the iteration till $\chi_{\omega, \ang}^{(n-1)}(\rtor)$. Then the Green's function has a singularity which, at worst, goes like $(\rtorp)^{{1 \over 2} -\nu}$ near the boundary. On the other hand, by assumption $\chi_{\omega, \ang}^{(n-1)}$ dies off like $(\rtorp)^{\nu + {1 \over 2} + 4(n-1)}$ near the boundary and, in addition, we can check that $\vbd(\rtorp) \underset{\rtorp \rightarrow 0}{\longrightarrow} (\rtorp)^2$. So the integrand scales like $(\rtorp)^{4 n - 1}$ as we approach the boundary. We see that after
we do the integral up to $\rtor$ and note that the Green's function has a leading factor of $(\rtor)^{\nu + {1 \over 2}}$ in \eqref{greenseries}, we find that  $\chi^{(n)}(\rtor)$ dies off  like $\rtor^{\nu + {1 \over 2} + 4 n}$, which is what we need.

Second, since we are integrating a finite integrand over a finite region, we see that the iteration \eqref{serieschin} produces no poles at finite $\omega$ and finite $\rtor$. Therefore the full solution \eqref{chisoln} has no poles or branch cuts at finite $\omega$ and finite $\rtor$. 

\paragraph{Transfer function using boundary-side modes \\}
We now describe how these modes can be combined with the modes of boundary operators to write down a 
bulk field. Since the function  $r^{1 - d \over 2} \chi_{\omega, \ang}$ is normalized to die off like $r^{-\Delta}$ near the boundary, we can immediately write an expression for the bulk field as
\be
\phi(t, \rtor, \Omega) = \sum_{\ang, \omega}  \op_{\omega, \ang} r^{1 - d \over 2}\chi_{\omega, \ang}(\rtor) e^{-i \omega t} Y_\ang(\Omega) + \text{h.c},
\ee
where the boundary modes $\op_{\omega, \ang}$ were defined in \eqref{opmodedef}.

However, we would like to write the bulk field in terms of ordinary creation and annihilation operators that have simple commutators. To do this, we consider the expansion of the radial
mode near the horizon
\be
\label{d0deltadef}
r^{1 - d \over 2}\chi_{\omega, \ang}(\rtor) \underset{\rtor \rightarrow -\infty}{\longrightarrow}  \dzero \left(e^{i \omega \rtor}  +  e^{2 i \delta_{\omega, \ang}} e^{-i \omega \rtor} \right),
\ee
where the functions $D^{0}$ and $\delta_{\omega, \ang}$ are defined by the relation above. We can now alternately define
\be
\zeta_{\omega, \ang}(\rtor) = {1 \over \dzero} r^{1 - d \over 2}\chi_{\omega, \ang}(\rtor),
\ee
where $\zeta_{\omega, \ang}(\rtor)$ is a sum of plane waves near the horizon. 
Using these wave-functions, the bulk field can also be written as
\be
\label{transferoutside}
\phi(t, \rtor, \Omega) = \sum_{\ang, \omega} {a_{\omega, \ang} \over \sqrt{\omega}}  \zeta_{\omega, \ang}(\rtor) e^{-i \omega t} Y_\ang(\Omega) + \text{h.c},
\ee
where, 
\be
\label{aomegaop}
a_{\omega, \ang} = \dzero \sqrt{\omega} \op_{\omega, \ang}.
\ee
With a little algebra, it is not difficult to see that (with an appropriate $\omega$-independent and $\ell$-independent normalization for the spherical harmonics) $\phi$ obeys the canonical commutation relations if the operators $a_{\omega, \ang}$ satisfy
\be
[a_{\omega, \ang}, a^{\dagger}_{\omega', \ang'}] = \delta_{\omega, \omega'} \delta_{\ang, \ang'}.
\ee
Note that this implies that\footnote{This relates a property of the bulk wave-function specified in \eqref{d0deltadef} to the commutator of boundary operators in a thermal state; this is a prediction of AdS/CFT and can be verified in several cases. (See, for example, appendix A of \cite{Papadodimas:2012aq}.)}
\be
\label{commutatordzero}
\langle \Psi | [\op_{\omega, \ell}, \op_{\omega', \ell'}^{\dagger}] | \Psi \rangle= {1 \over Z(\beta)} \tr\left(e^{-\beta H}[\op_{\omega, \ell}, \op_{\omega', \ell'}^{\dagger}] \right) =   {1 \over \omega |\dzero|^2} \delta_{\omega, \omega'} \delta_{\ell,\ell'}.
\ee

Now we turn to  the analyticity properties of $\zeta_{\omega, \ang}$. At finite $\omega$, since $\chi_{\omega, \ang}$ has no singularities, the singularities of $\zeta(\omega, \ang)$ can only come from the zeroes of  $\dzero$ since. But these zeroes have a nice physical interpretation. They correspond to the {\em quasinormal} modes of the black hole. This is because $\dzero$ is defined as the coefficient of the outgoing wave, $e^{i \omega \rtor}$, in \eqref{d0deltadef}. So, when $\dzero$ vanishes,  the mode becomes purely ingoing at horizon. 

Note that the vanishing of $\dzero$ does not mean that the entire mode in \eqref{d0deltadef} vanishes. The phase factor can be written as $e^{2 i \delta_{\omega, \ell}} = {\dminzero \over \dzero}$, and so it has poles corresponding to the complex zeroes of $\dzero$.

The quasinormal modes have the property that they all have  $\text{Im}(\omega) < 0$. (See \cite{Horowitz:1999jd} for a proof.) Therefore we see that $\zeta_{\omega, \ang}$ has poles when $\omega$ is in the lower-half plane, and it is regular in the upper-half plane. 

One subtle point has to do with the behaviour of $\dzero$  when $\omega$ approaches infinity through the upper-half plane. In this limit, where $|\omega| \gg 1$, while $\text{Im}(\omega) > 0$, the potential $\vbd$ becomes irrelevant and we can read off $\dzero$ from $\chi^0_{\omega, \ang}(\rtor)$ and we see that it approaches a constant and does not go to 0.

\subsection{Analysis near the horizon  \label{subsechor}}
We now consider the bulk modes from a second perspective starting with the horizon and solving outwards. This allows us to define what we mean by the ``right moving'' modes and analyze their analytic properties. 

Near the horizon, it is convenient to write
\be
\vhor(\rtor) = -\left[{(d-3)(1-d) \over 4} {f(r)^2 \over r^2} - m^2 f(r) - l(l+d-2){f(r) \over r^2} + {1 - d \over 2} {f'(r) f(r) \over r} \right],
\ee
and write the differential equation for the radial mode as
\be
\label{horequation}
{\partial \zeta^{\pm}_{\omega, \ang}(\rtor) \over \partial \rtor^2} + \omega^2 \zeta^{\pm}_{\omega, \ang}(\rtor) = \vhor(\rtor) \zeta^{\pm}_{\omega, \ang} (\rtor),
\ee
where the superscript $\pm$ distinguishes the two independent solutions of this equation.

We now construct $\zeta^{+}_{\omega, \ang}(\rtor)$, which corresponds to the right-moving mode. The construction of the left-movers is similar. In the absence of the potential, the equation \eqref{horequation} is solved by
\be
\zeta^{+0}_{\omega, \ang}(\rtor)  = e^{i \omega \rtor},
\ee
and $\zeta^{+}_{\omega, \ang}(\rtor)$ is the completion of this solution. 

The Green's function for \eqref{horequation} can be written as
\be
\ghor(\rtorp, \rtor)={1 \over \omega} \sin(\omega(\rtor - \rtorp)) \theta(\rtor - \rtorp).
\ee
We then again write an iterative solution for this wave-function as
\be
\zeta^{+,(n)}_{\omega, \ang}(\rtor) =  \int_{-\infty}^{\rtor} \zeta^{+,(n-1)}_{\omega, \ang}(\rtorp) \vhor(\rtorp) \ghor(\rtorp, \rtor) d \rtorp,
\ee
with the full solution being given by
\be
\zeta^+_{\omega, \ang}(\rtor) = \sum_{n=0}^{\infty} \zeta^{+,(n)}_{\omega, \ang}(\rtor).
\ee

Now, we note that the potential can be expanded as
\be
\label{potentialaplace}
\vhor(\rtor) = \sum_{m > 0} V_m e^{4 \pi m \rtor \over \beta},
\ee
To see this, it is most convenient to switch to the Kruskal coordinates. The potential only depends on the product $\ukrus \vkrus$ and not on the ratio $\vkrus/\ukrus$. Near the horizon, we can explicitly derive an expansion of the form \eqref{potentialaplace}. We now expect this horizon to converge till the nearest pole of $\vhor$. The closest poles of $\vhor$ occur at the singularity where $|\ukrus \vkrus| = e^{4 \pi \rtorsin \over \beta}$, and we expect this series expansion to be valid till this point.

In the region where \eqref{potentialaplace} is valid, we can easily determine the analytic form of $\zeta^{+, (n-1)}_{\omega, \ang}(\rtor)$. For example, $\zeta^{+,(1)}_{\omega, \ang}(\rtor)$ is simply given by
\be
\begin{split}
\zeta^{+, (1)}_{\omega, \ang}(\rtor) &= {1 \over 2 i \omega} \int_{-\infty}^{\rtor} d \rtorp e^{i \omega \rtorp} \left(e^{i \omega(\rtor - \rtorp)} - e^{i \omega (\rtorp - \rtor)}  \right) V_m e^{4 \pi m \rtorp \over \beta} \\
&= {e^{i \omega \rtor}  \over 2 i \omega} \sum_{m >0} V_m e^{4 \pi m \rtor \over \beta} \left({\beta  \over 4 \pi m} - {1 \over 2 i \omega + {4 \pi m \over \beta}} \right)  \\
&= \sum_m {\beta V_m e^{(i \omega + {4 \pi m \over \beta} ) \rtor} \over 4 \pi m (2 i \omega + {4 \pi m \over \beta})}.
\end{split}
\ee
Thus, we only get poles at $\omega = {2 \pi i m \over \beta}$ for $m$ a positive integer. We can now easily prove this by induction for all $m$. Let us say that we have proved that $\zeta^{+, (n)}_{\omega, \ang}(\rtor) = \sum D^{+, (n)}_m(\omega, \ang) e^{(i \omega + {4 \pi m \over \beta}) \rtor}$ where $D^{+, (n)}_m(\omega, \ang)$ has poles only for $\omega = {2 \pi i m \over \beta}$. Then we see that
\be
\zeta^{+, (n+1)}_{\omega, \ang}(\rtor) = \sum_{m, m'} D^{+, (n)}_m(\omega, \ang) {\beta V_{m'} e^{(i \omega + {4 \pi m \over \beta} + {4 \pi m' \over \beta}) \rtor} \over 4 \pi (m + m') (2 i \omega + {4 \pi (m + m') \over \beta})},
\ee
and therefore it also has poles only for $\omega = {2 \pi i q \over \beta}$, where $q$ is a positive integer. 

The conclusion is that the full solution $\zeta^{+}_{\omega, \ang}(\rtor)$  can be written as
\be
\zeta^{+}_{\omega, \ang}(\rtor) = e^{i \omega \rtor} D^{+}_{\omega, \ang}(\rtor),
\ee
where the function $D^{+}_{\omega, \ang}(\rtor)$ 
has poles only when ${-i \omega \beta \over 2 \pi}$ is a positive integer. 

We can obtain another solution $\zeta^{-}_{\omega, \ang}(\rtor)$ by repeating the procedure above, through 
\be
\begin{split}
&\zeta^{-0}_{\omega, \ang}(\rtor) = e^{-i \omega \rtor}; \\
&\zeta^{-,(n)}_{\omega, \ang}(\rtor) =  \int_{-\infty}^{\rtor} \zeta^{-,(n-1)}_{\omega, \ang}(\rtorp) \vhor(\rtorp) \ghor(\rtorp, \rtor) d \rtorp; \\
&\zeta^{-}_{\omega, \ang}(\rtor) = \sum_{n=0}^{\infty} \zeta^{-,(n)}_{\omega, \ang}(\rtor).
\end{split}
\ee
This is simply related to the solution above through $\zeta^{-}_{\omega, \ang}(\rtor) = \left( \zeta^{+}_{\omega, \ang}(\rtor) \right)^*$ and the mode $\zeta^{-}_{\omega, \ang}$ only has poles when ${i \omega \beta \over 2 \pi}$ is a positive integer.

\paragraph{Transfer function using horizon modes \\}
We can now write a transfer function using the modes $\zeta^{\pm}_{\omega, \ang}(\rtor)$  above. They are related quite simply to the mode $\zeta_{\omega, \ang}(\rtor)$ that we introduced above through \footnote{The careful reader may note that, in terms of the factors of $e^{i \delta_{\omega, \ang}}$ we are using a different convention from \cite{Papadodimas:2013jku,Papadodimas:2013wnh,Papadodimas:2015xma}. This is because we want to normalize the coefficient of the ``right-moving'' wave-function to 1. As a result our treatment of the left and right movers is not symmetric.}
\be
\zeta_{\omega, \ang}(\rtor) = \zeta^{+}_{\omega, \ang}(\rtor) + e^{2 i \delta_{\omega, \ang}} \zeta^{-}_{\omega, \ang}(\rtor).
\ee
The field outside the horizon can therefore be written as
\be
\label{leftrightdecomp}
\phi( t, \rtor, \Omega) =  \phi_{l}(r_*, t, \Omega) + \phi_{r}(r_*, t, \Omega), 
\ee
where
\be
\label{defleftrightmov}
\begin{split}
\phi_{r}(t, \rtor, \Omega) &= \sum_{\ang, \omega}  { a_{\omega, \ang} \over \sqrt{\omega}} \zeta^+_{\omega, \ang}(\rtor) e^{-i \omega t} Y_\ang(\Omega) + \text{h.c}, \\
\phi_{l}(t, \rtor, \Omega) &= \sum_{\ang, \omega}  {a_{\omega, \ang}  \over \sqrt{\omega}}  e^{2 i \delta_{\omega, \ang}}  \zeta^-_{\omega, \ang}(\rtor) e^{-i \omega t} Y_\ang(\Omega) + \text{h.c}.
\end{split}
\ee

We would like to make two important comments. First, while the factor of $e^{2 i \delta_{\omega, \ang}}$ can be fixed by comparing the solutions above to 
$\zeta_{\omega, \ang}(\rtor)$, it can also be fixed, simply by requiring the normalizability condition
\be
\lim_{\rtor \rightarrow 0} \big[\phi_l(t, \rtor, \Omega) + \phi_r(t, \rtor, \Omega)\big] = 0.
\ee
Second, note that the functions $\zeta^{+}_{\omega, \ang}(\rtor)$ and $\zeta^{-}_{\omega, \ang}(\rtor)$ individually have different analytic properties from $\zeta_{\omega, \ang}(\rtor)$. While the analytic properties of $\zeta^{+}_{\omega, \ang}(\rtor)$ will also important below, the reader should keep in mind that the analytic properties of the full wave-function outside are controlled by those of $\zeta_{\omega, \ang}(\rtor)$ and not by the components $\zeta^{\pm}_{\omega, \ang}(\rtor)$.

\subsection{Crossing the horizon \label{subsecbehindhor}}
Finally, we turn to the construction of the bulk field behind the horizon. 

As we reviewed in detail in \cite{Papadodimas:2015jra,Papadodimas:2013jku}, behind the horizon the field again consists of left and right movers. By continuity, the left movers from outside the horizon cross over smoothly to the black hole interior. More precisely, we can write the field as 
\be
\phi(t, \rtor, \Omega) = \tphi_{l}(r_*, t, \Omega) + \widetilde{\phi}_{r}(r_*, t, \Omega) , \quad \text{behind~the~horizon,}
\ee
where, the left and right moving elements are
\be
\begin{split}
&\tphi_{l}(t, \rtor, \Omega) =  \sum_{\omega, \ang} {a_{\omega, \ang} \over \sqrt{\omega}} e^{2 i \delta_{\omega,\ang}} e^{-i \omega t} Y_\ang(\Omega) \tzeta^{-}_{\omega, \ang} (\rtor) + \text{h.c},\\
&\widetilde{\phi}_{r}(t, \rtor, \Omega) =  \sum_{\omega, \ang} {\ta_{\omega, \ang} \over \sqrt{\omega}}e^{i \omega t} Y^*_\ang(\Omega) \tzeta^{-}_{\omega, \ang} (\rtor) + \text{h.c}.
\end{split}
\ee
 
Note that the operator modes $a_{\omega, \ang}$ that appear in $\tphi_l(t, \rtor, \Omega)$ are the same as those that appear outside the horizon. This is because the left-moving part of the field is continuous across the horizon.  Given the operators defined in \eqref{tocausdef}, these operators are defined through
\be
\ta_{\omega, \ang} =  (\dzero)^* \sqrt{\omega} \tO_{\omega, \ang}.
\ee
The $\tO_{\omega, \ang}$ operators are defined and discussed in detail in \cite{ Papadodimas:2013wnh, Papadodimas:2013jku, Papadodimas:2015jra, Papadodimas:2015xma}.

The functions $\tzeta^{-} = (\tzeta^{+})^* $ that appear in the field above have the near-horizon expansion from 
behind the horizon
\be
\tzeta^{-}_{\omega, \ang}(\rtor) = (\tzeta^{+}_{\omega, \ang})^*(\rtor) \underset{\rtor \rightarrow -\infty}{\longrightarrow} e^{-i \omega \rtor}.
\ee

We can analyze these modes using the same techniques as the previous section, and the same Green's functions. This is because the ``free wave equation'' both in front and behind the horizon is the same. So, the starting point of the iteration is just given by the functions $e^{\pm i \omega \rtor}$ and the Green's function that is used to construct the full solution is the same both in front and behind the horizon.

The potential can again be expanded in a series of the form \eqref{potentialaplace}. As we mentioned above this is really an expansion in the variable $(\ukrus \vkrus)$; the value of $\ukrus \vkrus$ at a point $(t, \rtor)$ behind the horizon is simply the negative of its value at a point with the same Schwarzschild coordinates outside the horizon. Since $\vhor$ is analytic across $\ukrus \vkrus= 0$, we can simply analytically continue the series to find that
\be
\label{potentialaplacebehind}
\vhor(\rtor) = \sum_{m > 0} (-1)^m V_m e^{4 \pi m \rtor \over \beta}, \quad \text{behind~the~horizon,}
\ee
where the coefficients $V_m$ are the same as those that appear in \eqref{potentialaplacebehind} but we have an additional factor of $(-1)^m$.

From this point the analysis of the analytic properties of $\tzeta^{-}$ and $\tzeta^{+}$ is exactly the same as in the previous subsection. The additional factors of $(-1)^m$ above make no difference to the analytic properties. So we find that  
\be
\tzeta^{\pm}_{\omega, \ang}(\rtor)  = e^{\pm i \omega \rtor} \tD^{\pm}_{\omega, \ang}(\rtor),
\ee
where the function $\tD^{-}_{\omega, \ang}(\rtor)$ 
has poles only when ${i \omega \beta \over 2 \pi}$ is a positive integer and the function $\tD^{+}_{\omega, \ang}(\rtor)$ 
  only has poles when ${-i \omega \beta \over 2 \pi}$ is a positive integer.

The key property that we will need here is the following. Consider a correlator involving an insertion of a right moving part of the field behind the horizon. Then using the definition of $\ta_{\omega, \ang}$ and the results above we find
that 
\be
\label{tilderesult}
\begin{split}
&\langle \Psi | \widetilde{\phi}_r(t, \rtor, \Omega)  \al_{\alpha} | \Psi \rangle = \langle \Psi | \al_{\alpha} \widetilde{\phi}_r(t, \rtor, \Omega) | \Psi \rangle  =
\langle \Psi | \al_{\alpha} e^{-{\beta H \over 2}} \phihat(t, \rtor, \Omega)  e^{\beta H \over 2} | \Psi \rangle,
\end{split}
\ee
where 
\be
\label{hatdef}
\phihat(t, \rtor, \Omega) =  \sum_{\omega, \ang} {a_{\omega, \ang} \over \sqrt{\omega}} \tzeta_{\omega, \ang}^+(\rtor) e^{-i \omega t} Y_{\ang}(\Omega) + \text{h.c.}
\ee
Note that $\phihat$ is made up entirely of ordinary operators.
To derive this we simply used the definition of
the mirror operators in \eqref{tocausdef} to first commute the mirror field through the ordinary field and then substitute the modes with a conjugated version of the ordinary modes. Note that whenever we use the field $\phihat(t, \rtor, \Omega)$ it is understood that the coordinates are localized in $\vcaus$.

\subsection{Commutators  \label{subseccommut}}
We are now in a position to prove an important intermediate technical result. If $\ucaus$ is a unitary made out of simple operators localized in $\bcaus$  and  $(t_1, \rtor_1, \Omega_1) \in \vcaus$ then inside any correlator
\be
\ucausd \phihat(t_1, \rtor_1, \Omega_1) \ucaus =  \phihat(t_1, \rtor_1, \Omega_1).
\ee
This can be proved by showing that for any point $(t_2, \Omega_2) \in \bcaus$  we have
\be
\label{commutlemma}
[{\phihat}(t_1, \rtor_1, \Omega_1), \op(t_2, \Omega_2)] = 0.
\ee

To see \eqref{commutlemma} we use the field expansions \eqref{defleftrightmov} and \eqref{transferoutside} and the mode commutators to find that the commutator of two field operators 
can be written as
\be
\label{c12def}
\begin{split}
&[\phihat(t_1, {\rtor}_1, \Omega_1), O(t_2, \Omega_2)] = \lim_{{\rtor}_{2} \rightarrow 0} ({\rtor}_2)^{-\Delta} [\phihat(t_1, {\rtor}_1, \Omega_1), \phi(t_2, {\rtor}_2, \Omega_2)]
\\ &=   \lim_{{\rtor}_{2} \rightarrow 0} ({\rtor_2})^{-\Delta} \sum_{\ang, \omega } 
e^{-i \omega(t_1 - t_2)} Y_{\ang}(\Omega_1) Y_{\ang}^*(\Omega_2) \omega^{-1} \tzeta^{+}_{\omega, \ang}(\rtor_1) \zeta_{\omega, \ang}(\rtor_2) - \text{h.c.}
\end{split}
\ee
As a consequence of the  definitions of $\zeta_{\omega, \ang}$ and   $\tzeta^{+}_{\omega, \ang}$, we can substitute
\be
\label{omeganegdef}
\zeta_{-\omega, \ang}(\rtor) \equiv \big(\zeta_{\omega, \ang}(\rtor)\big)^*, \quad \tzeta^{+}_{-\omega, \ang}(\rtor)  \equiv \big(\zeta^{+}_{\omega, \ang}(\rtor)\big)^*.
\ee
Second recall that the spherical harmonic can be conjugated by reversing some of the angular momentum quantum numbers, which we denote by $Y_{-\ang}(\Omega)=Y_{\ang}(\Omega)^*$. We also note that
\be
\label{lnegdef}
\zeta_{\omega, -\ang}(\rtor) = \zeta_{\omega, \ang}(\rtor), \quad \tzeta^{+}_{\omega, -\ang}(\rtor)  = \tzeta^{+}_{\omega, \ang}(\rtor).
\ee
Using  \eqref{omeganegdef} and \eqref{lnegdef} and by converting the sum over $\omega$ to an integral, we obtain an integral that runs over $\omega \in (-\infty, \infty)$:
\be
\begin{split}
[\phihat(t_1, {\rtor}_1, \Omega_1), O(t_2, \Omega_2)] &=  \lim_{\rtor_2 \rightarrow 0} \rtor_2^{-\Delta} \sum_{\ang} Y_{\ang}(\Omega_1) Y^*_{\ang}(\Omega_2) \int {d \omega \over \omega} \tzeta^+_{\omega, \ang}(\rtor_1) \zeta_{-\omega, \ang}(\rtor_2) e^{-i \omega(t_1 - t_2)}  \\
&= \sum_{\ang}   Y_{\ang}(\Omega_1) Y^*_{\ang}(\Omega_2) \int {d \omega \over \omega} e^{-i \omega(t_1 - t_2 - {\rtor}_1)} { \tD^+_{\omega, \ang}(\rtor) \over D^0(-\omega, \ang)}.
\end{split}
\ee
 The integral around $\omega = 0$ should be understood in the sense of a principal value; alternately the reader may simply wish to differentiate with respect to $t_1$ or $t_2$ to make the integrand regular there.  In either case we can neglect the potential pole there. Now we recall that $\tD^{+}_{\omega, \ang}(\rtor)$ has singularities only when $\omega$ is in the upper-half plane. $\big[\dzero\big]^{-1}$ has poles at the quasinormal mode frequencies when $\omega$ is in the lower-half plane, and therefore $\big[\dminzero]^{-1}$ which appears above has poles only in the upper-half plane. 

Moreover, with the points localized so that $(t_1, {\rtor}_1, \Omega_1) \in \vcaus$ and $(t_2, \Omega_2) \in \bcaus$  we also have
\be
t_1 - t_2 - {\rtor}_1 \geq 0.
\ee
So the integrand is regular in the lower half $\omega$ plane and we can complete the contour through that region to prove \eqref{commutlemma}.

\subsection{Stability of correlators}
We are now in a position to analyze and prove the stability of position space AdS correlators under perturbations in the causal patch. Consider a correlator involving the mirror
operators in a state that has been excited by a unitary $\ucaus$ localized in the causal patch and another insertion of some local operators that we denote by $\al_{\alpha}$ below.
By the definition of the mirror operators, we have
\be
\label{mirrorwithunit}
\begin{split}
&\langle \Psi | \ucausd \widetilde{\phi}_r(t, \rtor, \Omega) \al_{\alpha}  \ucaus | \Psi \rangle = \langle \Psi | \ucausd \al_{\alpha} \widetilde{\phi}_r(t, \rtor, \Omega) \ucaus | \Psi \rangle \\ &=
\langle \Psi | \ucausd  \al_{\alpha}  \ucaus e^{-\beta H \over 2} \phihat(t, \rtor, \Omega) e^{\beta H \over 2} | \Psi \rangle.
\end{split}
\ee

To analyze this correlator, we first note that
\be
\label{unitarythrough}
e^{-{\beta H \over 2}} U e^{\beta H \over 2} |\Psi \rangle = U |\Psi \rangle + \Or[\beta \delta E],
\ee
where $\delta E$ is the change in the expectation value of the Hamiltonian under the action of $U$ that was defined in \eqref{lowenergyexc}. As the notation indicates, this holds whether or not $U$ is localized on $\bcaus$.  Although this is an intuitive relation, we prove this relation in appendix \ref{applowenergy}. 

Since the relation above is an equality on states, using \eqref{typicalthermal} we see that for any coarse grained operator $\al_{\gamma}$
\be
\begin{split}
&\tr(e^{-\beta H} \al_{\gamma} U) = Z(\beta) \langle \Psi| \al_{\gamma} U |\Psi \rangle \\
&= Z(\beta) \langle \Psi| \al_{\gamma} e^{-\beta H \over 2} U e^{\beta H \over 2} | \Psi \rangle + \Or[\beta \delta E] = \tr(e^{-{\beta H \over 2}} \al_{\gamma} e^{-\beta H \over 2} U) + \Or[\beta \delta E],
\end{split}
\ee
where we have used the cyclicity of the trace in the last step. 

Turning to the correlator in \eqref{mirrorwithunit}, and dropping terms of $\Or[\beta \delta E]$ we see that
\be
\begin{split}
  & Z(\beta)\langle \Psi | \ucaus \al_{\alpha} \ucausd e^{-{\beta H \over 2}} \phihat(t, \rtor, \Omega) e^{\beta H \over 2}  | \Psi \rangle =  \tr(e^{-{\beta H \over 2}} \ucaus \al_{\alpha} \ucausd e^{-{\beta H \over 2}} \phihat(t, \rtor, \Omega)) \\ &= \tr(e^{-\beta H} \phihat(t, \rtor, \Omega) e^{-{\beta H \over 2}} \ucaus \al_{\alpha} \ucausd e^{\beta H \over 2})   = \tr(e^{-\beta H} \phihat(t, \rtor, \Omega) e^{-{\beta H \over 2}} \ucaus \al_{\alpha} e^{-{\beta H \over 2}} \ucausd e^{\beta H})   \\ &= \tr(e^{-{\beta H \over 2}} \ucaus \al_{\alpha} e^{-{\beta H \over 2}} \ucausd \phihat(t, \rtor, \Omega))    = \tr(e^{-\beta H} \ucaus \al_{\alpha} e^{-\beta H \over 2} \phihat(t, \rtor, \Omega) \ucausd e^{\beta H \over 2})  \\ &= \tr(e^{-\beta H} \ucaus \al_{\alpha} e^{-{\beta H \over 2}} \phihat (t, \rtor, \Omega) e^{\beta H \over 2} \ucausd).
\end{split}
\ee
In this sequence of manipulations we have repeatedly used \eqref{unitarythrough}, the cyclicity of the trace and \eqref{commutlemma}. The end result means that
\be
\langle \Psi | \ucaus \al_{\alpha} \widetilde{\phi}_r(t, \rtor, \Omega) \ucausd  | \Psi \rangle = \langle \Psi | \ucaus \al_{\alpha} e^{-{\beta H \over 2}}  \phihat(t, \rtor, \Omega) e^{\beta H \over 2} \ucausd | \Psi \rangle + \Or[\beta \delta E].
\ee

In particular the {\em change} in the correlator under the excitation $\ucausd$ is given by
\be
\label{tildeinhatnoneq}
\begin{split}
&\langle \Psi | \ucaus \al_{\alpha} \widetilde{\phi}_r(t, \rtor, \Omega) \ucausd  | \Psi \rangle - \langle \Psi |\al_{\alpha} \widetilde{\phi}_r(t, \rtor, \Omega) | \Psi \rangle \\ &= \langle \Psi | \ucaus \al_{\alpha} e^{-{\beta H \over 2}} \phihat(t, \rtor, \Omega) e^{\beta H \over 2} \ucausd  | \Psi \rangle - \langle \Psi |\al_{\alpha} e^{-{\beta H \over 2}} \phihat(t, \rtor, \Omega) e^{\beta H \over 2} | \Psi \rangle  + \Or[\beta \delta E].
\end{split}
\ee
It is possible to derive a similar relation by moving the mirror operators to the left and replacing them with a hatted operator
\be
\label{tildeinhatadjnoneq}
\begin{split}
&\langle \Psi | \ucaus \al_{\alpha} \widetilde{\phi}_r(t, \rtor, \Omega) \ucausd  | \Psi \rangle - \langle \Psi |\al_{\alpha} \widetilde{\phi}_r(t, \rtor, \Omega) | \Psi \rangle \\ &= \langle \Psi | \ucaus e^{\beta H \over 2} \phihat(t, \rtor, \Omega) e^{-{\beta H \over 2}} \al_{\alpha}  \ucausd  | \Psi \rangle - \langle \Psi |e^{\beta H \over 2} \phihat(t, \rtor, \Omega) e^{-{\beta H \over 2}} \al_{\alpha} | \Psi \rangle  + \Or[\beta \delta E].
\end{split}
\ee

But the right hand sides of \eqref{tildeinhatnoneq} and \eqref{tildeinhatadjnoneq}  represent the change in ordinary (i.e. state-independent) operators. These obviously obey the constraints of \eqref{bornruleineq}. We now show using these constraints that the original correlator involving the mirrors also obeys these constraints. The last factor we need to check is the ``deviation'' that appears in \eqref{bornruleineq}.

To obtain a finite deviation, we need to smear the position space operator. We define
\be
\phihat(g) = \int \phihat(t, \rtor, \Omega) g(t, \rtor, \Omega) \, d t \, d \rtor \, d \Omega; \quad \phihat^{\dagger}(g) = \int \phihat (t, \rtor, \Omega) g^*(t, \rtor, \Omega) \, d t \, d \rtor \, d \Omega,
\ee
where $g$ is some function. 

If we replace the operator on the right hand side of \eqref{tildeinhatnoneq} with its smeared version, then its deviation is not larger than $\widehat{\sigma}_1$ where
\be
(\widehat{\sigma}_1)^2 =  {1 \over Z(\beta)} \tr\left(\phihat^{\dagger}(g) e^{-{\beta H \over 2}} \al^{\dagger}_{\alpha} \al_{\alpha} e^{-\beta H \over 2} \phihat(g) \right) - \left| {1 \over Z(\beta)} \tr\left(e^{-{\beta H \over 2}} \al_{\alpha} e^{-{\beta H \over 2}} \phihat(g) \right) \right|^2.
\ee
The deviation could be smaller than this due to the other possible ordering in \eqref{stddev}.

Similarly, by replacing the operator on the right hand side of \eqref{tildeinhatadjnoneq} with its smeared version, we see that its deviation is not larger than $\widehat{\sigma}_2$ where
\be
(\widehat{\sigma}_2)^2 = {1 \over Z(\beta)} \tr\left( \phihat(g) e^{-{\beta H \over 2}} \al_{\alpha} \al^{\dagger}_{\alpha} e^{-{\beta H \over 2}} \phihat^{\dagger}(g) \right) -  \left| {1 \over Z(\beta)} \tr\left(e^{-{\beta H \over 2}} \al_{\alpha} e^{-{\beta H \over 2}} \phihat(g) \right) \right|^2.
\ee

The minimum of $\widehat{\sigma}_1$ and $\widehat{\sigma}_2$  matches precisely with the deviation of the operator $\al_{\alpha} \widetilde{\phi}_r(g)$. That deviation is given by $\text{min}\big(\widetilde{\sigma}_1, \widetilde{\sigma}_2 \big)$ where
\be
\begin{split}
&(\widetilde{\sigma}_1)^2 = \langle \Psi|   \tphi_r^{\dagger}(g) \al^{\dagger}_{\alpha} \al_{\alpha} \tphi_r(g) | \Psi \rangle - \big|\langle \Psi | \al_{\alpha} \tphi_r(g) | \Psi \rangle\big|^2  = (\widehat{\sigma}_1)^2, \\
&(\widetilde{\sigma}_2)^2 = \langle \Psi|  \al_{\alpha} \tphi_r(g) \tphi_r^{\dagger}(g) \al^{\dagger}_{\alpha}| \Psi \rangle -  \big| \langle \Psi | \al_{\alpha} \tphi_r(g) |\Psi \rangle \big|^2
= (\widehat{\sigma}_2)^2.
\end{split}
\ee
The relations above follow immediately by applying \eqref{tilderesult}.

Now applying \eqref{bornruleineq} to the right hand sides of \eqref{tildeinhatadjnoneq} and \eqref{tildeinhatnoneq} we see that as $\beta \delta E \rightarrow 0$ we have
\be
\langle \Psi | \ucaus \al_{\alpha} \widetilde{\phi}_r(t, \rtor, \Omega) \ucausd  | \Psi \rangle - \langle \Psi |\al_{\alpha} \widetilde{\phi}_r(t, \rtor, \Omega) | \Psi \rangle \leq 2 \text{min}(\widetilde{\sigma}_1, \widetilde{\sigma}_2) \sqrt{\beta \delta E},
\ee
which is exactly what we need!

 To summarize, the conclusion is as follows. If we consider correlators and boundary excitations within a single causal patch, then the correlators involving the tildes can be converted to correlators of ordinary operators both in equilibrium and in non-equilibrium states up to errors that are negligible when the non-equilibrium state has only slightly higher energy than the equilibrium state. Since the correlators of ordinary operators clearly obey the ordinary constraints of statistical mechanics, so do the correlators of the tildes. 

This brings us to the end of our somewhat intricate proof of the stability of AdS correlators. 
 in the causal patch when the excitations are also localized on the boundary of the patch. We have therefore proved that correlators about an AdS black hole obey \eqref{bornruleineq} and the low-energy paradoxes that one might naively expect go away when we carefully analyze the constraints from causality.

We remind the reader that as an immediate corollary, if we start with an equilibrium state and then turn on arbitrary sources, bulk AdS correlators are stable and \eqref{bornruleineq} holds.

\section{The Born rule paradox and its resolution \label{secparadox}}
The calculations of section \ref{secfullads} are rather involved. So, as an example, we now consider a concrete situation
where we can examine a specific low energy excitation and calculate the response of the black hole geometry. This example demonstrates
the key aspects of the paradox and our resolution. We will make some simplifications to facilitate the analysis and make it more accessible than the analysis of the previous section; the trade off is that
our results will be less precise . 

We will consider the specific paradox described in \cite{Marolf:2015dia} by Marolf and Polchinski (MP), who argued as follows. We see from the relation \eqref{tocausdef} that 
\be
\label{taaentang}
\langle \Psi | \ta_{\omega', \ang'} a_{\omega, \ang} | \Psi \rangle = {e^{-{\beta \omega \over 2}} \over 1 - e^{-\beta \omega}}  \delta_{\omega \omega'} \delta_{\ang \ang'}.
\ee
This value of the correlator is a precondition for a smooth horizon. MP now considered the unitary operator $\ump = e^{i \theta N}$ where $N \approx a_{\omega, \ang} a^{\dagger}_{\omega, \ang}$ is proportional to the number operator for a given mode outside the horizon. Then, $[N, H] \approx 0$, but yet $\ump$ introduces a phase $e^{i \theta}$ in the correlator above because it rotates $a_{\omega, \ang}$ but not $\ta_{\omega', \ang'}$.  Thus we seem to have a contradiction with \eqref{bornruleineq}.

MP termed this paradox  a violation of ``Born rule'' because  \eqref{bornruleineq} holds only for linear operators. So, MP suggested that to ensure that generic states have smooth horizons, the operators \eqref{tocausdef} would have to be significantly non-linear,  which would create difficulties for the quantum mechanical interpretation of the interior. The other alternative, that MP have advocated in previous papers,  was that generic black hole states in AdS/CFT do not have an interior at all, or have a firewall,  so that the operators \eqref{tocausdef} are not relevant for generic states.

The results of section \ref{secfullads} imply that if we only consider  observables that respect the constraints of bulk causality, the violations of the ``Born rule'' should vanish. To see this directly,  we translate the 
paradox above to position space by considering a unitary operator
\be
\begin{split}
&\ump = e^{i \theta N}; \\
&N = \int_{\bcaus} \op(t_1, \Omega_1) \op(t_2, \Omega_2) G(t_1, \Omega_1) G^*(t_2, \Omega_2)  \, d t_1 d t_2 d \Omega_1 d \Omega_2,
\end{split}
\ee
where the integral runs only over $\bcaus$ because the excitation is supported only on the boundary of the causal patch. We demand that the function $G$ be sharply peaked in frequency space; for example  $G(t, \Omega) = |\bcaus|^{-1 \over 2} e^{i \omega_0 t} Y_\ang(\Omega)$ where the normalization ensures that with $\theta = \Or[1]$, the unitary increases the energy only by a small amount. The specific form of $G$ will not be important below. 

In this section we will check that if we consider near-horizon correlators of bulk fields, $\phi(t, \rtor, \Omega)$, in the causal patch, which in particular implies that $t - \rtor > \tcaus$, then such correlators are {\em not} affected strongly by $\ump$ and change only by order $\sqrt{\beta \delta E}$ as predicted by \eqref{bornruleineq}. This result, of course, already follows from the general analysis of section \ref{secfullads} but here we show it through a simple and direct calculation.

As above, this has two related but distinct implications. The main result it that starting with an equilibrium state, it is impossible for any combination of observers to act with $\ump$ and also observe the strong bulk excitation that would naively be predicted by examining the frequency-space effect of this operation. Second, if we consider a state where the excitation $\ump$ appears spontaneously and we constrain the infalling  observer's worldline with the requirement that at least one other observer should be able to detect the excitation fully on the boundary, and report it to the infalling observer, we find that the infaller finds no violations of \eqref{bornruleineq}.

Before proceeding we wish to point out another significant difficulty with the argument of \cite{Marolf:2015dia}. The excitation $\ump$ cannot be produced in an active sense by simply modifying the Hamiltonian on the boundary as shown in \eqref{hdeformed}. The reason for this is that $N$ is a double-trace excitation on the boundary that is {\em non-local} in time. So,  ``acting'' with such an operator would require the observer at one point of time to have the ability to act with operators localized at a different point of time.  We might imagine that such an excitation could be produced by coupling the CFT to an external system. But then such a coupling would generate entanglement with that system, and not leave the CFT in a pure state.

Nevertheless, we proceed with $\ump$ in the form above, since this is the unitary operator considered in \cite{Marolf:2015dia}. It does not affect our resolution, the important aspect in the analysis above is causality---namely that an observer, or army of observers who interact with the boundary till a time $\tcaus$ can produce boundary excitations supported on the past of $\tcaus$ but not those supported on the future of $\tcaus$.

We define
\be
F(\omega, \ang ) = \int_{\bcaus}  G(t, \Omega) e^{-i \omega t} Y_{\ang}(\Omega) \, d t d \Omega; \quad F(-\omega, \ang ) = \int_{\bcaus}  G(t, \Omega) e^{i \omega t} Y_{\ang}(\Omega) \, d t d \Omega.
\ee
Note that we have reversed the sign conventions in the Fourier transform compared to \eqref{opmodedef} to avoid having to carry some minus signs below. 
We consider the case where 
where $G(t, \Omega)$ is largely made up of positive frequencies, and has only a negligible negative frequency component. This means that we approximate
\be
\label{positivefreq}
G(t, \Omega) \approx \sum_{\ang, \omega} F(\omega, \ang) e^{i \omega t} Y^*_{\ang}(\Omega).
\ee
where the sum runs only over positive frequencies as per our summation conventions. 
Conversely, $G(t, \Omega)^*$ effectively comprises only negative frequencies. 
Note that since $G(t, \Omega)$ vanishes outside $\bcaus$, then we cannot make $F(-\omega, \ang)$ exactly zero since the Fourier transform of a function with compact support always has both positive and negative frequencies. But even in that case, we can make ${|F(-\omega, \ang)| \over |F(\omega, \ang)|}  \sim  |\bcaus|^{-1}$ for the typical frequencies on which $F$ has support. This is small enough for our purposes. The second approximation that we will need below is that $F(\omega, \ang)$ is sharply peaked about some particular frequency $\omega_0$ and dies off rapidly as we move away from $\omega_0$. This ensures that $N$ behaves like a number operator for $\omega_0$. We will denote the spread in $F(\omega, \ang)$ by $\delta \omega$. 

Within this approximation, we can write
\be
\begin{split}
&N= \sum_{\ang_{i}, \omega_i}  \,  \op_{\omega_{1},\ang_{1}} \op_{\omega_{2},\ang_{2}}^{\dagger} F(\omega_{1}, \ang_{1}) F^*(\omega_{2}, \ang_{2}),
\end{split}
\ee
where we remind the reader that following our summation conventions $\omega_1 > 0, \omega_2 > 0$ in the sum above.

First, let us work out the change in the expectation value of the Hamiltonian. To first order we are interested in 
\be
\begin{split}
[N, H] &= \sum_{\ang_{i}, \omega_{i}} F(\omega_{1}, \ang_{1} ) F^*(\omega_{2}, \ang_{2}) \left( [\op_{\omega_{1}, \ang_{1}}, H] \op_{\omega_{2}, \ang_{2}}^{\dagger} + \op_{\omega_{1}, \ang_{1}} [\op_{\omega_{2}, \ang_{2}}^{\dagger}, H] \right) \\
&= \sum_{\ang_{i}, \omega_{i}}  F(\omega_{1}, \ang_{1}) F^*(\omega_{2}, \ang_{2}) (\omega_{1} - \omega_{2}) \op_{\omega_{1}, \ang_{1}} \op_{\omega_{2}, \ang_{2}}^{\dagger}.
\end{split}
\ee
In a typical state, by time-translational invariance,
\be
\langle \Psi | \op_{\omega_{1}, \ang_{1}} \op_{\omega_{2}, \ang_{2}}^{\dagger} | \Psi \rangle = G_{\omega_{1}, \ang_1} \delta_{\omega_1, \omega_2} \delta_{\ang_{1} \ang_{2}}, 
\ee
where $G_{\omega, \ang}$ --- the two point function of boundary modes --- is defined by the equation above.
The expectation value of the commutator above vanishes because when $\omega_{1} = \omega_{2}$ the summand vanishes.
\be
\langle \Psi | [N,H]| \Psi \rangle = 0.
\ee

To work out the second order term we need to compute
\be
[N, [N, H]] = \sum_{\omega_i,\ang_i }  F(\omega_1, \ang_1) F(\omega_2, \ang_2)^* (\omega_3 - \omega_4) F(\omega_3, \ang_3) F^*(\omega_4, \ang_4)  [\op_{\omega_1, \ang_1} \op_{\omega_2, \ang_2}^{\dagger}, \op_{\omega_3, \ang_3} \op_{\omega_4, \ang_4}^{\dagger}].
\ee
The commutator evaluates to
\be
\begin{split}
[\op_{\omega_1, \ang_1} \op_{\omega_2, \ang_2}^{\dagger}, \op_{\omega_3, \ang_3} \op_{\omega_4, \ang_4}^{\dagger}] &= \op_{\omega_1, \ang_1} [\op_{\omega_2, \ang_2}^{\dagger}, \op_{\omega_{3}, \ang_{3}}] \op_{\omega_{4}, \ang_{4}}^{\dagger} + [\op_{\omega_{1}, \ang_{1}}, \op_{\omega_{3}, \ang_{3}}] \op_{\omega_{2}, \ang_{2}}^{\dagger} \op_{\omega_{4}, \ang_{4}}^{\dagger} \\ & + \op_{\omega_{3}, \ang_{3}}[\op_{\omega_{1}, \ang_{1}}, \op_{\omega_{4}, \ang_{4}}^{\dagger}] \op_{\omega_{2}, \ang_{2}}^{\dagger} + \op_{\omega_{3}, \ang_{3}} \op_{\omega_{1}, \ang_{1}} [\op_{\omega_{2}, \ang_{2}}^{\dagger}, \op_{\omega_{4}, \ang_{4}}^{\dagger}] \\
&= C_{\omega_1, \ang_1} \op_{\omega_{3}, \ang_{3}} \op_{\omega_{2}, \ang_{2}}^{\dagger} \delta_{\omega_1, \omega_4} \delta_{\ang_1, \ang_4}  -C_{\omega_{2}, \ang_{2}} \op_{\omega_{1}, \ang_{1}} \op_{\omega_{4}, \ang_{4}}^{\dagger} \delta_{\omega_2, \omega_3}\delta_{\ang_2, \ang_3} ,
\end{split}
\ee
where we have used
\be
[\op_{\omega_1, \ang_1}, \op_{\omega_2, \ang_2}^{\dagger}] \equiv C_{\omega_1, \ang_1} \delta_{\omega_1, \omega_2} \delta_{\ang_1, \ang_2},
\ee
and all other commutators vanish. The form of $C_{\omega, \ang}$ was given in \eqref{commutatordzero}, but the precise expression
will not be important here. 
We now find that
\be
\langle \Psi | [N,[N,H]] | \Psi \rangle = \sum_{\omega_i, \ang_i} |F(\omega_{1}, \ang_{1})|^2 |F(\omega_{3}, \ang_{3})|^2 (\omega_{3} - \omega_{1}) \left[C_{\omega_{1}, \ang_{1}} G_{\omega_{3}, \ang_{3}} -C_{\omega_{3}, \ang_{3}} G_{\omega_{1}, \ang_{1}} \right].
\ee 
The KMS relation relates the expectation value of the commutator and the two point function through
\be
\begin{split}
\tr\Big(e^{-\beta H} \op_{\omega_1, \ang_1} \op_{\omega_2, \ang_2}^{\dagger} \Big) &= \tr\Big(\op_{\omega_2, \ang_2}^{\dagger} e^{-\beta H} \op_{\omega_1, \ang_1} \Big) = e^{\beta \omega_2} \tr\Big(e^{-\beta H} \op_{\omega_2, \ang_2}^{\dagger}  \op_{\omega_1, \ang_1} \Big) \\
&=  e^{\beta \omega_2} \tr\Big(e^{-\beta H}  \op_{\omega_1, \ang_1} \op_{\omega_2, \ang_2}^{\dagger} \Big) - Z(\beta) e^{\beta \omega_2} C_{\omega_1, \ang_1} \delta_{\omega_1, \omega_2}.
\end{split}  
\ee
Therefore
\be
G_{\omega, \ang} = {1 \over 1 - e^{-\beta \omega}} C_{\omega, \ang}.
\ee
This allows us to evaluate 
\be
\langle \Psi | \ump^{\dagger} H \ump | \Psi \rangle = \langle \Psi | H | \Psi \rangle - i \theta \langle \Psi | [N,H] | \Psi \rangle  - {\theta^2 \over 2} \langle \Psi| [N, [N, H]] | \Psi \rangle + \Or[\theta^3],
\ee
and using the relations above we see that 
\be
\begin{split}
\label{changeenergy}
\delta E &= \langle \Psi| \ump^{\dagger} H \ump |\Psi \rangle - \langle \Psi | H | \Psi \rangle \\ &= \sum_{\omega_1, \omega_3} C_{\omega_{1}, \ang_{1}} C_{\omega_{3}, \ang_{3}} |F(\omega_{1}, \ang_{1})|^2 |F(\omega_{3}, \ang_{3})|^2 (\omega_{3} - \omega_{1})  \left({1 \over 1 - e^{-\beta \omega_{1}}} - {1 \over 1 - e^{-\beta \omega_{3}}} \right),
\end{split}
\ee
neglecting terms of $\Or[\theta^3]$ and higher.

We see that this change in energy is rather small. In particular, when $F(\omega)$ is a sharply peaked function, with a spread in frequencies given by $(\delta \omega)$,  we have $\delta E \propto \beta (\delta \omega)^2$.

As the observable in \eqref{bornruleineq}, we take the two point function of the field. In the large-N limit this is the key observable, since all local correlators factorize into products of two point functions.  As we explained in the introduction, and in the previous section, the only two-point function that could be  potentially problematic has one point behind the horizon, and another point in front of the horizon.  This requires entanglement between the degrees of freedom outside and inside, which the unitary operator above is intended to break. 

We will work in the near-horizon region in this section, since that is where the effect of the unitary is the strongest. Of course, we emphasize that the proof of stability will work even without these approximations as we showed in section \ref{secfullads} and here we take this physically relevant limit only to simplify the algebra and permit explicit computations.

In the near horizon limit it is convenient to move to Kruskal coordinates \eqref{kruskaldef}. Then we see from the previous section that in the near-horizon limit the mode functions of section \ref{secfullads} simplify greatly and become proportional to $\zeta^{+}(\rtor) e^{-i \omega t} \rightarrow |\ukrus|^{i \beta \omega \over 2\pi}; \zeta^{-}(\rtor) e^{-i \omega t} \rightarrow \vkrus^{-i \beta \omega \over 2 \pi}$. The physics in the near-horizon limit becomes effectively two dimensional and so we consider derivatives of the scalar field to get smooth correlators. The product of two field operators near the horizon but on opposite sides, so that their Kruskal $\ukrus$ coordinates satisfy $U_1 < 0$ and $U_2 > 0$, can be written as 
\be
\label{mixedexpand}
 \partial_{U_1} \phi(U_1, V_1, \Omega_1) \partial_{U_2} \phi(U_2, V_2, \Omega_2)  = {\beta^2 \over 4 \pi^2 U_1 U_2} \sum_{\omega_i, \ang_i} (\omega_1 \omega_2)^{1 \over 2}  {\cal I}_{\omega_1, \omega_2, \ang_1, \ang_2} ,
\ee
with
\be
\label{mixedexpandintegrand}
\begin{split}
{\cal I}_{\omega_1, \omega_2, \ang_1, \ang_2}  &\equiv  a_{\omega_1, \ang_1} \widetilde{a}_{\omega_2, \ang_2} Y_{\ang_1}(\Omega_1) Y_{\ang_2}^*(\Omega_2) (-U_1)^{i \beta \omega_1 \over 2 \pi} (U_2)^{-i \beta \omega_2 \over 2 \pi}  \\ &+   a^{\dagger}_{\omega_1, \ang_1} \widetilde{a}^{\dagger}_{\omega_2, \ang_2}  (-U_1)^{-i \beta \omega_1 \over 2 \pi} (U_2)^{i \beta \omega_2 \over 2 \pi}  Y^*_{\ang_1}(\Omega_1) Y_{\ang_2}(\Omega_2)  + \ldots.
\end{split}
\ee
Here the $\ldots$ indicate terms with products like $a_{\omega, \ang} \ta^{\dagger}_{\omega', \ang'}$ that have no expectation value, and will
not pick up one under the action of $\ump$. 

Let us examine the change in the value of this correlator under the rotation above. In principle, this correlator is in danger of changing significantly because the ordinary mode gets rotated by $U_{\text{MP}}$ whereas the mirror mode does not. We need to verify that this effect vanishes within the causal patch.
Notice that
\be
\label{oancommut}
\begin{split}
&[N, a_{\omega_{1}, \ang_{1}} \widetilde{a}_{\omega_{2}, \ang_{2}}] = \sum_{\omega_3} F^*(\omega_{1}, \ang_{1}){-1\over D^0(\omega_1, \ang_1)^*  \sqrt{\omega_1}}  F(\omega_{3}, \ang_{3}) \op_{\omega_{3}, \ang_{3}} \widetilde{a}_{\omega_{2}, \ang_{2}}, \\
&[N, a^{\dagger}_{\omega_{1}, \ang_{1}} \widetilde{a}^{\dagger}_{\omega_{2}, \ang_{2}}] = \sum_{\omega_4} F(\omega_{1}, \ang_{1}) {1  \over D^0(\omega_1, \ang_1) \sqrt{\omega_1}} F(\omega_{4}, \ang_{4})^* O^{\dagger}_{\omega_{4}, \ang_{4}} \widetilde{a}_{\omega_{2}, \ang_{2}}^{\dagger}. \\
\end{split}
\ee
Here we have used the commutators that follow from \eqref{aomegaop}
\be
[\op_{\omega, \ang}, a^{\dagger}_{\omega', \ang'}] = {1 \over \dzero \sqrt{\omega}} \delta_{\omega \omega'} \delta_{\ang \ang'}.
\ee
Moreover, we see that
\be
\begin{split}
&\langle \Psi | \op_{\omega_{3}, \ang_{3}} \ta_{\omega_{2}, \ang_{2}}  | \Psi \rangle = {1 \over D^0(\omega_2, \ang_2) \sqrt{\omega_2}} {e^{-{\beta \omega_2 \over 2}} \over 1 - e^{-\beta \omega_2}} \delta_{\omega_2, \omega_3}  \delta_{\ang_2,\ang_3},\\
&\langle \Psi | \op_{\omega_{4}, \ang_{4}}^{\dagger} \ta_{\omega_{2}, \ang_{2}}^{\dagger} | \Psi \rangle =  {1 \over D^0(\omega_2, \ang_2)^* \sqrt{\omega_2}}  {e^{-{\beta \omega_2 \over 2}} \over 1 - e^{-\beta \omega_2}} \delta_{\omega_2, \omega_4}  \delta_{\ang_2, \ang_4}.\\
\end{split}
\ee

Therefore we see that, to first order in $\theta$,  the change in the correlator under the action of $\ump$ is given by
\be
\label{changecorr}
\begin{split}
& \langle \Psi| \ump \partial_{U_1} \phi(U_1, V_1,  \Omega_1) \partial_{U_2} \phi(U_2, V_2, \Omega_2) \ump^{\dagger} | \Psi \rangle -  \langle \Psi | \partial_{U_1} \phi(U_1, V_1,  \Omega_1) \partial_{U_2} \phi(U_2, V_2, \Omega_2) | \Psi \rangle   \\ &={i \beta^2 \theta \over 4 \pi^2 U_1 U_2}  \sum_{\omega_1, \ang_1}{Y_{\ang_1}^*(\Omega_1) \over D^0(\omega_1, \ang_1)}  F(\omega_{1}, \ang_{1}) (-U_1)^{-i \beta \omega_1 \over 2 \pi}  \sum_{\omega_2, \ang_2} {Y_{\ang_2}(\Omega_2) \over D^0(\omega_2, \ang_2)^* } F^*(\omega_{2}, \ang_{2}) (U_2)^{i \beta \omega_2 \over 2 \pi} {e^{-{\beta \omega_2 \over 2}} \over 1 - e^{-\beta \omega_2}} \\ &\hphantom{=} - \text{h.c}.
\end{split}
\ee
The relative minus sign between the term and its Hermitian conjugate comes from
the relative minus sign in the first and second line of \eqref{oancommut}.

Now, we recall from the previous section that $D^0(\omega, \ell)$ has zeroes only for $\text{Im}(\omega) < 0$ corresponding to the quasinormal frequencies of the black hole and is also bounded at large positive $\text{Im}(\omega)$. As a result we can write
\be
{1 \over D^0(\omega, \ell)} = \int_{x > 0} d(x, \ang) e^{i x \omega} d x,
\ee
for some $d(x, \ell)$. Using this, the first sum, over $\omega_1$, above is easy to deal with. 
\be
\begin{split}
&\sum_{\omega_1}{1 \over D^0(\omega_1, \ang_1)}  F(\omega_{1}, \ang_{1}) (-U_1)^{-i \beta \omega_1 \over 2 \pi} = \int_{x > 0} d(x,\ang_1)  \sum_{\omega_1} F(\omega_1, \ell_1) (-U_1)^{-{i \beta \omega_1 \over 2 \pi} } e^{i x \omega_1} \\ &= \int_{x > 0} \, d x  d \Omega \,  d(x, \ang_1) G(x - {\beta \ln(-U_1) \over 2 \pi} , \Omega) Y_{\ang_1}(\Omega)\end{split}
\ee
We see that this term vanishes unless $x - {\beta \ln(-U_1)  \over 2 \pi}  < \tcaus$. Since $x > 0$, this requires $-U_1 > e^{-2 \pi \tcaus \over \beta}$; therefore point $1$ must be in $\wcaus$ --- the causal wedge of $\bcaus$. This result, in fact, follows from a simple extension of the results obtained in the previous section. Since the fields $\phi_r$ have analytic properties very similar to $\phihat$, the effect of a unitary localized on $\bcaus$ vanishes on right movers that are localized between the causal wedge and the horizon.

The second sum, over $\omega_2$, is more interesting. To understand this sum, recall that we have assumed that $F(\omega)$ is narrowly peaked about some frequency $\omega_0$ with a spread $\delta \omega$. 
Now in this sum we see that since we are in the near-horizon region so that $U_2 \rightarrow 0$, the function $U_2^{i \omega_2 \beta \over 2 \pi}$ is rapidly varying, but on the other hand we can substitute the slowly varying $e^{-\beta \omega} \rightarrow e^{-\beta \omega_0}$ within the sum while making an error of only $\Or[\beta \delta \omega]$. So, 
\be
\begin{split}
&\sum_{\omega_2} { 1 \over D^0(\omega_2, \ang_2)^* } F^*(\omega_{2}, \ang_{2}) (U_2)^{i \beta \omega_2 \over 2 \pi} {e^{-{\beta \omega_2 \over 2}} \over 1 - e^{-\beta \omega_2}}  \\ &=  {e^{-\beta \omega_0} \over 1 - e^{-{\beta \omega_0}}} \int_{x > 0} \, d x  d \Omega \,  \Big[d(x, \ang_2) G(x-{\beta \over 2 \pi} \ln(U_2), \Omega) Y_{\ang_2}(\Omega) \Big]^* + \Or[\beta (\delta \omega)].
\end{split}
\ee
But now we see that since the second point is also in the causal patch, therefore  $U_2 < e^{-{2 \pi  \tcaus \over \beta}}$. Since $G$ has no support for times larger than $\tcaus$, the change in the correlator vanishes up to $\Or[\beta \delta \omega]$!

This leads us to the central result that we wanted to demonstrate: an observer in the causal patch cannot access the deep region where the effect of the unitary operator $\ump$ is strong if the observer also has time to observer the excitation on the boundary.  An immediate
corollary is that  if  an observer, or a set of observers, in an equilibrium state, attempt to act with $\ump$ by turning on a source and then jump into the bulk, they do not observe any anomalously large effect. 

We  would like to emphasize that it is not as if $U_{\text{MP}}$ has no effect on the correlator. As we see above, there is an effect at order $\beta \delta \omega = \sqrt{\beta \delta E}$.\footnote{We remind the reader that in the previous section, we proved that for arbitrary unitary excitations, which include $U_{\text{MP}}$,  that the coefficient of this $\Or[\sqrt{\beta \delta E}]$ term is also what is expected for the change in the correlation function of  ordinary operators. But we will not repeat this calculation in this simplified setting. } This is precisely what is expected from \eqref{bornruleineq}.  So, there is no issue of a ``frozen vacuum'' \cite{Bousso:2013ifa}: a low energy unitary on the boundary does have a non-zero effect on correlators in the causal patch, but this effect is bounded by the square root of the energy that the unitary injects into the system.

\section{Causal patch complementarity \label{seccomplement}}

In previous sections, we have proved that if one considers observables and excitations within a single causal patch, the paradoxes associated with low energy states vanish. As we explained, this has two implications. First, it implies that,  within effective field theory, starting with a black hole formed from collapse, no observer can excite the black hole with a local source and then observe such a paradox. But, more interestingly, our results also apply to states that are spontaneously excited. In such states, if we consider a setup where a combination of observers attempts to both confirm the presence of the excitation on the boundary and also observe bulk correlators, then the only bulk correlators accessible to such observers precisely obey the limit \eqref{bornruleineq} derived from statistical mechanics.

In this section---which is exploratory in nature and somewhat outside the main line of this paper--- we again focus on states that display a spontaneous excitation but  the important new element is that we consider observers who jump into the black hole and reach the singularity before the excitation appears on the boundary. More precisely, we  consider states of the form
\be
\label{neqstate}
|\psine \rangle = U(\tout) |\Psi \rangle,
\ee
where $|\Psi \rangle$ is a typical equilibrium state and $U(\tout)$ indicates a unitary that is localized  around  $\tout > \tcaus$. In the state \eqref{neqstate}, we now want to consider correlation functions of fields localized at some set of points $(t_i, \rtor_i, \Omega_i) \in {\cal C}$. 

In such states, it is easy to see that the description of physics in terms of the usual fields $\phi$ is problematic; the proof of the stability of correlators of $\phi$ that we provided in section \ref{secfullads} breaks down since its  assumptions are not met in this situation. In fact, the reader can check, by extending the calculation in  section \ref{secparadox}, 
that if $\tout > \tcaus$ then the difference,
\be
\langle \psine | \phi(t_1, \rtor_1, \Omega_1) \ldots \phi(t_n, \rtor_n, \Omega_n) |\psine \rangle - \langle \Psi | \phi(t_1, \rtor_1, \Omega_1) \ldots \phi(t_n, \rtor_n, \Omega_n) |\psi \rangle,
\ee
may not obey \eqref{bornruleineq} with  $\delta E = \langle \Psi| U^{\dagger}(\tout) H   U(\tout) |\Psi \rangle - \langle \Psi | H | \Psi \rangle \ll {1 \over \beta}$.

Now, if a superobserver has the ability to prepare arbitrary states in the CFT and can compare the experience of the infaller in the state $|\Psi \rangle$ with the experience of the infaller in the state $U(\tout) |\Psi \rangle$, then such a superobserver would detect that black holes violate the bound \eqref{bornruleineq}. One possibility, of course, is to simply ignore this problem since it requires the superobserver to either have the ability to prepare the exponentially unlikely states $U(\tout) |\Psi \rangle$, or else have the ability to act with $U(\tout)$  before the time $\tout$. Such a superobserver would also observer a violation of the second law of thermodynamics and so we may neglect the violation of \eqref{bornruleineq} as a relatively minor issue. 

However, states where the spontaneous excitation is inside the causal patch are also exponentially unlikely, but we were able to find a consistent description of physics in the causal patch in such states. So it is natural to ask whether we can extend this to states where the excitation is outside the causal patch.

\paragraph{\bf The Idea \\}
The idea is as follows. For states of the form \eqref{neqstate}, we postulate that the correct mapping between the bulk and the boundary requires us to specify not only the state but also the {\em causal patch} that we wish to examine in the bulk. We make this precise by  describing a new set of bulk observables $\phicaus$, whose relation to the boundary depends on the causal patch ${\cal C}$. These observables have the virtue that  \eqref{bornruleineq} is met in states of the form \eqref{neqstate} as well. On the other hand, since these operators $\phicaus$ depend on the causal patch, the price that we pay is that we lose the global picture of the black hole geometry that could simultaneously describe all causal patches. 

We do not prove, in this paper, that the operators $\phicaus$ are the correct operators for the infalling observer to use, and therefore we do not prove that this obstruction in the global reconstruction of the geometry exists. But, at the end of this section, we discuss briefly why we might expect such an obstruction to appear in spontaneously excited states.

\subsection{The Proposal}
To make the idea above precise, first, we define the algebra of simple polynomials localized on $\bcaus$. 
\be
\label{alsetcausdef}
\alsetcaus =  \text{span of}\{\op_{\ell_1}(t_1), \op_{\ell_2}(t_2)\op_{\ell_3}(t_3),  \ldots ,\op_{\ell_4}(t_4)\op_{\ell_5}(t_5) \cdots \op_{\ell_{\dmax}} (t_{\dmax}) \}, \quad t_i \in \bcaus.
\ee
This is the set of simple polynomials of single trace operators that are localized on $\bcaus$.  The cutoff $\dmax$ is the same as the cutoff that appeared in \eqref{alsetdef}. However, the important difference between $\alsetcaus$ and $\alset$ is that the times $t_i$ that appear above are restricted to lie in $\bcaus$. The reader will note that this is very similar to the small algebra ${\cal A}_{\text{small}}(B)$ that was defined in \cite{Banerjee:2016mhh}. 

\begin{figure}[!h]
\begin{center}
\includegraphics[height=0.3\textheight]{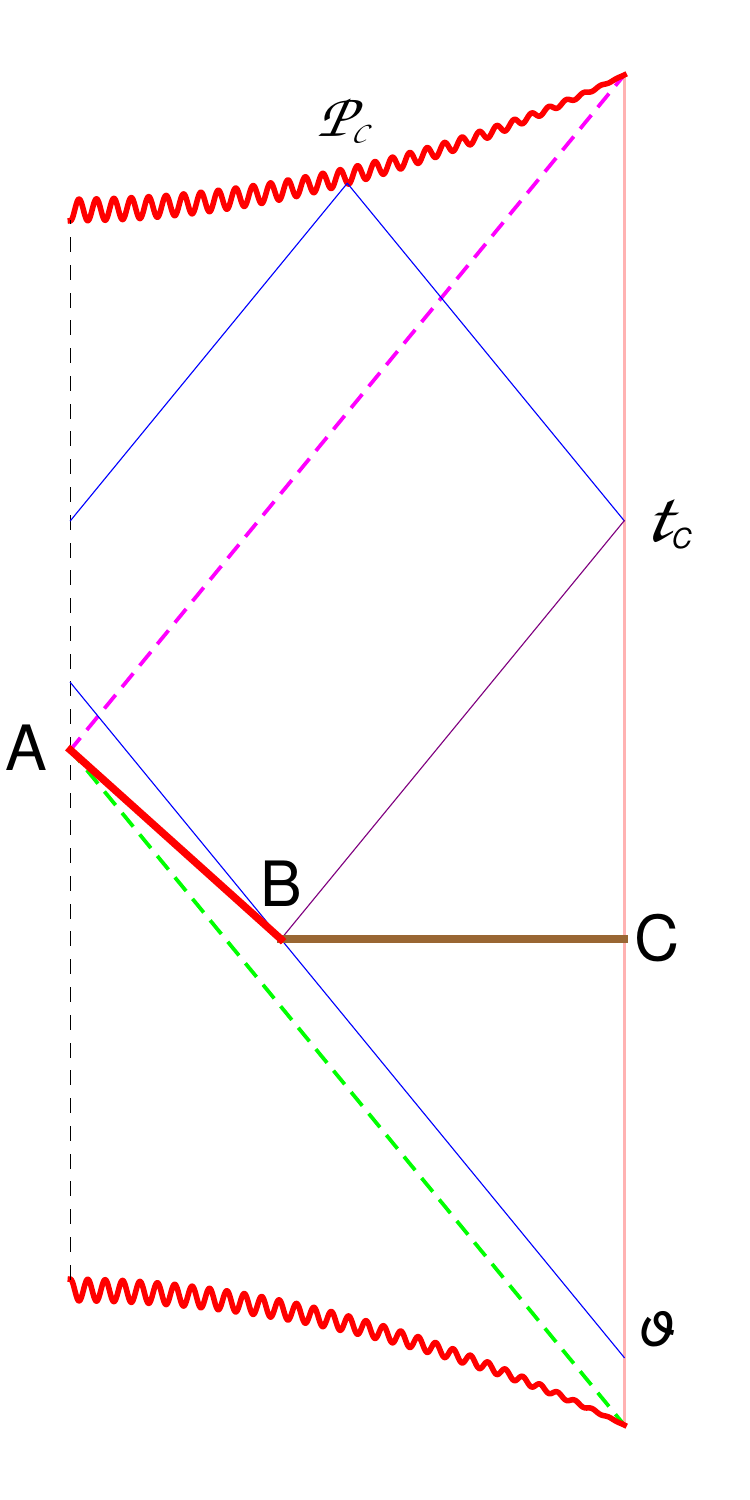}
\caption{\em The slice ABC gives a Cauchy slice for the entire geometry. BC provides complete initial data for the causal wedge $\wcaus$. In the large-N limit, operators on AB and BC commute. \label{figbulkdecom}}
\end{center}
\end{figure}

Next, we note that to leading order in ${1 \over N}$ the full algebra $\alset$ can be decomposed into two commuting subalgebras:
\be
\alset = \alsetcaus \otimes \albarsetcaus, \quad [\alsetcaus, \albarsetcaus] = 0.
\ee
The simplest way to understand the decomposition is through the bulk geometry.  Note that the relation \eqref{transout} gives us an explicit and $1-1$ mapping of all operators in $\alset$ to bulk operators outside the horizon. Now, using the bulk equations of motion, all bulk operators can be mapped to the Cauchy slice, $ABC$, outside the horizon in Figure \ref{figbulkdecom}. The operators in the segment $AB$ all commute with operators in $\alsetcaus$ because $AB$ is spacelike to $\bcaus$. On the other hand the algebra of operators on $BC$ is isomorphic to $\alsetcaus$. This gives us a decomposition of the boundary algebra into two commuting subalgebras.
As emphasized in \cite{Banerjee:2016mhh},  this decomposition holds only to leading order in ${1 \over N}$. So, our discussion here is only a leading order discussion.

We caution the reader that if we think of  elements of $\alset$ as smeared versions of local operators, then the decomposition of an operator into an element of $\alsetcaus$ and $\albarsetcaus$ does {\em not} correspond to simply selecting the part of the smearing function with support on $\bcaus$. This is because the commutator of single-trace operators is non-zero even for finite time-separation on the boundary.

For example, consider a single trace operator $\op(f) = \int O(t, \Omega) f(t, \Omega) d t d \Omega$, where $f(t, \Omega)$ has support both on $\bcaus$ and outside $\bcaus$. Then this operator can be decomposed into a sum of an element of $\alsetcaus$ and another element of $\albarsetcaus$ as follows. In figure \ref{figbulkdecom}, let the coordinates of $B$ be $(t_B, \rtor_B)$. ($B$ represents an entire $S^{d-1}$ sphere.) Consider the time slice $t_B$ that passes through the point $B$. 

Then we can write
\be
\op(f) = \int_{\rtor_B}^0 d \rtor \int d \Omega \phi(t_B, \rtor, \Omega) g(\rtor, \Omega) + \int_{-\infty}^{\rtor_B} d \rtor \int d \Omega \phi(t_B, \rtor, \Omega) g(\rtor, \Omega),
\ee
which represents the decomposition of this operator as a sum of an operator in $\alsetcaus$ and $\albarsetcaus$ respectively. Here, 
\be
g(\rtor, \Omega) =-i r^{d-1} [\op(f), {\partial \phi \over \partial t}(t_B, \rtor, \Omega)],
\ee
which follows just from using the canonical commutation relations for $\phi$.

Corresponding to the decomposition of algebras above, we consider a unitary operator that can be decomposed as
\be
U(\tout) = \ucaus_{\tout} \ucauscomp_{\tout},
\ee
where $\ucaus_{\tout} \in \alsetcaus$ and $\ucauscomp \in \albarsetcaus$. We will restrict ourselves to such unitaries in this section. This does not cover all unitaries $U(\tout)$ but it provides a broad class and the proposal below can easily be generalized to other cases. Note that as we explained above, just because the unitary $U(\tout)$ is largely localized outside $\bcaus$ does not mean that $\ucaus_{\tout} = 1$.

The unitary $\ucaus_{\tout}$ is the ``part'' of the unitary that is visible in the causal patch. This is because
\be
\langle \Psi | U(\tout)^{\dagger} \alcaus_{\alpha} U(\tout) | \Psi \rangle = \langle \Psi | \ucausd_{\tout} \alcaus_{\alpha} \ucaus_{\tout} |\Psi \rangle, \quad \forall \alpha.
\ee
So for an observer who conducts experiments on $\bcaus$, before jumping into the bulk, it {\em appears} as if the state has been excited just with $\ucaus_{\tout}$. 

Now the idea of causal patch complementarity is that in the causal patch ${\cal C}$ it is consistent to simply neglect $\ucauscomp_{\tout}$. More precisely, we suggest that physics in the state $|\psine \rangle$ in the causal patch ${\cal C}$ can be described through correlation functions of operators $\phicaus(t, \rtor, \Omega)$ that satisfy
\be
\label{causalpatchdef}
\langle \psine | \phicaus(t_1, \rtor_1, \Omega_1) \ldots \phicaus(t_n, \rtor_n, \Omega_n)  | \psine \rangle = \langle \Psi | \ucausd_{\tout}\phi(t_1, \rtor_1, \Omega_1) \ldots \phi(t_n, \rtor_n, \Omega_n)  \ucaus_{\tout} | \Psi \rangle.
\ee
It is possible to solve \eqref{causalpatchdef} to obtain the action of $\phicaus$ on states of ${\cal H}_{\Psi}$. Note that the operators $\phicaus$ are now state-dependent even outside the horizon.

If any of the points $(t_i, \rtor_i, \Omega_i) \notin {\cal C}$, then our proposal has nothing to say about such observables.   In such a situation, it may be possible to shift to a different causal patch ${\cal C'}$ so that all the points fit in that patch and use the field $\phi_{\cal C'}$ to describe these correlators. But,  it is important to note that it is impermissible to mix field operators defined with respect to different causal patches. So, the observables
\be
\langle \psine | \phicaus(t_1, \rtor_1, \Omega_1) \phi_{\cal C'}(t_2, \rtor_2, \Omega_2) \ldots | \psine \rangle,
\ee
where $C$ and $C'$ are two different causal patches are meaningless, except in some special cases that we describe below.

Let us mention some properties of this proposal, which also serve as basic consistency checks.  First, it is evident that in all equilibrium states, or states with an excitation only on $\bcaus$,  correlators of $\phicaus$  coincide with correlators of $\phi$. We write this as
\be
\phicaus(t, \rtor, \Omega) \doteq \phi(t, \rtor, \Omega), \quad \text{in~equilibrium~states,}
\ee
where the symbol $\doteq$ implies that correlators of $\phicaus$ coincide with correlators of $\phi$.   
Second, we note that even in arbitrary non-equilibrium states, the two operators coincide in the causal wedge of $\bcaus$.
\be
\phicaus(t, \rtor, \Omega) \doteq \phi(t, \rtor, \Omega), \quad (t, \rtor, \Omega) \in \wcaus.
\ee
This is because when the point $(t, \rtor, \Omega)$ is in $\wcaus$, the local field  $\phi(t, \rtor, \Omega) \in \alsetcaus$. So, it automatically commutes with $\ucauscomp_{\tout}$ in \eqref{causalpatchdef}. 

Next, we see that as a trivial consequence of this, $\phicaus$ obeys the correct boundary conditions:
\be
\label{causalgkpw}
\lim_{r \rightarrow \infty} \rtor^{-\Delta} \phicaus(t, \rtor, \Omega) \doteq O(t, \Omega), \quad \text{for}~(t, \Omega) \in \bcaus.
\ee
It is also clear  from \eqref{causalpatchdef} that $\phicaus(t, \rtor, \Omega)$ also obey the same equations of motion as $\phi(t, \rtor, \Omega)$.
\be
\label{causaleom}
(\Box - m^2) \phicaus(t, \rtor, \Omega) \doteq 0.
\ee
Perturbatively, if we add local interactions or sources in the action for $\phi$ then the equation of motion for $\phicaus$ will also be modified appropriately to match the equation of motion of $\phi$. 

The other property that is important to note is that the use of the fields $\phicaus$ may preclude a global description of the geometry in states $|\psine \rangle$. If we only want to describe the exterior of the black hole then, as we mentioned in section \ref{secsetup}, we may use the special causal patch ${\cal C}_{\infty}$ that contains only the exterior of the black hole. For this causal patch, the algebra $\alcaus$ coincides with $\alset$, and the field $\phicaus$ coincides with $\phi$. 

However no single causal patch accommodates the entire black hole geometry. In an equilibrium state or in an equilibrium state deformed with a source,  since $\phicaus \doteq \phi$ it is possible to consider correlation functions across different causal patches. But in a general non-equilibrium state, our proposal does not extend to correlators that do not fit in a single causal patch. In sacrificing a global description, we do not sacrifice any quantity that is observable for a bulk observer since, by construction, correlators that do not fit in any single causal patch cannot be measured by any combination of observers.

\subsection{Stability of causal-patch correlators in spontaneously excited states \label{subsecstability}}
An important result is that if we use the operators specified by \eqref{causalpatchdef}, we find that causal patch correlators are stable even in spontaneously excited states of the form \eqref{neqstate}. 

We note that the change in the energy induced by the unitary $U(\tau)$ in \eqref{neqstate} is given by
\be
\label{sumenergy}
\delta E = \delta E_{\cal C}  + \delta \widehat{E}_{\cal C},
\ee
where
\be
\begin{split}
&\delta E =  \langle \Psi | U(\tout)^{\dagger} H U(\tout) |\Psi \rangle - \langle \Psi | H  | \Psi \rangle, \\
&\delta E_{\cal C} = \langle \Psi | \ucausd_{\tout} H \ucaus_{\tout} |\Psi \rangle - \langle \Psi | H |\Psi \rangle, \\
&\delta \widehat{E}_{\cal C} = \langle \Psi | \ucauscompd_{\tout} H \ucauscomp_{\tout}  | \Psi \rangle -  \langle \Psi | H |\Psi \rangle. \\
\end{split}
\ee
This relation is rather intuitive. The action of unitaries on a typical state always tends to inject a small positive amount of energy. The two unitaries acting together above simply inject energy one
after the other. Nevertheless, we check the relation \eqref{sumenergy} explicitly in appendix \ref{appsumenergies}.

But we have already proved in section \ref{secfullads} that 
\be
\begin{split}
&\langle \Psi | \ucausd_{\tout} \phicaus(t_1, \rtor_1, \Omega_1) \ldots \phicaus(t_n, \rtor_n, \Omega_n) \ucaus_{\tout} | \Psi \rangle - \langle \Psi | \phicaus(t_1, \rtor_1, \Omega_1) \ldots \phicaus(t_n, \rtor_n, \Omega_n)  | \Psi \rangle \\ &= \langle \Psi | \ucausd_{\tout} \phi(t_1, \rtor_1, \Omega_1) \ldots \phi(t_n, \rtor_n, \Omega_n) \ucaus_{\tout} | \Psi \rangle - \langle \Psi | \phi(t_1, \rtor_1, \Omega_1) \ldots \phi(t_n, \rtor_n, \Omega_n)  | \Psi \rangle \\ &\leq 2 \sqrt{\beta \delta E_{\cal C}} \sigma,
\end{split}
\ee
where $\sigma$ is the deviation of the product of fields as defined in \eqref{stddev}. Here we have used the fact that for correlators in an equilibrium state involving insertions only in the causal patch, correlators of $\phicaus$ are the same as correlators of $\phi$.

But since $\delta E_{\cal C} \leq \delta E$, we see using \eqref{causalpatchdef} that we automatically  have
\be
\label{causalbornrule}
\begin{split}
&\langle \Psi | U(\tout)^{\dagger} \phicaus(t_1, \rtor_1, \Omega_1) \ldots \phicaus(t_n, \rtor_n, \Omega_n) U(\tout) | \Psi \rangle - \langle \Psi |  \phicaus(t_1, \rtor_1, \Omega_1) \ldots \phicaus(t_n, \rtor_n, \Omega_n) | \Psi \rangle \\ &\leq 2 \sqrt{\beta \delta E} \sigma.
\end{split}
\ee

This proves that if we use the fields $\phicaus$ then  the inequality \eqref{bornruleineq} is satisfied  even in spontaneously excited states, and correlators in the causal patch are stable under low energy excitations. The result \eqref{causalbornrule} 
is the one of the main reasons to believe that the operators \eqref{causalpatchdef} are the right operators to use in a causal patch.

An example may help to clarify the mechanism by means of which this stability under low-energy excitations is achieved. In the state $|\psine \rangle$, let us consider three causal patches ${\cal C}, {\cal C}', {\cal C}''$ where $\tcaus \ll \tcaus' < \tau < \tcaus''$.

The description using $\phi_{\cal C''}$ is appropriate for an observer who jumps in after the time $\tau$.\footnote{Here we are using the ``experience of the observers'' as shorthand to describe the results of correlators computed using \eqref{causalpatchdef} for the three causal patches.} This observer sees the full unitary $U(\tau)$ but he can only reach a limited ``depth'' (in terms of the maximum $\ukrus$ coordinate) inside the black hole, and so the correlators he measures all obey \eqref{bornruleineq}. Such an observer  may also naively {\em infer}  that the observer who jumped in at time $\tcaus$ had a strong interaction with a high-energy shock wave. 

But the description using the fields $\phi_{\cal C''}$ is a poor description for the observer who jumps into the bulk around $\tcaus$ because ${\cal C}''$ contains only a very small part of his worldline behind the horizon.   In the description provided by the fields $\phicaus$, the state $|\psine \rangle$ is essentially an equilibrium state and the observer simply falls through the horizon without encountering any particles.

The description using $\phicausprime$ is appropriate for an observer who jumps into the black hole, a little before the excitation appears on the boundary.  This observer sees only part of the full unitary. So, the description using $\phicausprime$ is slightly out of equilibrium, but this observer also does not observe any strong excitation at the horizon.

Thus we see that all three descriptions obey \eqref{bornruleineq} but for different reasons. This is similar to the spirit of the original proposal of  black hole complementarity \cite{'tHooft:1984re, Susskind:1993mu,Susskind:1993if,Kiem:1995iy} that suggested that we should use different  descriptions for the observer who stays outside the black hole and the infalling observer. Here, we have attempted to sharpen this idea by specifying the precise local correlation functions that are appropriate for a specific causal patch.

\subsection{Overlapping patches \label{subsecoverlap}}
\begin{figure}[!h]
\begin{center}
\includegraphics[height=0.3\textheight]{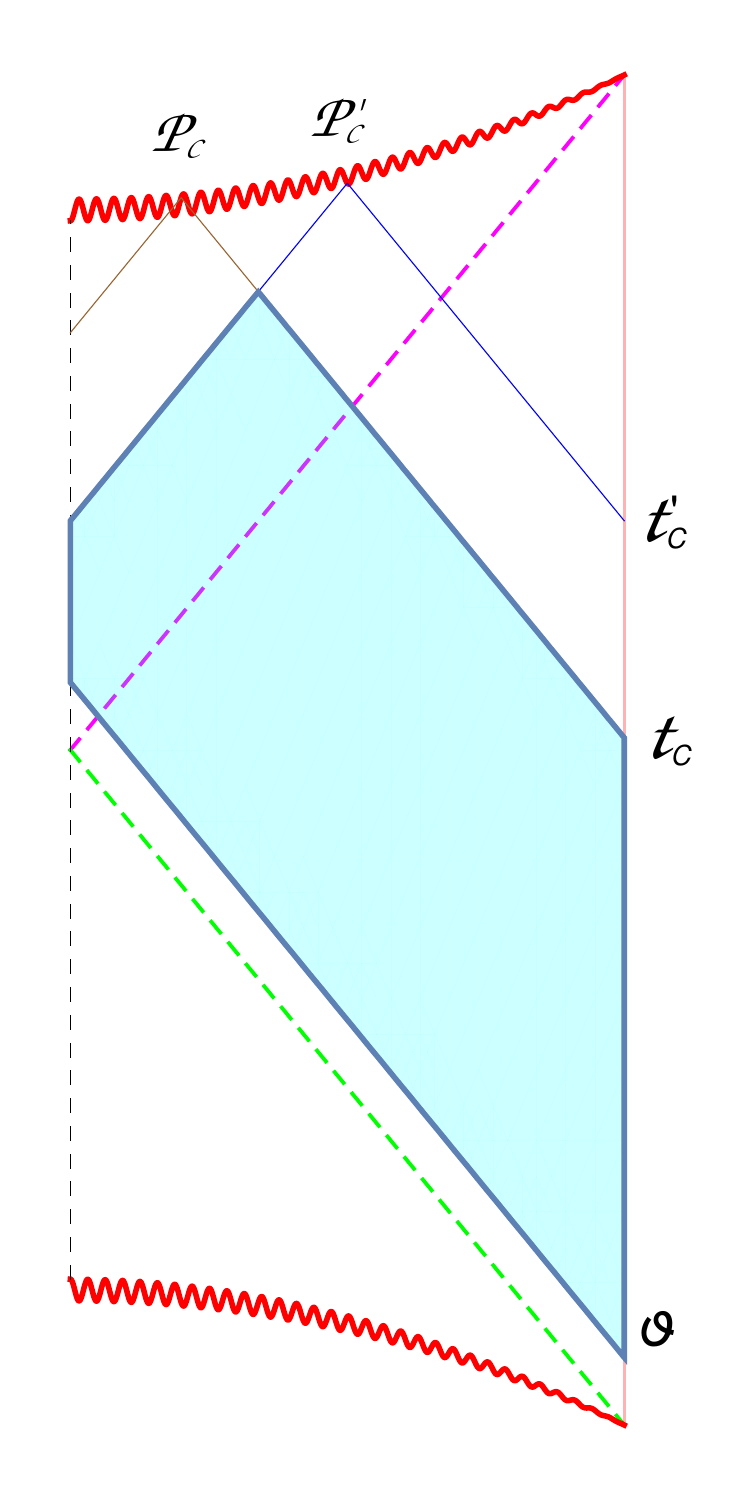}
\caption{\em The overlap of two causal patches \label{figoverlap}}
\end{center}
\end{figure}

We now turn to another aspect of our proposal: a single spacetime point belongs to multiple causal patches: figure \ref{figoverlap} shows the overlap of two causal patches ${\cal C}$ and ${\cal C'}$. Now if a point  belongs to both patches, we may ask:  ``what is the correct field to use at such a point? Is it $\phicaus$ or $\phicausprime$?'' 

We remind the reader that this question is not relevant in equilibrium states or even in non-equilibrium states where the point $(t, \rtor, \Omega) \in  \wcaus \cap \wedge_{C'}$; correlators of the fields $\phicaus$ and $\phicausprime$ coincide in such cases because both of them coincide with $\phi$.

But these correlators do differ in a class of states. For example if we consider a state that has a spontaneous excitation on the boundary in between $\tcaus$ and $\tcaus'$,  and also consider points that are not in the intersection of the causal wedges $\wedge_C \cap \wedge_{C'}$ then the correlators given by the two fields may differ.  Note that this is not a contradiction: even though notionally both these operators correspond to the ``field'', $\phicaus$ and $\phicausprime$ are different operators and can have different correlators. We say that the operators $\phicaus$ and $\phicausprime$ are giving complementary, but internally self-consistent descriptions of the same physics. So, as long as we consistently use the description provided by ${\cal C}$ or consistently use the one provided by ${\cal C}'$ both descriptions are valid about the point $P$. We emphasize that cross-correlators of the form $\langle \psine| \phicaus(t_1, \rtor_1, \Omega_1) \phicausprime(t_2, \rtor_2, \Omega_2) | \psine \rangle$ do {\em not} have any physical significance in this situation.

For example, if we act on  the state $|\psine \rangle$ with an additional unitary so that $U_{\text{obs}} |\psine \rangle$ can be interpreted as the state plus an observer, then our proposal allows us to reconstruct the bulk geometry including the observer in a causal patch. The description in terms of $\phicaus$ and the description in terms of $\phicausprime$ give internally consistent but complementary descriptions of the experience of the observer. 

What is important for us, is that provided we use each set of correlators in its respective domain of validity then {\em each} of these descriptions obeys effective field theory because of \eqref{causaleom}, the right relation with boundary fields because of \eqref{causalgkpw} and the constraints of statistical mechanics because of \eqref{causalbornrule}.

The reader may be a little concerned because in empty AdS, or flat-space quantum field theory, we do not usually have multiple possible descriptions of bulk physics.
The reason that this ambiguity appears here is as follows. Usually, in quantum field theory we proceed with the following tacit operational understanding. First, we do a set of experiments to fix the operators that we intend to use. Then in future experiments, we always tacitly refer back to this calibration. 

The contrast of the setting that we are considering with  empty AdS may help clarify this point. In empty AdS, if a bulk observer observes the boundary for a time band of length larger than $\pi$ (in units where the light-crossing time of AdS is $\pi$), and calibrates his fields to satisfy \eqref{causalgkpw} then this completely removes the ambiguity in any of his future measurements. We can make this precise by recalling the transfer function in empty AdS.
\begin{equation}
\label{globalrec}
\phi(t, r, \Omega) = \sum_{n, \ang}  \op_{n, \ang} e^{-i (2 n + \ang + \Delta) t} Y_{\ang}(\Omega) \chi^{\Omega}_{n, \ang}(r) + \text{h.c}, 
\end{equation}
where $\chi_{n, \ang}^{\Omega}(r)$ is a specific hypergeometric function that the reader can find in section 4 of \cite{Banerjee:2016mhh}. Now the key point here is that the modes $\op_{n,\ang}$ may be defined entirely in the causal past of the point $(t, r, \Omega)$ through
\be
\op_{n, \ang} = \int_{I} d t \int_{S^{d-1}} d^{d-1} \Omega \,  \op(t, \Omega) e^{i ( 2 n + \ang + \Delta) t} Y_{\ang}^*(\Omega),
\ee
where $I$ is any interval of length $\pi$ that can be placed entirely in the past of $(t, r, \Omega)$. So, if the bulk observer has ensured that his field variable obeys \eqref{causalgkpw} on the interval $I$ then this leaves no ambiguity in his future measurements.

On the other hand, the transfer function \eqref{transout} is not {\em causal}. Unlike empty AdS it cannot be written in a form that involves boundary modes only in the past. Therefore the definition of $\phi$ in terms of boundary operators is necessarily teleological and not causal in the presence of a black hole. And we see from \eqref{causalpatchdef} that the fields $\phicaus$ and $\phi$ differ only in their response to {\em future} spontaneous excitations.  So, at a point in the bulk from a causal perspective there is no reason to privilege the description in terms of $\phi$ over the description in terms of $\phicaus$ of $\phicausprime$.

Nevertheless, we could still ask: ``what will the bulk observer see subjectively?'' This is an imprecise question. 
Often, in situations like this, the question of what the observer will ``see'' is answered by simply picking the most convenient description. For example, consider an observer who couples himself to a system that is described by $N=4$ Yang-Mills theory at large 't Hooft coupling. Then, one description, of course, is that the observer turns into some mixture of strongly coupled glueballs. But usually we simply declare that the observer will ``perceive himself'' as falling into AdS$_5$. Both these descriptions are correct, but the description in terms of gravity is more convenient.

This happens in the case of causal patches as well.  For example, consider once again the non-equilibrium state $|\psine \rangle$ with $ \tout \gg \tcaus$. Then 
the correlators of $\phicaus$ are  equilibrium correlators.  On the other hand, if we consider the causal patch ${\cal C}'$ with $\tcaus' \sim \tout$ then correlators of the field $\phicausprime$ are out of equilibrium since in this patch the effect of the unitary is easier to see. If we consider a freely falling observer jumps into the bulk around $\tcaus$ and continues onto the singularity, then clearly the description in terms of ${\cal C}$ is more convenient since it allows us to describe a larger part of the observer's worldline and involves an undisturbed Schwarzschild black hole.

The general question of which description is the ``preferred'' one  in an arbitrary state goes beyond the scope of this paper. However, we mention that there is an interesting sense in which this question may be made well-defined.\footnote{I am grateful to R. Loganayagam for several discussions on this point.} Given a state $|\psine \rangle$ and a point $P$ can we find the causal patch ${\cal C}_{P}$ so that the physics described using the fields $\phi_{\cal C_P}$ is as ``classical'' as possible \cite{joos2013decoherence} around $P$? If so, it may be useful to declare that the fields $\phi_{\cal C_P}$ define the ``subjective experience'' of the observer and that this description is the ``preferred description.'' However to explore this question further requires a quantification of the notion of ``more classical'' and a more precise dynamical description of an observer, and we do not address it in this paper.

\section{Conclusion \label{secconclusion}}

In this paper we described and resolved the paradox of low energy excitations about black holes. The paradox arises because a robust property of statistical mechanics appears to be in conflict with properties of the black hole geometry. Statistical mechanics tells us that low energy excitations in thermal systems have small effects on physical observables. This property can be quantified, and we did so in \eqref{bornruleineq}. A paradox arises in the black hole geometry, because a naive analysis suggests that a locally large distortion of the geometry near the horizon should raise the AdM energy by only a small amount. Another way to understand this is that if we consider a state that displays an excitation on the boundary at some time, and back-calculate the geometry in the past, then this geometry has a blueshifted excitation near the horizon.

We showed how this paradox was resolved by considerations of causality. To implement this constraint, our analysis in this paper focused on the properties of AdS correlators in position-space. These correlators provide insights that are sometimes missed in the usual frequency-mode analysis of these problems. In our case, we found that if we consider a correlator where the excitation and insertions all fit in a single causal patch, then a delicate combination of factors ensures that this correlator obey the standard constraints of statistical mechanics. This not obvious, a priori, since the operators inside the black hole are state-dependent. Although  we established this by means of a detailed calculation in  in section \ref{secfullads},   it is not difficult to state the final result.

The most common reason that a black hole state may be excited at late times is because we start with an equilibrium state and then turn on a source on the boundary. In such a situation, the past of the state is just an ordinary equilibrium state and so the blueshifted excitation that could have created a problem does not exist. 

A far more unlikely but technically more interesting situation arises if we consider a state that displays  a spontaneous excitation at some point of time. Now we may consider two observers --- one of them jumps into the black hole at an early time, while the other waits near the boundary to confirm that there is indeed a late-time excitation and then jumps into the black hole. The remarkable fact that we pointed out in this paper is as follows. If the first observer follows a trajectory that allows him to meet with the second observer, then he cannot go too ``deep'' into the black hole. But, on the other hand, the first observer can only detect the blueshifted excitation by freely falling through the horizon for at least some time and probing the geometry in the interior to some depth. It turns out the constraint imposed by the fact that he must meet the second observer precisely prevents him from probing the black hole deep enough to be able to detect the excitation. 

These conclusions already resolve the paradox of low energy excitations in the vast majority of states. Starting in an equilibrium state, no combination of observers can excite the state on the boundary and observe a correlator that violates \eqref{bornruleineq}. But, even in spontaneously excited states no combination of observers can verify the presence of the excitation and also observe a bulk correlator that shows a violation of \eqref{bornruleineq}.  

Spontaneously excited states are exponentially unlikely. Nevertheless, we may ask what happens in such a state if an observer simply continues to fall freely through the horizon without bothering about meeting the second observer who waits on the boundary to verify the presence of the excitation. In this situation, the unitary operator that produces the excitation does not fit into the causal patch of the first observer. This is the situation that we considered in section \ref{seccomplement}. Here we proposed that it is consistent to use a description where the first observer only sees the part of the unitary that fits on the boundary of his causal patch and, therefore, using the results of section \ref{secfullads} does not see an anomalously large effect.  This description is not consistent with the experience of the second observer. But now since the worldlines of the two observers do not fit into a single patch, the descriptions that we must use for them are complementary and do not have to be reconciled.

In a sense, such a notion of complementarity is inbuilt into AdS/CFT and was used tacitly in \cite{Papadodimas:2013wnh,Papadodimas:2013jku}. This is because, if we wait for an exponentially long time, every pure state will eventually see a spontaneous excitation. If we take the bulk geometry in the presence of this excitation seriously,  then the geometry involves an extraordinarily blue-shifted shock wave that lurks close to the horizon for a very long time, only to emerge much later, hit the boundary and fall back into the black hole. Evidently, a more sensible perspective is simply to put an early and late time cutoff on the boundary and ignore these excitations that occur beyond these cutoffs.  Our proposal here simply makes this more precise and generalizes this notion of complementarity to settings where the spontaneous excitation is just slightly in the future of the causal patch of interest, rather than being exponentially far away. 

Although we have now proposed a boundary to bulk map that ensures that a description in terms of fields restricted to a single causal patch is consistent, how do we know that these are the fields that the infalling observer will ``observe''? This is also an open question for the construction of \cite{Papadodimas:2013wnh,Papadodimas:2013jku}. The mirror operators provide a consistent description of the interior, but why are they the right operators to use? So, in a strict sense, the results of this paper should be added to the results of \cite{Papadodimas:2013wnh,Papadodimas:2013jku} as an existence proof: it is possible to find a consistent description of the black hole interior that avoids various potential paradoxes and ensures that the horizon is smooth in generic states.

To go further, we may need to construct a dynamical model of the observer and then analyze the full system including the observer. In a rough sense, the situation here is analogous to the situation that arises in quantum field theory in curved spacetime. There it was realized that particle number was an ambiguous concept. However, by constructing a dynamical model of a particle detector, Unruh \cite{Unruh:1976db} and DeWitt \cite{DeWitt:1980hx} were able  to specify, among the infinity of possible definitions of particle number, which definition would be most convenient for an observer moving along a given trajectory. Here, we are asked to go a step further, and specify which ``field variable'' would be most convenient for an infalling observer to use. This is a very interesting question, and we do not address it in this paper, although we hope to return to it in future work.

There are various technical questions that should be amenable to simple analysis. For example, while we have dealt with pure states here, it would be interesting to understand situations where the CFT is entangled with another system. The only potential complication is the interplay between the superposition of geometries and the causal structure. However, geometries that have significant overlap differ only by quantum fluctuations, and so have the same causal structure; geometries with different causal structures have an exponentially small overlap and therefore behave like a classical statistical mixture.  So, it should not be difficult to extend these results to entangled states in the spirit of \cite{Papadodimas:2015jra}. 

Next, our analysis is currently only a leading order analysis, and it would be quite interesting to understand how it is modified by ${1 \over N}$ corrections. In gravity, these corrections prevent any notion of microcausality, beyond leading order,  and so it would be interesting to understand how our analysis here that relies strongly on the light-cone structure of the geometry must be generalized. We hope to address this question in future work.

\section*{Acknowledgements} I am grateful to  Ben Freivogel,  Daniel Harlow, Tom Hartman, Daniel Jafferis, R. Loganayagam, Juan Maldacena, Gautam Mandal, Shiraz Minwalla,  Donald Marolf, Lubos Motl,  David Poland,  Joseph Polchinski, Douglas Stanford, Sandip Trivedi, Erik Verlinde,  and all the members of the string theory groups at ICTS-TIFR and the Indian Institute of Science (Bangalore) for useful discussions. I am  especially grateful to Kyriakos Papadodimas for many helpful conversations and discussions. I am grateful to Tom Hartman, Juan Maldacena and Lubos Motl for comments on a draft of this manuscript. I would like to thank the Kavli Institute of Theoretical Physics (Santa Barbara), the Tata Institute of Fundamental Research (Mumbai), the University of Amsterdam, the National Tsing Hua University that hosted the 8\text{th} Taiwan String Workshop,    the Birla Institute of Technology and Science (Pilani), and the Harish-Chandra Research Institute (Allahabad) for their hospitality while this work was being completed.
\appendix
\section*{Appendix}
\section{Modes in the BTZ black hole}
Here, we explicitly work out the wave functions in the BTZ black hole and verify the claims made in section \ref{secfullads}. 

The metric of the BTZ black hole is just \eqref{bhmetric}, at $d = 2$
\be
ds^2 = -(r^2 - \rhor^2) dt^2 + (r^2 - \rhor^2)^{-1} dr^2 + r^2 d \theta^2.
\ee
Consider the bulk  wave equation $(\Box - m^2) \phi = 0$, and to be consistent with the conventions of section \ref{secfullads},  we take the ansatz $\phi_{\omega, \ang} = r^{1-d \over 2}\chi_{\omega, \ang}(r) e^{-i \omega t} e^{i \ang \theta}$. The equation for $\chi$ then becomes
\be
{1 \over r} \partial_r (r^2 - \rhor^2) r \partial_r r^{1 - d \over 2}\chi_{\omega, \ang}(r)  + \big[ {\omega^2 \over r^2 - \rhor^2} - {\ang^2 \over r^2} - m^2 \big] r^{1 - d \over 2} \chi_{\omega, \ang}(r) = 0.
\ee
It is convenient to work directly with $r$ rather than the tortoise coordinate for the BTZ black hole. Near the boundary, the radial wave equation is solved by a combination of hypergeometric functions
\be
\begin{split}
r^{1 - d \over 2} \chi_{\omega, \ang}(r) &=r^{-\Delta}  {\rhor}^{\Delta - {i \omega \over \rhor} } \left(1-{\rhor^2 \over r^2}\right)^{-\frac{i \omega }{2 \rhor}} \Big[A_1 \, _2F_1\big({\Delta \over 2} + i \frac{\ang - \omega }{2 \rhor},{\Delta \over 2} - i \frac{ (\ang+ \omega )}{2 \rhor};\Delta
   ;\frac{\rhor^2}{r^2}\big)  \\ &+ A_2 \rhor^2 r^{2 \Delta-2 } \, _2F_1\big(1 - {\Delta \over 2} + i \frac{ \ang- \omega }{2 \rhor},1 - {\Delta \over 2} -i \frac{ \ang+\omega }{2 \rhor};2-\Delta ;\frac{\rhor^2}{r^2}\big)\Big].
\end{split}
\ee
By imposing normalizability at the boundary, we see that  $A_2 = 0$. To ensure that the field scales like $r^{-\Delta}$ near the boundary with no other factors, we set  $A_1 = \rhor^{{i \omega \over \rhor} - \Delta}$

Next recall the series expansion of the hypergeometric function
\be
 _2F_1\big(a, b; c; x) = \sum_{n=0}^{\infty} {(a)_n (b)_n \over (c)_n} {x^n \over n!},
\ee
where $(a)_n \equiv a (a + 1) \ldots (a + n - 1)$. From this expansion we see that, as claimed, the function $\chi_{\omega, \ang}$ has no poles at finite values of $\omega$ and for $\rhor < r < \infty$.

By using hypergeometric transformations, we can put the solution in a form appropriate for the near-horizon expansion. We can rewrite
\be
\begin{split}
r^{1 - d \over 2} \chi_{\omega, \ang}(r) &= 
A_1 \Gamma (\Delta ) \rhor^{-i \omega \over \rhor} \left(\frac{r}{\rhor}\right)^{-\frac{i \ang}{\rhor}} \\
&\times   \Bigg[\frac{\left(\frac{r^2}{\rhor^2}-1\right)^{\frac{i \omega }{ 2 \rhor}} \Gamma \left(-\frac{i \omega }{\rhor}\right) \, _2F_1\left({\Delta \over 2} -i \frac{(\ang-\omega )}{2
   \rhor},1 - {\Delta \over 2} - i \frac{\ang +  \omega }{2 \rhor};\frac{i \omega }{\rhor}+1;1-\frac{r^2}{\rhor^2}\right)}{\Gamma \left(\frac{i \ang+\rhor \Delta -i
   \omega }{2 \rhor}\right) \Gamma \left(-\frac{i (\ang+i \rhor \Delta +\omega )}{2 \rhor}\right)} \\ &\hphantom{\times   \Bigg[}
+ \frac{\left(\frac{r^2}{\rhor^2}-1\right)^{\frac{-i \omega }{ 2 \rhor}} \Gamma \left(\frac{i \omega }{\rhor}\right) \, _2F_1\left(1 - {\Delta \over 2} -i \frac{
   \ang+\omega }{2 \rhor},-{\Delta \over 2} - i \frac{ (\ang+\omega )}{2 \rhor};1-\frac{i \omega }{\rhor};1-\frac{r^2}{\rhor^2}\right)}{\Gamma
   \left(-\frac{i (\ang+i \rhor \Delta -\omega )}{2 \rhor}\right) \Gamma \left(\frac{i (\ang-i \rhor \Delta +\omega )}{2 \rhor}\right)}\Bigg].
\end{split}
\ee

Recall that in this case the tortoise coordinate is given simply by
\be
\rtor = {1 \over 2 \rhor} \log{r - \rhor \over r + \rhor}.
\ee
Therefore we see that \eqref{d0deltadef} holds and we can write
\be
r^{1 - d \over 2}\chi_{\omega, \ang}(\rtor) \underset{\rtor \rightarrow -\infty}{\longrightarrow}  \dzero \left(e^{i \omega \rtor}  +  e^{2 i \delta_{\omega, \ang}} e^{-i \omega \rtor} \right),
\ee
with
\be
\dzero =  \left({r \over \rhor} \right)^{-{i \ang \over \rhor}} 2^{i \omega \over \rhor} {\Gamma(\Delta) \over \rhor^{\Delta}} {\Gamma(-{i \omega \over \rhor}) \over \Gamma({\Delta \over 2} - {i \omega \over 2 \rhor} - {i \ell \over 2 \rhor}) \Gamma({\Delta \over 2} - {i \omega \over 2 \rhor} + {i \ell \over 2 \rhor})},
\ee
where we have now substituted the value of $A_1$ calculated above. The phase factors in \eqref{d0deltadef} is just
\be
e^{2 i \delta_{\omega, \ang}} = {D^0(-\omega) \over D^0(\omega)}.
\ee

As advertised, the function $D^0$ has zeroes when the gamma functions in the denominator have poles: ${-i \omega \over 2 \rhor} \pm {i \ang \over 2 \rhor}  + {\Delta \over 2} \in {\cal Z}^{-}$. Therefore the zeros are 
\be
\omega = \pm \ell - {i \Delta \rhor} - 2 i n \rhor,
\ee
for $n$ a positive integer. Clearly these occur only for $\text{Im}(\omega) < 0$, and indicate the positions of the quasinormal modes.  The function $D^0$ is regular in the upper half $\omega$ plane; using Stirling's approximation we can check that this is also true as we approach infinity through the upper-half plane. And we also see that for $\omega \in {\cal R}$, the phase $\delta$ is real.

Finally, defining $\zeta(r) = {r^{1 - d \over 2} \chi_{\omega, \ang}(r) ( \dzero )^{-1}} $ we can write
\be
\zeta(r) = \zeta^+(r) + e^{2 i \delta_{\omega, \ang}} \zeta^{-}(r),
\ee
where
\be
\zeta^{+} =\left(\frac{r}{\rhor}\right)^{-\frac{i \ang}{\rhor}} 
   {\left({r - \rhor \over r+ \rhor} \right)^{\frac{i \omega }{ 2 \rhor}}  \, _2F_1\left({\Delta \over 2} -i \frac{(\ang-\omega )}{2
   \rhor},1 - {\Delta \over 2} - i \frac{\ang +  \omega }{2 \rhor};\frac{i \omega }{\rhor}+1;1-\frac{r^2}{\rhor^2}\right)}.
\ee
is the function relevant for  the ``right moving'' wave. This is again consistent with our analysis in section \ref{secfullads}. Using the expansion of the hypergeometric function, we see that $\zeta^{+}(r)$ has poles only for ${\omega \over \rhor} = {\beta \omega \over 2 \pi} \in i {\cal Z}^+$, which is exactly what we expect. The same function for $r < r_0$ also describes the interior right moving mode; as expected, this has the same analytic properties as the right movers outside the horizon.

\section{Low energy unitaries \label{applowenergy}}

In this appendix, we prove \eqref{unitarythrough}. We note that with $V = e^{-\beta H \over 2} U e^{\beta H \over 2}$,
\be
\langle \Psi| V^{\dagger} V |\Psi \rangle = \langle \Psi|  e^{\beta H \over 2} U^{\dagger} e^{-\beta H} U e^{\beta H \over 2} | \Psi \rangle = {1 \over Z(\beta)} \tr(e^{-\beta H} e^{\beta H \over 2} U^{\dagger} e^{-\beta H} U e^{\beta H \over 2} ) = 1.
\ee
So, we will be done if we can show that
\be
\label{innervwithu}
\langle \Psi | U^{\dagger} V |\Psi \rangle = 1 - \Or[\beta \delta E].
\ee

Our strategy is to expand the unitary
as 
\be
U = e^{i \theta \sum_{\omega} (\al_{\omega} + \al_{\omega}^{\dagger}}), 
\ee
where $\al_{\omega} \in \alset$ are elements of the algebra with definite energy so that $[H, \al_{\omega}] = - \omega \al_{\omega}$ plus possible $\Or[{1 \over N}]$ terms that we neglect.
First let us work out $\delta E$ under this unitary. We see that
\be
U^{\dagger} H U = H - i \theta  \sum_{\omega} \omega (\al_{\omega} - \al^{\dagger}_{\omega}) - {\theta^2 \over 2} \sum_{\omega', \omega} \omega [\al_{\omega'} + \al_{\omega'}^{\dagger}, \al_{\omega} - \al_{\omega}^{\dagger}] + \Or[\theta^3].
\ee
Taking the expectation value in a typical state, we see that 
\be
\delta E = \langle \Psi| U^{\dagger} H U| \Psi \rangle - \langle \Psi | H | \Psi \rangle = {\theta^2 } \sum_{\omega} \omega  C^{\al}_{\omega},
\ee
where $\langle \Psi | [\al_{\omega}, \al_{\omega'}^{\dagger}] | \Psi \rangle \equiv C^{\al}_{\omega} \delta_{\omega \omega'}$. This appears because in an equilibrium state only expectation values of operators with ``zero energy'' are non-vanishing and we neglect terms of order $\theta^3$ and higher. 

Now we compare this with \eqref{innervwithu}. We see that
\be
\begin{split}
V &=  e^{-{\beta H \over 2}} U e^{\beta H \over 2} \\ &= 1  + i \theta \sum_{\omega} (e^{{\beta \omega \over 2}} \al_{\omega} + e^{-{\beta \omega \over 2}} \al_{\omega}^{\dagger}) - {\theta^2 \over 2} \sum_{\omega, \omega'} (e^{{\beta \omega \over 2}}\al_{\omega} + e^{-{\beta \omega \over 2}} \al_{\omega}^{\dagger}) (e^{\beta \omega' \over 2}\al_{\omega'} + e^{-{\beta \omega' \over 2}}\al_{\omega'}^{\dagger}) \\ &+ \Or[\theta^3],
\end{split}
\ee
while
\be
U^{\dagger} = 1 - i \theta \sum_{\omega} (\al_{\omega} + \al_{\omega}^{\dagger}) - {\theta^2 \over 2} \sum_{\omega, \omega'} (\al_{\omega} + \al_{\omega}^{\dagger}) (\al_{\omega'} + \al_{\omega'}^{\dagger}) + \Or[\theta^3].
\ee
Now we see that to first order in $\theta$
\be
\langle \Psi | U^{\dagger} V | \Psi \rangle = 1 - i \theta  \sum_{\omega} \langle \Psi |  (e^{-\beta \omega} - 1)  \al_{\omega} + (e^{\beta \omega \over 2} - 1) \al_{\omega}^{\dagger} | \Psi \rangle.
\ee
the first order term vanishes because the only operators with a non-zero expectation value are those with $\omega = 0$, but at that point the coefficients of these operators in the sum vanish. 
Moving to the next term, which is  second order in $\theta$ we see that
\be
\langle \Psi | U^{\dagger} V | \Psi \rangle = 1 + \theta^2 \sum_{\omega} \Big[ G_{\omega}^{\al} (e^{-{\beta \omega \over 2}} - 1) + (G_{\omega}^{\al} - C_{\omega}^{\al}) (e^{\beta \omega \over 2} - 1) \Big],
\ee
where  $\langle \Psi | \al_{\omega} \al^{\dagger}_{\omega'} | \Psi \rangle = \langle \Psi | \al^{\dagger}_{\omega} \al_{\omega'} | \Psi \rangle  + C^{\al}_{\omega} \equiv G_{\omega}^{\al} \delta_{\omega, \omega'}$.
We can use the KMS relations to convert this to a form that is similar to $\delta E$. We note that, first
\be
\tr(e^{-\beta H} \al_{\omega}^{\dagger} \al_{\omega}) = e^{-{\beta \omega}} \tr(e^{-\beta H} \al_{\omega} \al_{\omega}^{\dagger}),
\ee
by using the cyclicity of the trace and the commutator of the negative frequency operator with the Hamiltonian. 
Second, from here we see that
\be
\tr(e^{-\beta H} [\al_{\omega}, \al_{\omega}^{\dagger}]) = (1 - e^{-\beta \omega}) \tr(e^{-\beta H} \al_{\omega} \al_{\omega}^{\dagger}),
\ee
or $C_{\omega} = (1 - e^{-\beta \omega}) G_{\omega}$. 
Therefore
\be
\begin{split}
&\langle \Psi | U^{\dagger} V |\Psi \rangle = 1 - \theta^2  \sum_{\omega} (1 - e^{-{\beta \omega \over 2}})^2 G_{\omega} \\
&= 1 - {\theta^2} \sum_{\omega}  {(1 - e^{-{\beta \omega \over 2}})^2 \over 1 - e^{-\beta \omega}} C_{\omega}.
\end{split}
\ee
Noting that $(1 - e^{-{\beta \omega \over 2}})^2 < \omega (1 - e^{-\beta \omega})$ for $\omega > 0$, we see that
\be
1 - \langle \Psi | U^{\dagger} V | \Psi \rangle < (\beta \delta E),
\ee
where we have used the fact that $\langle \Psi| [\al_{\omega}, \al_{\omega}^{\dagger}] |\Psi \rangle > 0$, and so both terms involve a sum over positive numbers, and the sum involving the change in energy is larger when examined term by term.

This proves the result \eqref{unitarythrough}. Since the result is intuitive and easy to state, it is possible that we have missed a simpler argument. However, we briefly mention one complication that leads us to reject some other seemingly simple arguments. We see that
\be
U^{\dagger} H U = H + \delta H,
\ee
where we already know that $\langle \Psi | \delta H | \Psi \rangle =  \delta E$, but we do not have much control over the operator $(\delta H)$ itself from this relation. As a result of this, even though the unitary $U$ raises the energy by $\delta E$, we see that  $e^{-{\beta H \over 2}} U e^{\beta H \over 2} \neq e^{-{\beta \delta E \over 2}} U$. We see that such a relation cannot hold because  unitary $U^{\dagger}$  {\em also} raises the energy, and so the conjugate of this relation is manifestly untrue $e^{{\beta H \over 2}} U^{\dagger} e^{-{\beta H \over 2}} \neq e^{{\beta \delta E \over 2}} U^{\dagger}$.

This tells us that the fact that $\delta H$ cannot be substituted by its expectation value is important. It is to get control over this operator that we have to examine the expansion of the unitary in terms of an exponential of components of definite frequency as we did above. 

\subsection{Multiple unitaries \label{appsumenergies}}
Using the results above, it is also easy to prove the result \eqref{sumenergy}. We expand both unitaries $\ucaus_{\tcaus}$ and $\ucauscomp_{\tcaus}$ 
as
\be
\begin{split}
\ucaus_{\tout} = e^{i \theta_1 \sum_{\omega} (\alcaus_{\omega} + \alcausd_{\omega})}; \quad\ucauscomp_{\tout} = e^{i \theta_2 \sum (\albarcaus_{\omega} + \albarcausd_{\omega})}.
\end{split}
\ee
We now see that
\be
\begin{split}
\delta E &= \langle \Psi | \ucauscompd_{\tout} \ucausd_{\tout} H \ucaus \ucauscompd | \Psi \rangle - \langle \Psi | H | \Psi \rangle\\
&= {\theta_1^2 } \sum_{\omega, \omega'} \omega \langle \Psi| \ucauscompd_{\tout} [\alcaus_{\omega}, \alcausd_{\omega'}]  \ucauscomp_{\tout}  | \Psi \rangle  +  {\theta_2^2 } \sum_{\omega, \omega'} \omega \langle \Psi| \ucausd_{\tout} [\albarcaus_{\omega}, \albarcausd_{\omega'}] \ucaus_{\tout} | \Psi \rangle + \Or[\theta_1^3] + \Or[\theta_2^3] \\ &= \delta E_{\cal C} + \delta \bar{E}_{\cal C}.
\end{split}
\ee
Here we have just used the fact that $\ucaus_{\tout}$ commutes with elements of $\alset$ whereas $\ucauscomp$ commutes with elements of $\bar{\alset}$.  
\bibliographystyle{JHEP}
\bibliography{references}
\end{document}